\documentclass[12pt, twoside]{uwthesis}[2012/06/19]
\usepackage{slashed}
\usepackage{braket}
\usepackage{amsmath}
\usepackage{bm}
\usepackage{float}
\usepackage{graphicx}
\usepackage{amsfonts}
\usepackage{simpler-wick}
\usepackage{subcaption}
\usepackage{url}

\begin{document}

\prelimpages

%
% ----- copyright and title pages
%

\Title{New Developments in Light-Front Nuclear Structure}
\Author{Dmitriy Nikolaevich Kim}
\Year{2026}
\Program{Physics}

\Chair{Gerald A. Miller}{Professor}{Physics}
\Signature{Aurel Bulgac}
\Signature{Silas Beane}

\copyrightpage

\titlepage

% ----- abstract

\setcounter{page}{-1}
\abstract{
Nuclear physics is the study of many-body systems of protons and neutrons and their constituent interactions. Historically, our understanding of
nuclear structure developed from low-energy scattering experiments, for which independent-particle (mean-field) models proved widely successful. Early
$A(e,e'p)$ measurements at NIKHEF and Saclay, however, revealed that this picture is incomplete, updating our theoretical description of the nucleus
toward a correlated many-body framework. In particular, these experiments---along with later high-momentum-transfer work at Jefferson
Lab---highlighted the importance of nucleon-nucleon short-range correlations (SRCs), two-nucleon configurations with high relative momentum and low
center-of-mass momentum. SRC phenomenology uses such configurations to explain the plateaus observed in ratios of inclusive electron-nucleus to
deuterium cross sections.

With current and forthcoming high-energy electron-nucleus experiments at Jefferson Lab and the Electron-Ion Collider, nuclear structure must once
again be updated---this time into a relativistic formulation suitable for such kinematics, which has not previously been carried out. This
dissertation develops nuclear structure in that direction, motivated by these future experiments. Building on existing tools from conventional nuclear
physics, we reformulate them in a relativistic, light-front-quantized framework. Our light-front nuclear structure calculations, adapted from density
functional theory, reproduce nuclear binding energies and shell structure well, and incorporate the physics of nucleon-nucleon SRCs through similarity
renormalization group techniques. Our results indicate that a purely nucleonic description of scattering is insufficient to capture inclusive electron-nucleus 
data and does not fully reproduce the plateaus seen at high Bjorken-$x_B$, pointing to the importance of inelastic final-state interactions that current SRC 
phenomenology does not account for.
}

% ----- contents & etc.

\tableofcontents
\listoffigures
%\listoftables  % I have no tables

% ----- acknowledgments

\acknowledgments{% \vskip2pc
  % {\narrower\noindent
I would like to express my sincerest appreciation to my advisor, Gerald A. Miller, who pushed me beyond what I believed was possible for myself. 
During the uncertain times of this research journey, his simple catchphrase—``keep going''—helped me realize that it is never over until I give up. 
I will hold onto this lesson for the rest of my life. His guidance fostered my critical thinking, and I can confidently say I now know what it takes 
to conduct independent research.

Beyond the academic realm, my family provided the foundation that made this journey possible. I owe a massive debt of gratitude to my mother, Yanna En, 
and my father, Nikolay Kim, for their unyielding support; I hope this achievement honors the countless sacrifices you have made for me. I also want to 
thank my older brother, Vladimir Kim, for being a guiding role model and for always putting up with my stubbornness. And to my grandparents—Leonid Kim, 
Ludmila Tegay, Leonid En, and Rima Kim—thank you for paving the initial road that started it all.
  % \par}
}

% ----- dedication

\dedication{\begin{center}To my family\end{center}}

% end of the preliminary pages

%
% ==========      Text pages
%

% ----- conventions
% !TEX root = ../uwthesis.tex
\chapter*{Conventions and Notation}
\addcontentsline{toc}{chapter}{Conventions and Notation}

Throughout this thesis the following conventions are adopted unless stated otherwise.

\paragraph{Units.} Natural units are used, with $\hbar = c = 1$. Energies, momenta, and masses are
expressed in fm or MeV/GeV, as appropriate.

\paragraph{Light-front coordinates.} Light-front coordinates follow the Lepage--Brodsky convention,
\begin{equation}
    x^\pm \equiv x^0 \pm x^3, \qquad \bm{x}^\perp = (x^1, x^2),
\end{equation}
with $x^+$ playing the role of light-front time and $x^-$ the longitudinal spatial coordinate. The
corresponding momenta are $p^\pm = p^0 \pm p^3$, with $p^-$ identified as the light-front energy.
The on-shell condition reads $p^+ p^- - (\bm{p}^\perp)^{2} = m^2$.

All Instant Form variables will be denoted with ``IF'' subscript or superscript. Boldface with the
subscript ``IF'' denotes spatial three-vectors, $\bm{p}_{\rm IF} = (p^1, p^2, p^3) = (\bm{p}^\perp,
p^3)$, while boldface symbols without subscript denote light-front three-vectors, $\bm{p} = (p^+,
\bm{p}^\perp)$. The energy component is conventionally listed first in both forms:
\begin{equation}
    p^\mu_{\rm IF} = (p^0, \bm{p}_{\rm IF}) \quad \text{(instant form)}, \qquad p^\mu = (p^-, \bm{p}) \quad \text{(front form)}.
\end{equation}

\paragraph{States.} Plane-wave nucleon states are defined by
\begin{equation}
    \sqrt{2p^+}\hat{a}^{\dagger}(\bm{p},\sigma,\tau)\ket{0} = \ket{\bm{p},\sigma,\tau}.
\end{equation}
When dealing with multi-nucleon states we will take the convention that the creation operators are
indexed from smallest to largest, i.e.
\begin{equation}
    \ket{\bm{p}_1,\sigma_1,\tau_1;\bm{p}_2,\sigma_2,\tau_2} = \left(\sqrt{2p_1^+}\hat{a}^{\dagger}(\bm{p}_1,\sigma_1,\tau_1) \right)\left(\sqrt{2p_2^+}\hat{a}^{\dagger}(\bm{p}_2,\sigma_2,\tau_2) \right) \ket{0}.
\end{equation}
Furthermore, momenta $\bm{p}$, spin $\sigma$, and isospin $\tau$ labels will sometimes all be
packaged into a non boldfaced momentum label. For instance,
\begin{equation}
    \begin{gathered}
        \ket{\bm{p},\sigma,\tau} =\ket{p}, \\
        \hat{a}(\bm{p},\sigma,\tau) = \hat{a}(p),\\ \ket{\bm{p}_1,\sigma_1,\tau_1;\bm{p}_2,\sigma_2,\tau_2} = \ket{p_1,p_2},
    \end{gathered}
\end{equation}

\paragraph{Integrals.} Momentum integrals carry factors of $(2\pi)^{-3}$. The light-front Lorentz
invariant phase space is
\begin{equation}
    \int [dp] = \int \frac{dp^+d\bm{p}^\perp}{(2\pi)^3 2p^+}\theta(p^+),
\end{equation}
the $\theta(p^+)$ inside the integral, which enforces $p^+\geq0$, will typically be dropped and it
is implicitly assumed that the domain of $p^+$ is positive definite, unless explicitly said so.

\paragraph{Metric.} The mostly-minus signature, $g_{\mu\nu} = \mathrm{diag}(+1,-1,-1,-1)$, of the
Minkowski metric is used. Greek indices $\mu,\nu,\ldots$ run over $0,1,2,3$ for instant form and
$-,+,1,2$ for front form.

\paragraph{Dirac spinors.} Lepage-Brodsky Dirac spinor representation is used
\begin{align}
    u(\bm{p}, \sigma) &= \frac{1}{\sqrt{p^+}} \left( p^+ + \beta m + \bm{\alpha}^\perp \cdot \bm{p}^\perp \right) \times
    \begin{cases}
        \chi(\uparrow), & \text{for } \sigma = +1, \\
        \chi(\downarrow), & \text{for } \sigma = -1,
    \end{cases} \\
    v(\bm{p}, \sigma) &= \frac{1}{\sqrt{p^+}} \left( p^+ - \beta m + \bm{\alpha}^\perp \cdot \bm{p}^\perp \right) \times
    \begin{cases}
        \chi(\downarrow), & \text{for } \sigma = +1, \\
        \chi(\uparrow), & \text{for } \sigma = -1.
    \end{cases}
\end{align}
The two $\chi$-spinors are
\begin{equation}
    \chi(\uparrow) = \frac{1}{\sqrt{2}} \begin{pmatrix} 1 \\ 0 \\ 1 \\ 0 \end{pmatrix},
    \qquad
    \chi(\downarrow) = \frac{1}{\sqrt{2}} \begin{pmatrix} 0 \\ 1 \\ 0 \\ -1 \end{pmatrix}.
\end{equation}

\paragraph{Dirac matrices.} The Dirac matrices satisfy $\{\gamma^\mu, \gamma^\nu\} = 2g^{\mu\nu}$,
with $\gamma^5 = i\gamma^0 \gamma^1 \gamma^2 \gamma^3$. The Dirac representation is used throughout.

\textpages

% ========== Chapter 1

\textpages

% ========== Chapter 1

% ── chapters ──────────────────────────────────────────
% !TEX root = ../uwthesis.tex
\chapter {Introduction}\label{chap:introduction}

Nuclear physics is the study of the atomic nucleus, its constituents, and the interactions between
them. The discipline is conventionally partitioned, on the basis of the energy scales involved, into
the two sub-fields of low- and high-energy nuclear physics. Our understanding of nuclear structure
began at low energies, with Rutherford's gold-foil experiment published in 1911~\cite{Rutherford:1911zz}. By
firing alpha particles at thin gold foil, Rutherford, Geiger, and Marsden discovered that at the
center of the atom lies a small, dense, positively charged object: the nucleus. Twenty-one years
later, Chadwick's discovery of the neutron~\cite{Chadwick:1932ma}, together with the birth of
quantum mechanics in the early twentieth century, set the stage for microscopic theoretical
descriptions of nuclear structure in which protons and neutrons (nucleons) serve as the fundamental
degrees of freedom. Early models---Gamow, Bohr, and Wheeler's liquid-drop model~\cite{Bohr:1939ej}
and the nuclear shell model of Mayer and 
Jensen~\cite{Mayer:1948zz,Mayer:1949pd,Haxel:1949fjd}---achieved remarkable success in describing a
wide range of nuclear observables, including binding energies, charge radii, and shell structure.
Modern low-energy nuclear physics builds on these foundations with sophisticated \emph{ab initio}
approaches such as Density Functional Theory \cite{Drut:2009ce,Shen:2019dls} and Quantum Monte Carlo
\cite{Carlson:2014vla}.

Theoretical progress has, throughout, depended on a close partnership with experiment. To study the
nucleus, Rutherford relied on alpha particles emitted by naturally radioactive substances such as
radium---a beam whose energy, intensity, and composition were fixed by the source and entirely
outside the experimenter's control. Recognizing this limitation, Rutherford began advocating 
for actively accelerating particles to interrogate the nucleus.
Inspired by this vision, his students John Cockcroft and Ernest Walton constructed the first
particle accelerator in 1932. By accelerating protons to 700--800~keV, they achieved the first
artificial nuclear disintegration, splitting a lithium atom into two alpha
particles~\cite{Cockcroft:1932zz,Cockcroft:1932vuu}. This milestone gave physicists unprecedented
control over subatomic probes and established the accelerator-driven experimental program that
continues to drive theoretical developments today.

The subsequent global enterprise of building ever more powerful accelerators culminated for nuclear
physics in 1968 at the Stanford Linear Accelerator Center (SLAC). Deep inelastic scattering (DIS)
of high-energy electrons off protons revealed the composite internal structure of the nucleon
itself~\cite{Bloom:1969kc}, establishing the field of high-energy nuclear physics with a new set of
fundamental degrees of freedom: quarks and gluons (partons). Within five years, Quantum
Chromodynamics (QCD) was formulated as the fundamental theory governing their
dynamics~\cite{Yang:1954ek,Gross:1973id,Politzer:1973fx,Gross:1973ju}. Understanding how nucleons
and complex nuclei emerge from these underlying quark and gluon interactions has since become an
overarching goal of the field. The short-distance behavior of the strong interaction is
perturbatively calculable and well understood; at the larger distances relevant to hadronic and
nuclear structure, however, QCD becomes highly non-perturbative. This non-perturbative regime---the
intersection of low- and high-energy nuclear physics---remains profoundly challenging and is a
primary focus of contemporary research.

The earliest experimental window into this regime was opened in 1983 by the European Muon
Collaboration (EMC). Performing DIS on iron and deuterium targets with
high-energy muons, they observed that the ratio of per-nucleon cross-sections exhibited a distinct
depletion at intermediate momentum fractions~\cite{EuropeanMuon:1983wih}. At the time, it was widely
assumed that high-energy partonic observables would be insensitive to low-energy nuclear
dynamics---that, at energies sufficient to resolve partons, the nucleus could be treated as a
collection of free nucleons. The ``EMC effect,'' as this depletion came to be known, sparked intense
theoretical investigation. Early approaches invoking conventional nuclear mechanisms such as Fermi
motion and pionic enhancements failed to fully reproduce the data, and despite more than forty years
of effort and over a thousand publications, the origin of partonic modifications in bound nucleons
remains a major open question.

Experimental momentum on this question is only growing. Future measurements at the Electron--Ion
Collider (EIC), which will accelerate nuclear targets to energies as high as
100~GeV/nucleon~\cite{Accardi:2012qut,AbdulKhalek:2021gbh}, and current experiments at Jefferson
Lab (JLab) colliding 12~GeV electrons against nuclei~\cite{Dudek:2012vr}, will probe quark
substructure in nuclei with unprecedented resolution. This experimental progress demands a parallel
theoretical advance: a description of the initial nuclear state suitable for high-energy processes.
Nuclear structure calculations have historically been performed by solving the Schr\"odinger
equation, but as the energy of the probe increases, relativistic effects can no longer be neglected.
Without a properly relativistic treatment, purely kinematic effects can easily be misinterpreted as
genuine signatures of new physics. Yet the calculation of nuclear wavefunctions within a consistent
relativistic framework remains largely unexplored.

Conventional instant form quantization, using $x^0$ as the time variable, does not adequately meet
this need because it yields frame-dependent descriptions of bound states \cite{Brodsky:2022fqy}. In
this formalism, a nucleus boosted to high momentum appears Lorentz contracted, and reproducing such
a deformed state requires a superposition over all excited states of the system. This contraction is
unphysical in the present context: it is a frame-dependent artifact rather than an intrinsic
property of the nucleus. The resulting calculations are both technically cumbersome---requiring the
explicit inclusion of excited nuclear states---and conceptually opaque, since frame-dependent
effects become entangled with the intrinsic dynamics one aims to study. Such ``fictitious dynamics''
and related pathologies of instant form-based bound-state calculations have historically led to
misleading conclusions, including apparent violations of the Gerasimov--Drell--Hearn sum rule in the
deuteron \cite{Brodsky:1968xc} and more recent efforts to extract time-independent,
three-dimensional densities for systems of relativistically moving constituents
\cite{Miller:2025zte}.

Light-front quantization, which uses $x^+ = x^0 + x^3$ as the time variable, offers a natural
resolution to these difficulties: its wavefunctions are invariant under boosts, enabling a
consistent, frame-independent description of nuclear structure. Despite this central advantage,
relativistic light-front descriptions of finite nuclei remain largely undeveloped. Developing such a
framework is essential for interpreting data from the high-energy nuclear experiments underway and
forthcoming at JLab and the EIC, and is the central motivation for the work presented in this
thesis.

The remainder of this thesis is organized as follows. \textbf{Chapter~\ref{chap:lightfront}}
provides an introduction into light-front quantization, with the guiding question: \emph{why should
low-energy nuclear physicists care?} Rather than offering a complete treatment of the mathematical
formalism---which is extensively documented in literature---the chapter aims to build qualitative
intuition for the importance of the light-front approach and to present the basic theoretical
machinery needed to perform calculations

The primary observable throughout this work is the inclusive electron--nucleus scattering
cross-section, and \textbf{Chapter~\ref{chap:scattering}} provides a theoretical overview of this
process, introducing the main theoretical objects of interest.

\textbf{Chapter~\ref{chap:dft}} begins the calculational portion of the thesis with light-front
density functional theory, the simplest realistic model of nuclear structure that provides a
pathway to computing inclusive electron--nucleus cross-sections. Here, we highlight a procedure for
obtaining light-front nuclear structure calculations directly from previously published instant form
results.

The independent-particle picture underlying DFT, however, proves inadequate for high-energy nuclear
processes. This shortcoming motivates \textbf{Chapter~\ref{chap:srg}}, which introduces the
similarity renormalization group (SRG) and its novel application to relativistic descriptions of
nuclear structure on the light-front. We show how incorporating the SRG modifies the cross-section
calculations of the previous chapter.

\textbf{Chapter~\ref{chap:fsi}} addresses the role of final-state interactions and finally,
\textbf{Chapter~\ref{chap:conclusion}} summarizes the results of this thesis.

% !TEX root = ../uwthesis.tex
\chapter{Light-Front Quantization}\label{chap:lightfront}

In 1949, Dirac published his foundational paper \emph{Forms of Relativistic Dynamics}, in which he
investigated distinct Hamiltonian formulations of relativistic quantum mechanics by exploring
different parameterizations of spacetime~\cite{Dirac:1949cp}. His central insight was that the
choice of spacetime foliation, while leaving physical observables unchanged, can offer significant
computational and conceptual advantages depending on the system under investigation---much as
spherical coordinates simplify spherically symmetric problems without altering their physical
content. Dirac identified three classes of foliations that cannot be related by Lorentz
transformations in any non-trivial way, which he termed ``Forms'': the Instant Form (IF), the Front
Form (FF), and the Point Form (PF). The IF, in which states are quantized on surfaces of constant
time $x^0=t$, is the most familiar and most widely adopted, aligning closely with our built-up
intuition about space and time. The FF, in which states are quantized on surfaces of constant time
$x^+ = x^0 + x^3$, departs from this intuition but has proven exceptionally powerful for high-energy
problems and serves as the primary framework of this dissertation. The PF, in which states are
quantized on a hyperboloid, has seen comparatively limited application.

This chapter aims to convey why formulating nuclear physics on the light front is essential for the
accurate study of high-energy nuclear reactions, which underpins the original calculations presented
later in this work. It highlights that the FF is the only relativistic Hamiltonian framework in
which (i) intrinsic bound-state wavefunctions are invariant under the boosts, and (ii) the off-shell
extrapolation of nucleon-level subprocesses is free of contamination from the probe. Both are
necessary to compute high-energy nuclear cross sections, and only the FF satisfies both. The
emphasis throughout is on qualitative motivation drawn from fundamental properties of special
relativity and quantum mechanics, together with the theoretical formulation needed to support the
calculations in this thesis. Dirac's FF has been independently rediscovered and rebranded across the
literature as the Infinite Momentum Frame, Null-Plane Quantization, and Light-Cone Quantization; we
adopt the name Light-Front (LF) Quantization here. For comprehensive pedagogical reviews, the reader
is directed to Ref.~\cite{Brodsky:1997de} and references therein; for applications to nuclear
physics specifically, see Refs.~\cite{Frankfurt:1981mk,Frankfurt:1988nt,Miller:2000kv}.

\section{Example: Lorentz Boosting the Deuteron}
\begin{figure}
    \centering
    \includegraphics[width=1\linewidth]{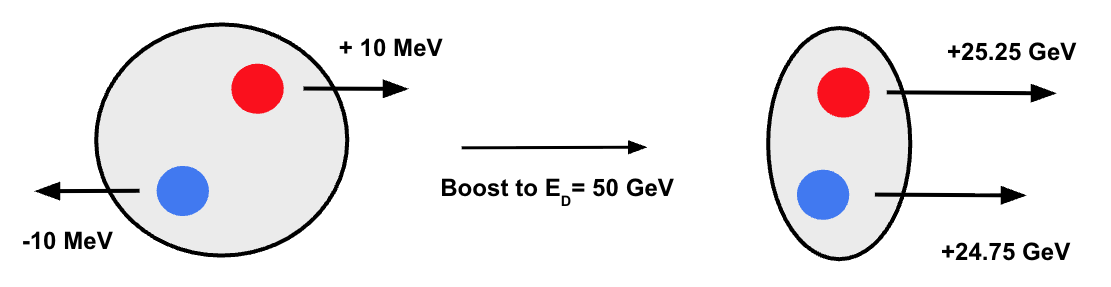}
    \caption{(color online) Cartoon picture of a deuteron configuration where the proton (red) and neutron (blue) have back-to-back momenta of 10 MeV along the $z$-axis. Boosting this configuration to an energy of 50 GeV increases both the absolute and the relative momenta of the constituents.}
    \label{fig:deuteron_cartoon}
\end{figure}

To motivate the LF formulation of nuclear dynamics, consider the interplay between quantum mechanics
and special relativity in a simple system under IF quantization: a deuteron at rest, with its proton
and neutron aligned along the $z$-axis at a relative momentum of 10 MeV
(Fig.~\ref{fig:deuteron_cartoon}). Anticipating the kinematics of the EIC,
where nuclear beams will reach hundreds of GeV per nucleon, apply a longitudinal Lorentz boost of 50
GeV. The individual nucleon momenta shift to approximately 24.75 and 25.25 GeV, and their relative
momentum jumps from 10 MeV to roughly 250 MeV.

This kinematic exercise exposes a fundamental incompatibility between boosts and the traditional
separation of bound-state dynamics into center-of-mass and relative motion. Non-relativistic nuclear
physics---built on the Schr\"odinger equation---relies on this separation, yet the boost has
entangled the two: the relative momentum has changed, despite us having done nothing to the internal
dynamics of the system. The entanglement carries severe physical consequences. The boost-induced
increase in relative momentum corresponds to a decrease in the relative distance between
constituents; equivalently, the deuteron appears Lorentz-contracted along its direction of motion.
Its spatial density profile in the boosted frame is therefore not the same as in the rest frame, and
no single wavefunction can describe both.

Within IF quantization, the moving state must instead be represented as a superposition of the
rest-frame Hamiltonian's eigensolutions. Although the rest-frame and boosted Hamiltonians are
related by a unitary Lorentz transformation, the change in the bound-state wavefunction means that
boosting effectively induces interactions. These ``fictitious dynamics'' are unphysical,
frame-dependent artifacts rather than intrinsic nuclear properties. Historically, their entanglement
with genuine dynamics in IF calculations has led to misleading conclusions, as noted in
Chapter~\ref{chap:introduction} (see Ref.~\cite{Brodsky:2022fqy} for further discussion).

Computationally, the entanglement creates a severe bottleneck: the entire nuclear wavefunction must
be recalculated for every reference frame. Restricting evaluation to the nuclear rest frame does not
help, since physical observables must remain frame-independent and the dynamical complexities simply
migrate elsewhere in the calculation---a point we will revisit in
Section~\ref{sec:relativistic_hamiltonian}, once the mechanism behind this migration has been made
explicit.

\section{General Considerations of Relativistic Hamiltonian Formulations}

The difficulties of the previous section stem from the choice of coordinates used to describe the
bound-state system. In the non-relativistic regime governed by the Schr\"odinger equation, Galilean
invariance enforces the equivalence of inertial frames and permits a clean separation of
center-of-mass and relative variables. As just demonstrated, conventional IF variables are
structurally unsuited for a relativistic description of bound states: they entangle overall motion
with internal structure, introducing fictitious, frame-dependent dynamics. The choice of time
variable lies at the heart of the puzzle. Understanding why requires examining how many-body quantum
mechanics is constructed in general, and revisiting why the non-relativistic case worked in the
first place.

In any canonical quantization framework---relativistic or non-relativistic---one foliates spacetime
into a family of surfaces labeled by a ``time'' coordinate and imposes commutation relations between
fields on each surface. Noether's theorem, applied to the symmetries of the action, then produces a
set of conserved charges that act as generators of unitary transformations on the Hilbert space. For
non-relativistic quantum mechanics, Galilean invariance yields ten such generators: the Hamiltonian
$\hat{H}$, the three momentum components $\hat{P}^i$, the three angular momenta $\hat{J}^i$, and the
three Galilean boost generators $\hat{K}_G^i$. The relativistic case yields ten generators of the
Poincar\'e group: $\hat{H}$, $\hat{P}^i$, $\hat{J}^i$, and the three Lorentz boost generators
$\hat{K}^i$.

The act of foliating exposes a subtlety: a symmetry of the full four-dimensional spacetime need not
act as a symmetry of any single lower-dimensional slice. A cylinder aligned with the $z$-axis is
rotationally symmetric about $z$, yet only certain slices inherit that symmetry---a perpendicular
cut yields a disk, still invariant under rotations about $z$, while a tilted cut yields an ellipse
that is not. This same logic partitions the ten generators into two classes. Those that map a
foliation surface into itself are \emph{kinematic} and contain no interactions; those that move
points off the surface are \emph{dynamical} and carry the interactions of the theory. Different
foliations preserve different subgroups of the full symmetry, and the resulting split between
kinematic and dynamical generators governs which transformations can be performed trivially and
which require solving the dynamics.

In non-relativistic quantum mechanics, the standard choice of $x^0$ as the time variable happens to
be optimal: of the ten Galilean generators, only the Hamiltonian is dynamical. The other
nine---spatial translations, rotations, and Galilean boosts---leave equal-time surfaces invariant
and are therefore kinematic. This is precisely why center-of-mass and relative coordinates separate
cleanly, and why a single non-relativistic wavefunction describes the bound state in every inertial
frame.

The relativistic case is more delicate. Promoting Galilean boosts to Lorentz boosts changes their
character: a Lorentz boost tilts an equal-time surface rather than mapping it to itself. The three
Lorentz boost generators therefore join the Hamiltonian as dynamical generators, shrinking the
kinematic subgroup to six---spatial translations and rotations alone. The non-intrinsic dynamics
encountered in the boosted deuteron are the direct consequence: boosting a bound state now requires
evaluating an interaction-dependent operator. This is visible explicitly in physical cross sections.
Schematically,
\begin{equation}\label{eq:IF_boost_invariance}
\begin{aligned}
d\sigma \sim & \sum_X \braket{A,\bm{P}|\hat{J}^\mu(0)|X}\braket{X|\hat{J}^{\nu}(0)|A, \bm{P}} \\
= & \sum_X \braket{A,\bm{P}=0|\hat{U}^{\dagger}(\Lambda)\hat{J}^\mu(0)|X}\braket{X|\hat{J}^{\nu}(0)\hat{U}(\Lambda)|A, \bm{P}=0},
\end{aligned}
\end{equation}
where restricting the bound state to its rest frame merely transfers the dynamical complexity into
the boost operators $\hat{U}(\Lambda)$ acting on the currents. The nuclear interactions embedded in
those operators are unavoidable so long as boosts are dynamical generators---they cannot be removed,
only relocated. The same argument also affects cases where the final state contains bound nuclear
states as well. In high-energy nuclear processes where you have moving composite objects in your
initial and final states, boosts are central to the kinematics, and ensuring that bound states
transform kinematically under them matters both for calculational efficiency and for physical
interpretation. The FF, introduced next, does precisely this.

\section{Light-Front Variables}
\label{sec:relativistic_hamiltonian}
\begin{figure}
    \centering
    \includegraphics[width=0.8\linewidth]{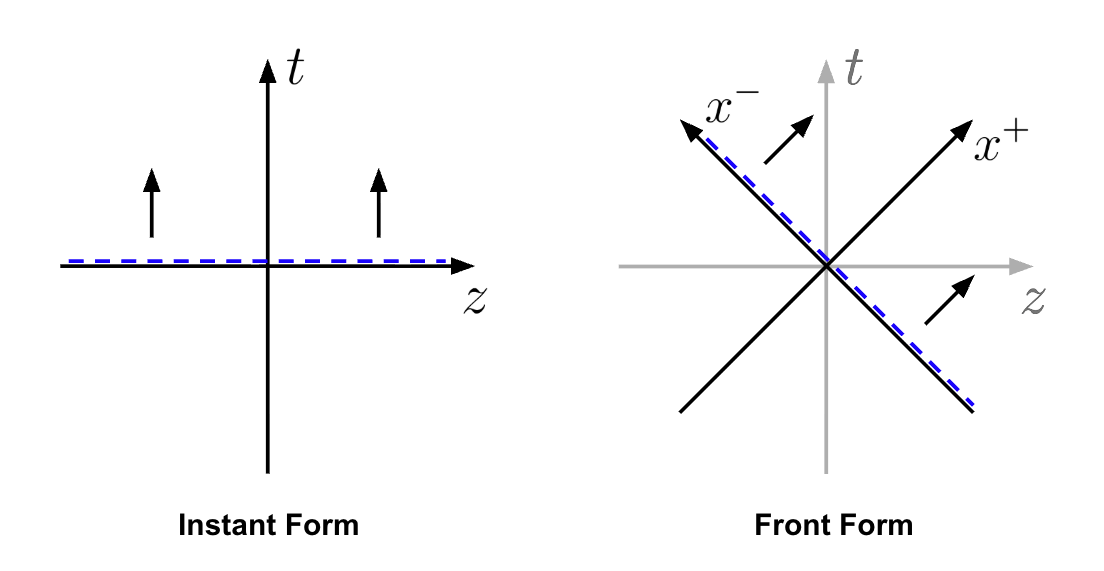}
    \caption{(color online) Minkowski space-time diagrams highlighting the differences between the Instant Form (IF) (left) and Front Form (FF) (right). Eigenstates of your Hamiltonian are determined at $t=$ const and $x^+=$ const for the IF and FF respectively. The Hamiltonian generates time evolution in the respective ``times'' as denoted by the arrows.}
    \label{fig:quantization_scheme}
\end{figure}
The pathologies identified in the previous section can be traced to the use of the equal-time
hyperplane $x^0 = 0$ as the quantization surface, on which the boost generators are dynamical and
entangle internal motion with the overall frame. The FF, used interchangeably with LF quantization,
instead quantizes on the $x^+ = 0$ hyperplane, see Fig. \ref{fig:quantization_scheme}. The variables
used in FF are called LF variables, they are defined as
\begin{equation}
\begin{gathered}
    x^\mu = (x^+, x^-, x^1, x^2), \\
    x^\pm \equiv x^0 \pm x^3.
\end{gathered}
\end{equation}
The choice of time variable is a matter of convention; throughout this thesis we take $x^+$ as the
LF time. The corresponding contraction between position and momentum is
\begin{equation}
    x_\mu p^\mu = g_{\mu \nu}x^\nu p^\mu = \tfrac{1}{2} x^+ p^- + \tfrac{1}{2} x^- p^+ - \bm{x}^\perp \cdot \bm{p}^\perp,
\end{equation}
with the LF metric tensor given in Table~\ref{tab:kinematics_comparison}. From this we see that
$x^+$ and $p^-$ are conjugate, and we accordingly identify $p^-$ as the LF energy. For a free
massive particle, the plus component is manifestly positive,
\begin{equation}
    p^+ = p^0 + p^z = \sqrt{m^2 + \bm{p}^2_{\rm IF}} + p^z > 0,
\end{equation}
and the mass-shell condition $p_\mu p^\mu = m^2$ takes the form
\begin{equation}
    p^+ p^- - (\bm{p}^\perp)^2 = m^2,
\end{equation}
which can be solved algebraically for the LF energy,
\begin{equation}
    p^- = \frac{m^2 + (\bm{p}^\perp)^2}{p^+} > 0.
\end{equation}
Unlike the equal-time dispersion relation $p^0 = \pm\sqrt{m^2 + \bm{p}^2_{\rm IF}}$, the LF
dispersion relation is rational in the kinematical variables $(p^+, \bm{p}^\perp)$ and admits no
negative-energy branch.

The decisive feature of the FF, however, lies in the structure of its boost generators, which are
kinematical rather than dynamical. To state these definitions, we recall that the Poincar\'e algebra
contains three rotation generators $J^i$ and three boost generators $K^i$, where $J^i$ generates
rotations about the $x^i$-axis and $K^i$ generates ordinary Lorentz boosts along the
$x^i$-direction. The longitudinal LF boost is simply $K^3$, the ordinary Lorentz boost along the
$z$-direction. The transverse LF boosts, however, are the combinations
\begin{equation}
    E^1 = K^1 + J^2, \qquad E^2 = K^2 - J^1,
\end{equation}
in which an ordinary transverse boost is paired with a rotation in such a way that the quantization
surface $x^+ = 0$ is preserved. Together with the longitudinal boost, the $E^i$ generate the full
set of LF boosts; we refer to these throughout as longitudinal and transverse LF boosts. Their
action on LF momenta makes their utility manifest.

Defining the rapidity $\omega$ through $e^\omega = \sqrt{(1-v)/(1+v)}$, where $v$ is the velocity of
the boost in natural units, a longitudinal LF boost acts as
\begin{equation}
    \begin{gathered}
        p^+ \rightarrow e^{\omega} p^+, \\
        p^- \rightarrow e^{-\omega} p^-, \\
        \bm{p}^\perp \rightarrow \bm{p}^\perp,
    \end{gathered}
\end{equation}
so that the plus and minus components are simply rescaled and the transverse momentum is left
untouched. A transverse LF boost with velocity $\bm{v}^\perp$ instead yields
\begin{equation}
    \begin{gathered}
        p^+ \rightarrow p^+, \\
        \bm{p}^\perp \rightarrow \bm{p}^\perp + p^+ \bm{v}^\perp, \\
        p^- \rightarrow p^- + 2\, \bm{v}^\perp \cdot \bm{p}^\perp + p^+ (\bm{v}^\perp)^2,
    \end{gathered}
\end{equation}
leaving the plus component invariant.

These transformation laws have important consequences for the identification of intrinsic variables
in nuclear physics. Because longitudinal LF boosts rescale every plus momentum by the same factor
$e^\omega$, the ratio
\begin{equation}\label{eq:LF_x_rel}
    x_i = \frac{p_i^+}{P^+},
\end{equation}
with $P^+ = \sum_i p_i^+$ the total plus momentum of the bound state and the index $i$ running over
its $A$ constituents, is invariant under both longitudinal and transverse LF boosts. Combined with
the positivity of $p^+$ established above, this further implies $0 < x_i < 1$ for each massive
constituent, so that the $x_i$ form a natural set of boost-invariant longitudinal momentum
fractions.

The transverse boost laws are equally suggestive---and reveal a remarkable feature of the FF. The
transformations of $p^-$ and $\bm{p}^\perp$ closely mirror the way non-relativistic kinetic energy
and momentum transform under a Galilean boost,
\begin{equation}
\begin{gathered}
    \left[ \bm{p} \rightarrow \bm{p} + m\bm{v} \right]_{\rm IF}, \\
    \left[ E_{\text{kin}} \rightarrow E_{\text{kin}} + \bm{v} \cdot \bm{p} + \tfrac{1}{2}m \bm{v}^{2} \right]_{\rm IF},
\end{gathered}
\end{equation}
with the plus momentum $p^+$ playing the role of the mass. The subscript IF on the square brackets
is there to emphasize that all variables are IF three-momenta. The very Galilean structure that
broke down under Lorentz transformations re-emerges, intact, in the transverse plane. This
correspondence motivates the definition of the LF relative transverse momenta
\begin{equation}
    \bm{k}_i^\perp = \bm{p}^\perp_i - \frac{p_i^+}{P^+} \bm{P}^\perp = \bm{p}^\perp_i - x_i \bm{P}^\perp,
\end{equation}
which takes precisely the form of their non-relativistic counterparts
\begin{equation}
\begin{gathered}
    \left[ \bm{k}_i = \bm{p}_i - (m_i/M)\bm{P} \right]_{\rm IF},
\end{gathered}
\end{equation}
with $M = \sum_i m_i$. By construction, the $\bm{k}_i^\perp$ are invariant under both longitudinal
and transverse LF boosts, and satisfy the constraint $\sum_i \bm{k}_i^\perp = 0$.

With $(x_i, \bm{k}_i^\perp)$ we have thus identified a set of intrinsic variables that do not mix
with the center-of-mass momentum under LF boosts. The utility of this separation is made manifest by
examining the total LF kinetic energy of an $A$-body system, multiplied by $P^+$,
\begin{equation}\label{eq:LF_kinetic_energy}
    P^+\sum_i p_i^- = P^+\sum_i \frac{m_i^2 + (\bm{p}_i^\perp)^2}{p_i^+} = (\bm{P}^\perp)^2 + \sum_i \frac{m_i^2 + (\bm{k}_i^\perp)^2}{x_i},
\end{equation}
in which the center-of-mass and intrinsic contributions separate cleanly, in direct analogy with the
non-relativistic decomposition $\left[ T = \bm{P}^2/2M + \sum_i \bm{k}_i^2/2m_i\right]_{\rm IF}$.
Since the intrinsic dynamics of the bound state must be independent of its overall center-of-mass
motion, any interaction term added to the Hamiltonian must depend only on the intrinsic variables
$(x_i, \bm{k}_i^\perp)$. Hence, the bound-state wavefunctions obtained by diagonalizing such a
Hamiltonian are then functions of $(x_i, \bm{k}_i^\perp)$ alone, and are accordingly invariant under
both longitudinal and transverse LF boosts.

This is precisely the structure that was absent in the IF formulation. The fictitious dynamics that
plagued the IF treatment of the boosted deuteron---the recalculation of the wavefunction in every
frame, the entanglement of internal structure with center-of-mass motion, the migration of dynamical
complexity into the boost operators of Eq.~\eqref{eq:IF_boost_invariance}---are absent here by
construction. The rest-frame and boosted descriptions of the bound state are connected by a
kinematical transformation acting only on $P^+$ and $\bm{P}^\perp$, while the intrinsic wavefunction
$\psi(x_i, \bm{k}_i^\perp)$ is left invariant. The LF variables thus provide the kinematic basis
suited to relativistic bound states that the previous section called for.

\section{Scattering Amplitudes}
The bound-state wavefunctions identified in the previous section are one ingredient in a cross
section; the other is a model of the elementary subprocess in which the probe interacts with a
constituent of the bound state. We now turn to this second ingredient and show that LF quantization
is equally indispensable there.

In modern quantum field theory, scattering processes are typically evaluated using the machinery of
Feynman diagrams, which offer the distinct advantage of manifest Lorentz invariance for every
individual graph. When considering scattering off composite targets, however, this covariant
approach becomes unwieldy: the bound-state input it requires---a fully covariant wavefunction---is
only available for two-body bound states, where it is given by the Bethe-Salpeter equation
\cite{Salpeter:1951sz}. Even there, its practical use is hampered both by prohibitive computational
costs and by well-known conceptual difficulties such as the treatment of relative time. To make
progress, one reverts to a Hamiltonian framework, in which bound-state wavefunctions are rigorously
defined and tractable. The cost of this shift is that scattering amplitudes must be computed using
old-fashioned, time-ordered perturbation theory, which increases the number of topological graphs to
be summed. The trade-off is favorable: a manageable proliferation of diagrams in exchange for access
to bound-state structure.

\begin{figure}
    \centering
    \includegraphics[width=0.7\linewidth]{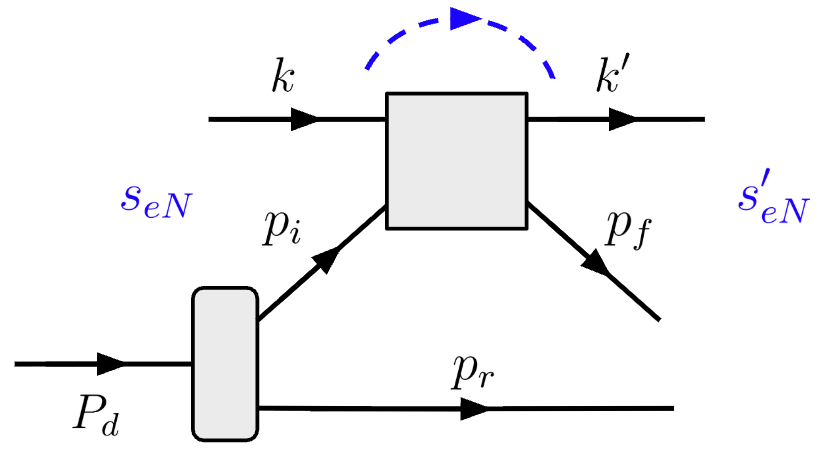}
    \caption{(color online) Feynman diagram for electron-deuteron scattering with an electron-nucleon subprocess. The invariant energy before and after the collision are labeled $s_{eN}$ and $s_{eN}'$ respectively.}
    \label{fig:nuclear_subprocess}
\end{figure}

For high-energy nuclear processes in which the probe resolves the individual nucleons inside the
nucleus, the central theoretical challenge, aside from nuclear structure, is modeling the electron-nucleon
subprocess. The standard practice is an off-shell extrapolation of the on-shell electron-nucleon
amplitude: the bound nucleon, unlike a free one, does not lie on its mass shell, and the
free-nucleon cross section is extended into the off-shell region using a prescription such as that
of de Forest~\cite{DeForest:1983ahx}. Under IF at relativistic kinematics, however, this
extrapolation acquires a contamination unrelated to the nuclear target. The IF and FF, as
Hamiltonian formulations, share a common feature: only three components of momentum are conserved at
each interaction vertex, with intermediate particles placed on their mass shell while the remaining
momentum component absorbs the binding-induced mismatch. They differ in which three components:
spatial three-momentum $\bm{p}_{\rm IF}$ in IF, with energy $p^0$ unconserved; LF three-momentum
$\bm{p}=(p^+, \bm{p}^\perp)$ in FF, with $p^-$ unconserved. This distinction has direct consequences
for the off-shellness of the struck nucleon.

Consider the Feynman diagram in Fig.~\ref{fig:nuclear_subprocess}, with $k$ the incident electron,
$k'$ the scattered electron, $P_d$ the deuteron, $p_r$ the spectator/recoil nucleon, $p_i$ the
initial nucleon, and $p_f$ the scattered nucleon. We quantify the off-shellness by the difference
between the invariant energy of the electron-nucleon subprocess before and after the interaction,
$s_{eN}' - s_{eN}$. Using overall four-momentum conservation, $k' + p_f = k + P_d - p_r$,
\begin{equation}\label{eq:soffshell_general}
\begin{aligned}
s_{eN}' - s_{eN} &= (k' + p_f)^2 - (k + p_i)^2 = (k + P_d - p_r)^2 - (k + p_i)^2 \\
                  &= \left[\, 2k \cdot (P_d - p_r - p_i) \,\right] + \left[\, (P_d - p_r)^2 - p_i^2 \,\right].
\end{aligned}
\end{equation}
The square brackets separate two physically distinct contributions. The right-hand bracket, $(P_d -
p_r)^2 - p_i^2$, measures the off-shellness of the struck nucleon itself: $(P_d - p_r)$ is the
four-momentum that would be assigned to it if four-momentum conservation were enforced at the
deuteron vertex, while $p_i$ is the four-momentum actually assigned by the quantization scheme (on
mass shell, $p_i^2 = m^2$). The two differ because only three momentum components are conserved at
the vertex; the remaining component is fixed instead by the mass-shell condition. The left-hand
bracket depends on the electron's four-momentum and is therefore a property of the probe rather than
the target. A consistent off-shell extrapolation should depend only on the right-hand bracket; the
left-hand bracket is the contamination.

Working in the deuteron rest frame with $(P_d^\mu)_{\rm IF} = (M_d, \bm{0})$ and an incident
electron with $k^\mu_{\rm IF} = (\omega, 0, 0, -\omega)$ in the high-energy limit $\omega > 1$ GeV,
the IF rules at the deuteron vertex give $(\bm{p}_i)_{\rm IF} = -(\bm{p}_r)_{\rm IF}$ and $p_i^0 =
\sqrt{m^2 + (\bm{p}_r)_{\rm IF}^2}$, while $p_r^0 + p_i^0 \neq M_d$---the energy mismatch that
defines the bound state. Evaluating the probe-dependent bracket to leading order in $|\bm{p}_r|_{\rm
IF}/m$ and neglecting the small deuteron binding energy,
\begin{equation}\label{eq:soffshell_IF}
\left[ s_{eN}' - s_{eN} \right]_{\rm IF}\approx -2\omega\,\frac{(\bm{p}_r)_{\rm IF}^2}{m} + \left[\, (P_d - p_r)^2 - p_i^2 \,\right]_{\rm IF}.
\end{equation}
The contamination grows linearly with the incident energy $\omega$ and depends explicitly on the
kinematics of the probe, even though the off-shellness is meant to be an intrinsic property of the
bound nucleon. At high energies the IF prescription cannot disentangle target structure from probe
kinematics.

LF quantization removes the contamination. Expanding the four-vector dot product in LF components,
\begin{equation}\label{eq:soffshell_LF}
\begin{aligned}
s_{eN}' - s_{eN} &= \left[\, k^+ (P_d - p_r - p_i)^- + k^- (P_d - p_r - p_i)^+ - 2\bm{k}^\perp \cdot (\bm{P}_d - \bm{p}_r - \bm{p}_i)^\perp \,\right] \\
&\quad + \left[\, (P_d - p_r)^2 - p_i^2 \,\right].
\end{aligned}
\end{equation}
With the chosen kinematics, $k^+ = 0$ and $\bm{k}^\perp = 0$, and plus-momentum conservation at the
deuteron vertex enforces $(P_d - p_r - p_i)^+ = 0$. The entire probe-dependent bracket vanishes,
leaving
\begin{equation}\label{eq:soffshell_LF_final}
s_{eN}' - s_{eN} = (P_d - p_r)^2 - p_i^2.
\end{equation}
The off-shellness is determined entirely by the nuclear target: probe kinematics no longer enter,
and the extrapolation has the clean physical interpretation it was intended to have.

\begin{table}[htpb]
    \centering
    \renewcommand{\arraystretch}{2.0}
    \begin{tabular}{lcc}
        \hline\hline
        \textbf{Property} & \textbf{Instant Form} & \textbf{Front Form} \\
        \hline
        Coordinates &
        $x^\mu = (x^0, \bm{x}_{\rm IF})$ &
        $\begin{gathered} x^\mu = (x^+, x^-, \bm{x}^\perp) \\ x^\pm \equiv x^0\pm x^3\end{gathered}$ \\[20pt]

        $g_{\mu \nu}$ &
        ${ \renewcommand{\arraystretch}{1.0} \begin{pmatrix} 1 & 0 & 0 & 0 \\ 0 & -1 & 0 & 0 \\ 0 & 0 & -1 & 0 \\ 0 & 0 & 0 & -1 \end{pmatrix} }$ &
        ${ \renewcommand{\arraystretch}{1.0} \begin{pmatrix} 0 & 1/2 & 0 & 0 \\ 1/2 & 0 & 0 & 0 \\ 0 & 0 & -1 & 0 \\ 0 & 0 & 0 & -1 \end{pmatrix} }$ \\[40pt]

        Free Particle &
        $\begin{gathered} p^\mu = (p^0,\bm{p}_{\rm IF}) \\ p^\mu p_\mu = (p^0)^2 - \bm{p}_{\rm IF}^2 = m^2 \\ p^0 = \sqrt{m^2 + \bm{p}^2_{\rm IF}} > 0  \end{gathered}$ &
        $\begin{gathered} p^\mu = (p^-,p^+,\bm{p}^{\perp}) \\ p^\mu p_\mu = p^+ p^- - (\bm{p}^\perp)^2 = m^2 \\ p^- = \frac{m^2 + (\bm{p}^{\perp})^2}{p^+} > 0 \\  \bm{!} \, p^+ > 0 \, \bm{!}\end{gathered}$ \\[40pt]

        Normalization &
        $\begin{gathered} \braket{\bm{p}'_{\rm IF},\sigma'|\bm{p}_{\rm IF},\sigma} = \\ (2 \pi)^3 2E(\bm{p}_{\rm IF})\delta^{(3)}(\bm{p}'_{\rm IF}-\bm{p}_{\rm IF}) \delta_{\sigma \sigma'}\end{gathered}$ &
        $\begin{gathered} \braket{\bm{p}',\sigma'|\bm{p},\sigma} = \\ (2 \pi)^3 2p^+ \delta(p'^+ - p^+)\delta^{(2)}(\bm{p}'^\perp-\bm{p}^\perp) \delta_{\sigma \sigma'}\end{gathered}$ \\[30pt]

        Phase Space &
        $\displaystyle \int d\Gamma_p = \int \frac{d\bm{p}_{\rm IF}}{(2 \pi)^3 2E(\bm{p}_{\rm IF})}$ &
        $\displaystyle \int d\Gamma_p = \int \frac{dp^+ d\bm{p}^\perp}{(2 \pi)^3 2p^+} \theta(p^+)$ \\[20pt]
        \hline\hline
    \end{tabular}
    \caption{Comparison of kinematic variables, the Minkowski metric, and Lorentz invariant phase space between the Instant Form and Front Form.}
    \label{tab:kinematics_comparison}
\end{table}

% \section{Discussion}
% This chapter was meant to motivate the necessary use of light-front quantization techniques for high-energy nuclear processes. To summarize, utilizing Lagrangian formulations to describe bound states are not computationally feasible and conceptually difficult. But in going back to Hamiltonian formulations we must carefully consider how we pick a time-variable, something that we didn't need to consider in the non-relativistic case. Foliating space-time into surfaces of constant $x^0$, the conventional choice, leads to intrinsic dynamics being entanged with center-of-mass motion of our composite system. This not only leads to computaitonal difficulties, but also conceptual ones as well. Foliating space-time into surfaces of constant $x^+$ solves this problem, allows a clean separation of center-of-mass and relative variables, and leads to boost invariant wavefunctions, which are important for calculations in high-energy processes. Apart from the description of the bound state, light-front quantization also allows for a composite description of high-energy scattering, where sub-processes can be modeled in closer connection to known on-shell phenomolology, without off-shell artifacts from the probe. In fact, light-front dynamics is the only scheme which avoids large energy off-shellness in the nucleon subprocess, hence its use is necessary, not optional in high-energy scattering.

\section{Minimal Relativity and the Front Form}
The nucleon-nucleon (NN) interaction plays a central role throughout this
dissertation. At present, no published work has developed phenomenological
relativistic LF NN potentials capable of describing elastic NN phase shifts to
high accuracy, and constructing such potentials lies outside the scope of this work. We therefore
adopt an alternative prescription: we recast an existing non-relativistic NN
potential---here the Argonne $v_{18}$ (AV18) potential \cite{Wiringa:1994wb}---into relativistic
form, perform a change of variables, and incorporate the effects of LF spinors to obtain an
approximate LF NN potential. For further reading, the reader is directed to Refs.
\cite{Frankfurt:1981mk,Frankfurt:1988nt,Strikman:2017koc,Cosyn:2020kwu}.

The starting point is the non-relativistic Lippmann-Schwinger equation, at zero center-of-mass
momenta, for a two-nucleon system in relative-momentum coordinates, neglecting spin and isospin,
\begin{equation}\label{lippmann_schwinger}
    \left [ \frac{k^2}{m_N} \psi(\bm{k}) + \int \frac{d\bm{k}'}{(2 \pi)^3} V(\bm{k},\bm{k}') \psi(\bm{k}') = \epsilon \psi(\bm{k}) \right]_{\rm IF},
\end{equation}
where again, the subscript IF on the square brackets is there to emphasize that all variables are IF
three-momenta. To recast this as an effective relativistic equation, we perform the following
sequence of manipulations: multiply throughout by $m_N$, add $m_N^2 \psi(\bm{k}_{\rm IF})$ to both
sides, multiply throughout by $4 \sqrt{E(\bm{k}_{\rm IF})}$, and insert a factor of $E(\bm{k}_{\rm
IF}')/E(\bm{k}_{\rm IF}')$ inside the integral. The result is
\begin{equation}\label{rel_lippmann_schwinger}
    \begin{gathered}
    \left[ 4(k^2+m_N^2) \tilde{\psi}(\bm{k}) + \int \frac{d\bm{k}'}{(2 \pi)^3 E(\bm{k}')} \tilde{V}(\bm{k},\bm{k}') \tilde{\psi}(\bm{k}') = M^2 \tilde{\psi}(\bm{k}) \right]_{\rm IF}, \\
    \left[ \tilde{\psi}(\bm{k}) = \sqrt{E(\bm{k})} \psi(\bm{k}) \right]_{\rm IF}, \\
    \left[ \tilde{V}(\bm{k},\bm{k}') = 4m_N\sqrt{E(\bm{k})}V(\bm{k},\bm{k}')\sqrt{E(\bm{k}')} \right]_{\rm IF},
\end{gathered}
\end{equation}
where $M^2 = 4(m^2_N + m_N \epsilon)$.

This procedure, known as minimal relativity, was first introduced by Brown, Jackson, and Kuo
\cite{Brown:1969tfp}. Their motivation was the following puzzle. Realistic potentials with strong
repulsive cores naturally generate high-momentum components in NN wavefunctions, yet the physical
validity---and the surprising success---of these components in describing high-momentum deuteron
phenomena is difficult to justify when they are derived solely from the non-relativistic
Schr\"odinger equation. To place them on firmer footing, the authors argued that such realistic
potentials should instead be regarded as effective solutions of a relativistic equation---what is
today understood as arising from an IF Hamiltonian in the NN rest frame. Minimal relativity then
supplies the necessary relativistic kinematics without altering the phenomenologically fitted
scattering phase shifts, thereby motivating the high-momentum components that emerge from realistic
NN potentials.

\begin{figure}
    \centering
    \includegraphics[width=1\linewidth]{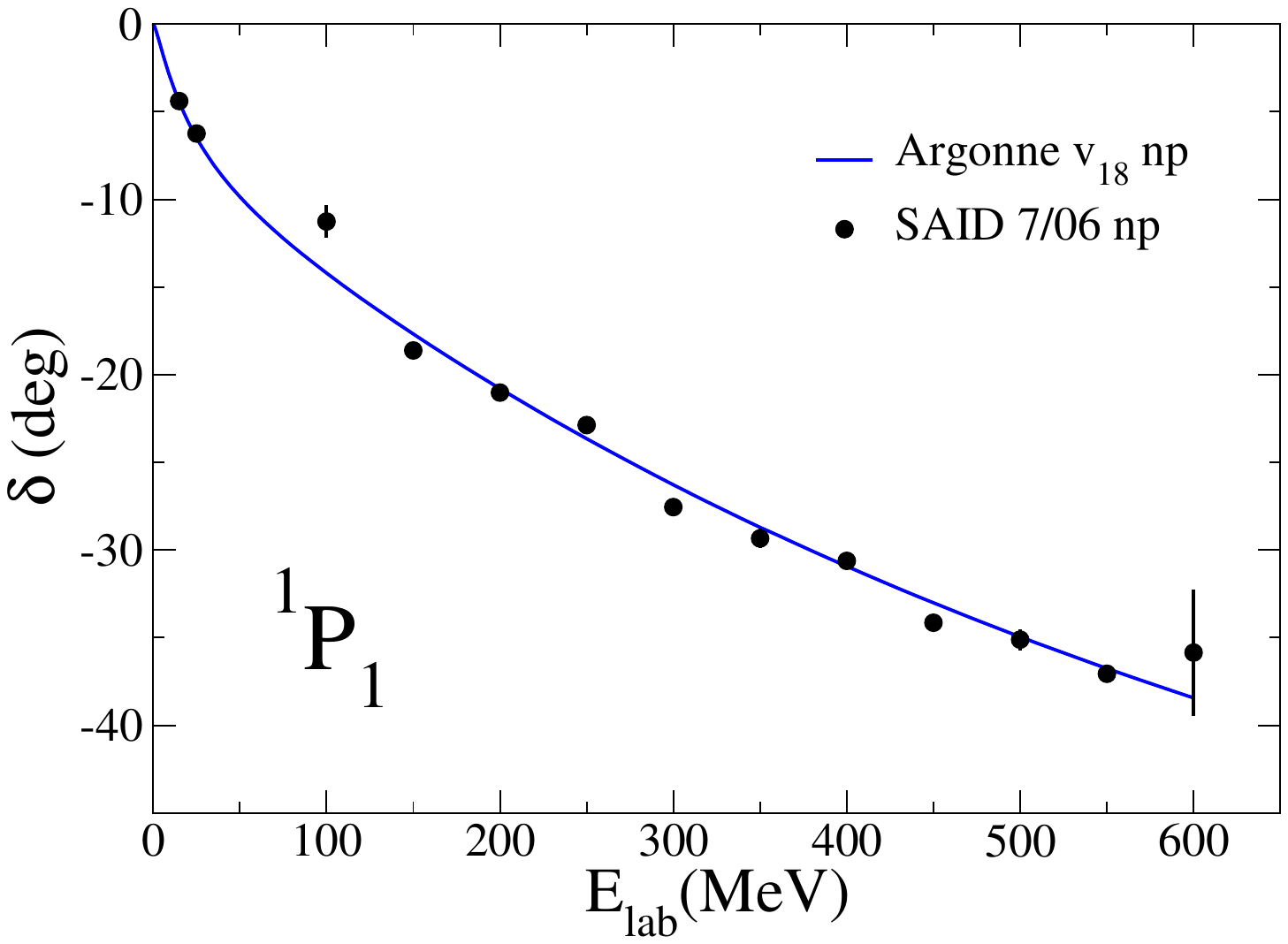}
    \caption[The $^1P_1$ nucleon-nucleon elastic phase shift calculated using the Argonne V18 potential.]{The $^1P_1$ nucleon-nucleon elastic phase shift, $\delta$, calculated using the Argonne V18 potential, as a function of lab energy, $E_{\text{lab}}$. The solid line represents the theoretical calculation, and the black dots represent experimental data from the SAID nucleon-nucleon database in July 2006. Figure obtained from Ref.~\cite{av18_website}.}
    \label{fig:av18_phase_shift}
\end{figure}

The performance of the AV18 potential offers further support. Although it was
originally fitted to reproduce nucleon-nucleon phase shifts up to a laboratory energy of 350~MeV, it
nonetheless reproduces them well up to 600~MeV in several partial-wave channels (see, e.g.,
Fig.~\ref{fig:av18_phase_shift}).

The connection to light-front quantization is established by the change of
variables,
\begin{equation}
    \left[ k^2 + m_N^2 \right]_{\rm IF} = \frac{m_N^2 + (\bm{k}^\perp)^2}{\alpha (2-\alpha)},
\end{equation}
where $\alpha$ is equal to Eq. \eqref{eq:LF_x_rel} multiplied by the nuclear mass number $A$. The integration measure becomes
\begin{equation}
    \int\frac{d\bm{k}'_{\rm IF}}{(2 \pi)^3 E(\bm{k}'_{\rm IF})} = \int\frac{d\alpha'd\bm{k}'^\perp}{(2 \pi)^3 \alpha'(2-\alpha')}.
\end{equation}
With this substitution, Eq.~\eqref{rel_lippmann_schwinger} takes the form of the two-body Weinberg
equation \cite{Weinberg:1966jm}, where we relabel $\psi(\alpha,\bm{k}^\perp) \equiv
\tilde{\psi}_{\rm IF}(\bm{k}_{\rm IF})$ and $V \equiv \tilde{V}_{\rm IF}$ under the new
coordinates,
\begin{equation}\label{weinberg_eq}
    \begin{aligned}
    \left( 4\frac{m_N^2 + (\bm{k}^\perp)^2}{\alpha (2-\alpha)} \right)\psi(\alpha,\bm{k}^\perp) + \int\frac{d\alpha'd\bm{k}'^\perp}{(2 \pi)^3 \alpha'(2-\alpha')} V(\alpha,\bm{k}^\perp;\alpha',\bm{k}'^\perp) & \psi(\alpha',\bm{k}'^\perp) \\
    & = M^2 \psi(\alpha,\bm{k}^\perp),
    \end{aligned}
\end{equation}
whose operator form is
\begin{equation}
\begin{gathered}
    \left[ \hat{P}^+\hat{P}^- - \hat{P}^\perp\cdot\hat{P}^\perp \right] \ket{\Psi_{NN}, \bm{P}} = M^2 \ket{\Psi_{NN}, \bm{P}}, \\
    \hat{P}^- = \hat{P}^-_{\rm kin} + \hat{V}.
\end{gathered}
\end{equation}
Given the operator form, one can trace back kinetic energy term in Eq. \eqref{weinberg_eq} from Eq.
\eqref{eq:LF_kinetic_energy} and see that $V(\alpha,\bm{k}^\perp;\alpha',\bm{k}'^\perp)$ comes from
$\hat{P}^+\hat{V}$.

One subtlety remains once spin is incorporated. The manipulations above act on the IF
potential---which is, in principle, a matrix element taken between IF Dirac spinors---and merely
change variables to bring it into LF form, leaving the spinors themselves untouched. What remains is
to transform the IF Dirac spinors into LF Dirac spinors, which is accomplished through a Melosh
rotation, a unitary relation between the two (for the original study see Ref. \cite{Melosh:1974cu},
for a detailed derivation see. Ref. \cite{Cosyn:2020kwu}):
\begin{equation}
    u(\bm{p},\sigma) = \sum_{\sigma'} u_{\rm IF}(\bm{p}_{\rm IF},
    \sigma') U(\bm{p}_{\rm IF},\sigma',\sigma)
\end{equation}
\begin{equation}
\begin{gathered}
    U(\bm{p}_{\rm IF},\sigma',\sigma) = \chi^\dagger(\sigma') \mathcal{U}(\bm{p}_{\rm IF}) \chi(\sigma) \\
    \mathcal{U}(\bm{p}_{\rm IF}) \equiv \frac{p^+ + m_N + \bm{p}^\perp \cdot \bm{\sigma}^\perp \sigma^3}{\sqrt{2p^+} \sqrt{E(\bm{p}_{\rm IF})+m_N}} \\ p^+ \equiv E(\bm{p}_{\rm IF}) + p^3
\end{gathered}
\end{equation}
where $(\bm{\sigma}^\perp,\sigma^3)$ are the Pauli matrices and $\chi(\sigma)$ are the Pauli
spinors,
\begin{equation}
    \chi(+1) = \begin{pmatrix} 1 \\ 0 \end{pmatrix},
    \qquad
    \chi(-1) = \begin{pmatrix} 0 \\ 1\end{pmatrix}.
\end{equation}
Including spin, the bound and scattering-state wavefunction solutions of Eq.~\eqref{weinberg_eq} are
therefore
\begin{equation}
    \psi_\beta(\alpha,\bm{k}^\perp,\sigma_1,\tau_1,\sigma_2,\tau_2) = \sum_{\sigma_1' \sigma_2'} \tilde{\psi}^\beta_{\rm IF}(\bm{k}_{\rm IF},\sigma_1',\tau_1,\sigma_2',\tau_2) U^*(\bm{k}_{\rm IF},\sigma_1',\sigma_1)U^*(-\bm{k}_{\rm IF},\sigma_2',\sigma_2),
\end{equation}
where $\beta$ indexes the eigenvalues of the Hamiltonian. The half-off-shell $T$-matrix is
analogously,
\begin{equation}\label{eq:T_matrix_ET_LF}
\begin{gathered}
    T(\alpha_{\rm off}, \bm{k}^\perp_{\rm off}, \sigma_1, \tau_1, \sigma_2, \tau_2;\alpha, \bm{k}^\perp, \sigma_3, \tau_3, \sigma_4, \tau_4) = \\
    \sum_{\sigma_1' \sigma_2'}  \sum_{\sigma_3' \sigma_4'}U^*(\bm{k}_{\rm IF}^{\rm off},\sigma_1',\sigma_1)U^*(-\bm{k}^{\rm off}_{\rm IF},\sigma_2',\sigma_2)  U(\bm{k}_{\rm IF},\sigma_3',\sigma_3)U(-\bm{k}_{\rm IF},\sigma_4',\sigma_4) \\
    \times \tilde{T}_{\rm IF}(\bm{k}^{\rm off}_{\rm IF}, \sigma_1', \tau_1, \sigma_2', \tau_2;\bm{k}_{\rm IF}, \sigma_3', \tau_3, \sigma_4', \tau_4),
\end{gathered}
\end{equation}
where $\tilde{T}_{\rm IF}(\bm{k}_{\rm IF}^{\rm off},\bm{k}_{\rm IF}) = \sqrt{E(\bm{k}_{\rm IF}^{\rm
off})} T_{\rm IF}(\bm{k}_{\rm IF}^{\rm off},\bm{k}_{\rm IF}) \sqrt{E(\bm{k}_{\rm IF})}$.

\begin{figure}[H]
    \centering
    \includegraphics[width=0.7\linewidth]{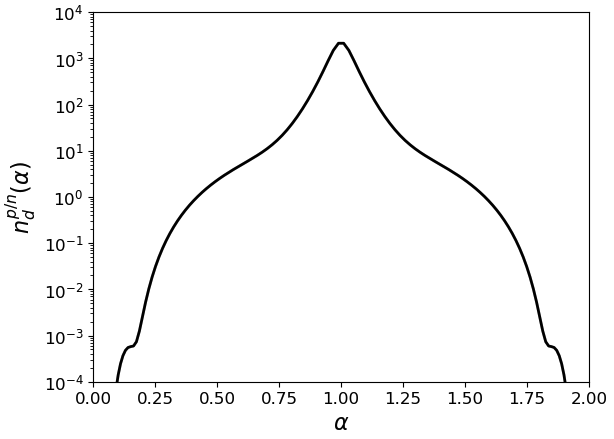}
    \caption{The light-front momentum distribution, integrated over transverse momenta, of deuterium as a function of the Lorentz-invariant relative variable $\alpha=2p^+/P_d^+$.}
    \label{fig:deut_lf_dist}
\end{figure}

Taken together, minimal relativity, the transformation to relative LF
coordinates, and the inclusion of LF spinors allow us to employ
well-established non-relativistic potentials within the LF quantization
framework. This prescription has been widely used in high-energy nuclear
physics, for instance in deuteron electrodisintegration \cite{Sargsian:2009hf} and in semi-exclusive
electron-deuteron reactions in both the quasi-elastic and DIS regimes
\cite{Sargsian:2001ax,Strikman:2017koc,Cosyn:2020kwu,Cosyn:2026vpc}. Although
the theoretical uncertainty of the prescription is difficult to quantify, these studies find that it
performs well for relative momenta up to roughly
600~MeV---large enough that we can reliably describe short-range
nucleon-nucleon dynamics within LF quantization. For a concrete example, Fig. \ref{fig:deut_lf_dist}
presents the LF momentum distribution, integrated over transverse momenta, of deuterium. The
deuteron state is labeled as
\begin{equation}
    \ket{d, M_J; \bm{P}_d}, \qquad P_d^- = \frac{m_d^2 + (\bm{P}_d^\perp)^2}{P_d^+},
\end{equation}
the overlap with an anti-symmetric two-nucleon state being
\begin{equation}
\begin{gathered}
    \braket{\bm{p}_1,\sigma_1,\tau_1;\bm{p}_2,\sigma_2,\tau_2|d, M_J; \bm{P}_d} = \\(2 \pi)^3 2P_d^+ \delta(p_1^+ + p_2^+ - P_d^+)\delta^{(2)}(\bm{p}_1^\perp + \bm{p}_2^\perp - \bm{P}_d^\perp) \psi_d(\alpha,\bm{k}^\perp,\sigma_1,\tau_1,\sigma_2,\tau_2|M_J),
\end{gathered}
\end{equation}
with wavefunction normalization
\begin{equation}
    \frac{1}{3}\sum_{M_J} \frac{1}{2}\sum_{\substack{\sigma_1 \tau_1 \\\sigma_2 \tau_2}}
    \int\frac{d\alpha d\bm{k}^\perp}{(2 \pi)^3\alpha(2-\alpha)} |\psi_d(\alpha,\bm{k}^\perp,\sigma_1,\tau_1,\sigma_2,\tau_2|M_J)|^2 = 1.
\end{equation}

\section{Summary}
This chapter has served to motivate the study of nuclear physics within the FF (LF quantization),
the framework on which the original calculations of this dissertation rest. Its central conclusion
is that, for high-energy nuclear reactions, the FF is not merely a convenient choice but a necessary
one: it is the only relativistic Hamiltonian formulation in which the bound-state wavefunction
admits a frame-independent treatment and in which the off-shellness of the electron-nucleon subprocess
remains free of probe contamination; together, these properties permit a composite, factorized
description of electron-nucleus scattering. The origin of this conclusion lies in the choice of time
variable, which dictates which Poincar\'e generators are kinematical and which are dynamical.

Quantization on surfaces of constant $x^0$ places the three Lorentz boosts in the dynamical set,
entangling intrinsic structure with center-of-mass motion, so that a single wavefunction can no
longer describe a bound state across different frames. The scattering cross section is itself a
Lorentz-invariant quantity and may be evaluated in any frame; however, this invariance does not
eliminate the dynamical complexity. Evaluating the cross section in the rest
frame merely relocates that complexity into the boost operators acting on the currents, as in
Eq.~\eqref{eq:IF_boost_invariance}. Quantization on surfaces of constant $x^+$ instead places the
longitudinal and transverse LF boosts in the kinematical set. The intrinsic variables $(x_i,
\bm{k}_i^\perp)$ then separate cleanly from the center-of-mass coordinates, and the resulting
bound-state wavefunctions are invariant under LF boosts.

The choice of quantization surface also governs the modeling of the elementary subprocess. In any
Hamiltonian formulation, only three momentum components are conserved at each interaction vertex,
and the off-shell extrapolation required to describe bound-nucleon kinematics depends on which
component is left unconserved. In the IF, the probe-dependent contribution to the off-shellness
grows linearly with the probe energy, contaminating the very target structure one seeks to isolate.
In the FF, this contribution vanishes for the relevant kinematics, so that the off-shellness
is determined entirely by the nuclear target. The FF is therefore the only relativistic Hamiltonian
formulation free of large, probe-induced off-shellness in the electron-nucleon subprocess.

Finally, as phenomenological LF NN potentials capable of reproducing elastic phase shifts to high
accuracy are not yet available, we use the minimal-relativity prescription, along with a change of variables
and the inclusion of LF spinors, to recast well established non-relativistic IF potentials 
into approximate LF form. Taken together, these results establish LF quantization as the
Hamiltonian framework required for high-energy nuclear scattering, while the minimal-relativity
prescription supplies the LF two-body NN interactions on which the calculations of the following
chapters are built.

% !TEX root = ../uwthesis.tex
\chapter{Inclusive Electron-Nucleus Scattering}\label{chap:scattering}

Across both low- and high-energy regimes, electron scattering serves as an indispensable tool for
probing nuclear and partonic structure. For experimentalists, the electron acts as a high-resolution
microscope; its low mass allows it to be readily accelerated to relativistic speeds. Furthermore,
modern tracking detectors, electromagnetic calorimeters, and reconstruction techniques enable the
measurement of scattered electrons with exceptional precision. From a theoretical perspective, the
scattering process is governed by Quantum Electrodynamics and can therefore be modeled
accurately. Consequently, electron-nucleus scattering provides a clean separation between the
internal target structure and reaction mechanisms.

Inclusive electron-nucleus scattering $A(e,e')$, the main subject of this work, provides two
independent kinematic ``dials''—the energy transfer, $\nu$, and the momentum transfer, $q$—that can
be tuned to isolate specific physical regimes. The momentum transfer determines the spatial
resolution of the probe, governed by the de Broglie wavelength $\lambda = \frac{\hbar}{q}$, which
establishes the effective target the electron interacts with. Conversely, the energy transfer
dictates the excitation energy transferred to that target. At low momentum transfers (on the order
of tens of MeV), the electron probes the nucleus as a whole; increasing $\nu$ in this regime reveals
nuclear excited states and giant resonances. At intermediate momentum transfers (hundreds of MeV to
a few GeV), the spatial resolution sharpens to resolve individual nucleons within the nucleus. This
kinematic region is known as quasi-elastic (QE) scattering. Here, increasing $\nu$ accesses nucleon
excited states, such as the Roper resonance. Finally, at momentum and energy transfers on the order
of hundreds of GeV, the electron resolves the partonic structure within the nucleons and deposits
enough energy to break a nucleon apart. Such processes are known as deep inelastic scattering (DIS),
and they dominate the reaction cross section in this regime.

In order to develop nuclear models applicable to future high-energy nuclear physics experiments at
the EIC and JLab, one must first understand QE scattering---a process that probes the nucleonic
structure of the nucleus---and then build on that baseline to understand how nucleonic
configurations in the nucleus interplay with their partonic dynamics. The experimental data
considered in this dissertation is from JLab E08-018, an inclusive electron-nucleus scattering
measurement with an electron beam energy of $3.356$~GeV and a scattered-electron angle of $25^\circ$
\cite{Zhang:2025nst}. Restricting ourselves to $x_B = Q^2/(2 m_N \nu) > 1$ yields $Q^2 > 1$~GeV$^2$,
placing us in a kinematic window with momentum transfer high enough to resolve short-distance
nuclear structure, but energy transfer low enough that we need not consider the transition to quark
degrees of freedom in DIS.

\section{Collinear Frames}
\begin{figure}
    \centering
    \includegraphics[width=0.65\linewidth]{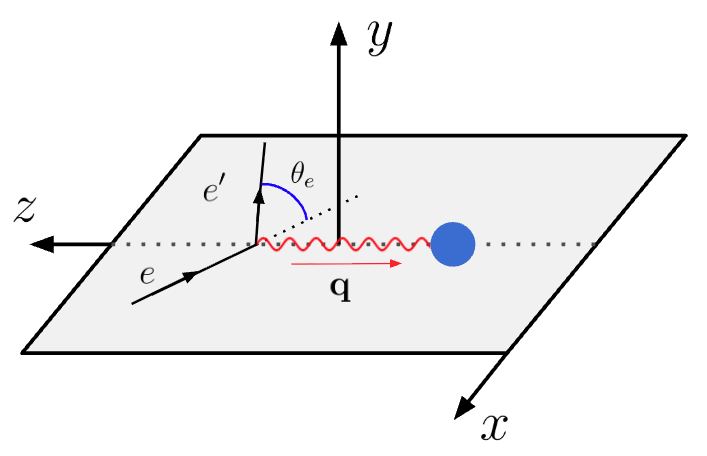}
    \caption{(color online) Nuclear rest frame, with the virtual photon momentum, $\bm{q}_{\rm IF}$, 
             defining the $z$-axis and the $x-z$ plane defined by the electron and scattered electron 
             momentum vectors. The nucleus is represented by the filled blue circle.}
    \label{fig:collinear_frame}
\end{figure}
Throughout this dissertation, all calculations are performed in the nucleus
rest frame, with the virtual photon's IF three-momentum defining the negative
$z$-axis (see Fig.~\ref{fig:collinear_frame}). The nucleus rest frame belongs to a broader
class of frames---known as \emph{collinear frames}---that are related to one another by boosts along
the $z$-axis. In any such frame, the LF momentum components of the nucleus and the virtual photon
take the form
\begin{equation}\label{collinear_lf}
\begin{gathered}
    P_A^+ > 0, \qquad P_A^- = \frac{m_A^2}{P_A^+}, \qquad \bm{P}_A^\perp = 0,\\[4pt]
    q^+ = -\xi_A P_A^+, \qquad q^- = \frac{q^2}{q^+} = \frac{Q^2}{\xi_A P_A^+}, \qquad \bm{q}^\perp = 0,
\end{gathered}
\end{equation}
where the parameter $\xi_A$ is fixed by the kinematic invariant
\begin{equation}
    2P_A \cdot q = P_A^+ q^- + P_A^- q^+ = \frac{Q^2}{x_A}.
\end{equation}
Solving this constraint for $\xi_A$ yields
\begin{equation}
    \xi_A = \frac{2x_A}{1 \pm \sqrt{1+4m_A^2 x_A^2/Q^2}}.
\end{equation}
We select the positive root, known as the Nachtmann variable \cite{Nachtmann:1973mr}, which is what
orients the virtual photon's momentum along the negative $z$-axis. In the scaling limit $Q^2 \gg m_A^2$ this
reduces to
\begin{equation}
    \xi_A = x_A + \mathcal{O}(m_A^2/Q^2).
\end{equation}
In the nucleus rest frame, $P_A^+ = P_A^- = m_A$, and $x_A$ is related to
the Bjorken variable $x_B = Q^2/(2 m_N \nu)$ by
\begin{equation}
    x_B = \frac{m_A}{m_N}\, x_A.
\end{equation}

\section{Unpolarized Inclusive Electron-Nucleus Cross-Section}

We start with the general parameterization of the cross-section for unpolarized inclusive
electron-nucleus scattering where the final states can be anything, $X$. Because the electromagnetic
interaction between the electron and nucleus is weak, one can model the interaction through the
exchange of a single virtual photon.

\begin{equation}\label{cs_parameterization}
    \frac{d\sigma}{dE'd\Omega'} = \frac{E'}{E} \frac{\alpha_e^2}{Q^4} L_{\mu \nu} W^{\mu \nu}_A,
\end{equation}

\noindent where $q^\mu$ is the virtual photon's four momentum and $Q^2 = -q^{\mu}q_{\mu}$, $E$ and
$E'$ are the energies of the incident and scattered electron respectively, and $\alpha_e$ is the
fine structure constant. Ignoring the mass of the electron, the leptonic tensor, $L_{\mu \nu}$,
is given by

\begin{equation}\label{leptonic_tensor}
    L_{\mu \nu} =  \frac{1}{2} Tr[ \slashed{k'} \gamma_{\mu} \slashed{k} \gamma_{\nu}] = 2 (k'_{\mu} k_{\nu} + k'_{\nu} k_{\mu} - k' \cdot k \, g_{\mu \nu}),
\end{equation}

\noindent where $k^{\mu}$ and $k'^{\mu}$ are the four momenta of the incident and scattered
electrons respectively, $\gamma^{\mu}$ are the Dirac gamma matrices, and $g^{\mu \nu}$ is the
Minkowski metric tensor. The unpolarized nuclear tensor, $W_A^{\mu \nu}$, is given by
\begin{equation}\label{nuclear_tensor}
    \begin{split}
    W_A^{\mu \nu} = \frac{1}{4 \pi m_A} \frac{1}{2J + 1}\sum_{M_J} \sum_{X} \int d\Pi_{X} \bra{\Psi_A, \bm{P}_A} \hat{J}^{\mu}_A (0) \ket{X} \bra{X} \hat{J}_A^{\nu}(0) \ket{\Psi_A, \bm{P}_A} \\ \times \, (2 \pi)^4 \delta^4(q^\mu + P_A^\mu - p_X^\mu),
\end{split}
\end{equation}
where $\Psi_A$ denotes the nuclear state which implicitly contains the mass $m_A$, 
total angular momentum $J$, and total angular momentum projection $M_J$ labels. $d\Pi_X$ is the Lorentz 
invariant phase space measure for the final states and $P_A^{\mu}$ and $p_X^\mu$ are the four momenta 
of the nucleus and final states.

It is worthwhile to re-write Eq. (\ref{cs_parameterization}) in terms of response functions, which
describes the scattering of the target by a virtual photon with a given polarization. This is done
by including explicit factors of the Minkowski metric in the contraction between the leptonic and
nuclear tensors, $L_{\mu \nu} W_A^{\mu \nu} = L^{\rho \sigma} g_{\rho \mu} g_{\sigma \nu} W_A^{\mu
\nu}$, and $\sum_{\lambda} \epsilon_{\mu}^\lambda(q) \epsilon_{\nu}^{*\lambda}(q) = -g_{\mu
\nu} + \frac{q_\mu q_\nu}{q^2}$. Using that the unpolarized nuclear tensor is symmetric and obeys
the Ward identity, $q_\mu W_A^{\mu \nu} = 0$, we get the following decomposition of Eq.
(\ref{cs_parameterization}):

\begin{equation}\label{cs_response_func}
     \frac{d\sigma}{dE'd\Omega'} = \left( \frac{d\sigma}{d\Omega'} \right)_{Mott} \left\{  \frac{Q^4}{|\bm{q}_{\rm IF}|^4} W_A^L(x_B, Q^2) + \left( \frac{1}{2}\frac{Q^2}{|\bm{q}_{\rm IF}|^2} + \mbox{tan}^2\frac{\theta_e}{2}\right) W_A^T(x_B,Q^2)   \right\},
\end{equation}

\begin{equation}\label{mott_cs}
     \left( \frac{d\sigma}{d\Omega'} \right)_{Mott} = \frac{\alpha_e \mbox{cos}^2\frac{\theta_e}{2}}{4 E^2 \mbox{sin}^4 \frac{\theta_e}{2}},
\end{equation}

\begin{equation}\label{transverse_response}
    W_A^T = W_A^{11} + W_A^{22},
\end{equation}

\begin{equation}\label{longitudinal_response}
    W_A^L = \frac{|\bm{q}_{\rm IF}|^2}{4Q^2} \left( \frac{Q^2}{(q^{+})^2}W_A^{++} + 2 W_A^{+-}+ \frac{(q^{+})^2}{Q^2}W_A^{--} \right),
\end{equation}

\noindent where $W_A^T$ and $W_A^L$ are the transverse and longitudinal nuclear response functions
respectively, and $\theta_e$ is the angle between the incident and scattered electron. With this, we
have reduced the problem down to calculating $W_A^T$ and $W_A^L$, the nuclear structure terms.

\section{Electromagnetic Current}
The following discussion closely follows Refs.~\cite{Vera:2018orr,Vera:2021rnw}
and is presented here for pedagogical purposes; readers are directed to those references for the
complete derivations.

This dissertation focuses on the QE regime, in which the
momentum transfer is large enough to resolve individual nucleons inside the nucleus. In this
kinematic domain the electromagnetic probe couples primarily to a single nucleon, which justifies
the replacement $\hat{J}_A^\mu(0) \to \hat{J}_N^\mu(0)$ in Eq.~(\ref{nuclear_tensor}). The operator
$\hat{J}_N^\mu(0)$ is not, however, the free-nucleon electromagnetic (EM) current, because the
nucleons inside the target are bound and therefore off-shell. Modeling this \emph{off-shell} EM
current operator is the central theoretical challenge of this section.

Historically, the off-shell problem rose to prominence in the early 1980s
with the advent of intermediate-energy $A(e,e'N)$ experiments at
SACLAY~\cite{Alde:1990im,Turck-Chieze:1984ido} and
NIKHEF~\cite{VanDenBrand:1988pv}. The most widely used prescriptions to this
day are those of de Forest~\cite{DeForest:1983ahx}, which extrapolate the
on-shell EM current to off-shell kinematics while retaining on-shell spinors
for the struck nucleon. This construction is, however, fundamentally
incompatible with the covariant Feynman-diagram calculations in which it is
typically embedded, since a bound nucleon in such a calculation is described
by a propagator rather than by on-shell spinors.

\begin{figure}
    \centering
    \includegraphics[width=0.5\linewidth]{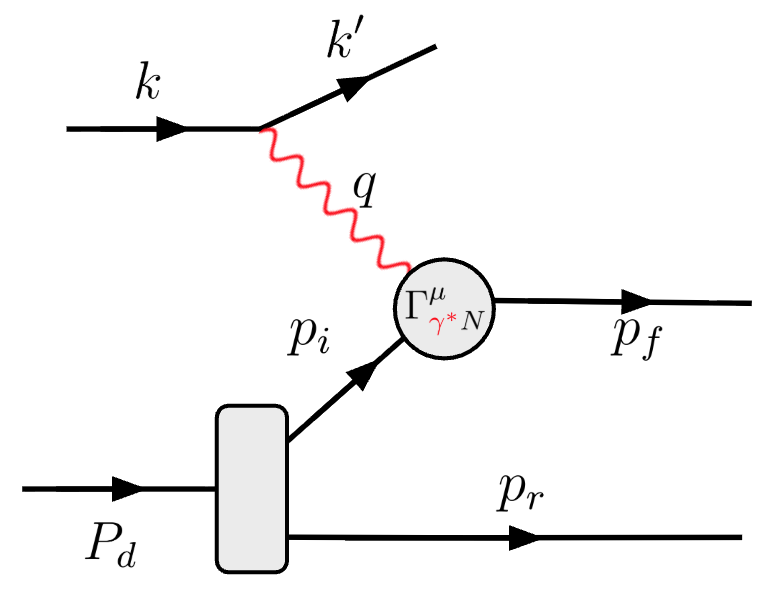}
    \caption{Covariant Feynman diagram describing electron--nucleon scattering inside a deuterium target.}
    \label{fig:e_d_scattering}
\end{figure}

To make this incompatibility concrete, consider the covariant Feynman diagram
for electron--deuteron scattering shown in Fig.~\ref{fig:e_d_scattering}.
Evaluating the diagram with the standard Feynman rules gives
\begin{equation}
\begin{aligned}
    \bar{u}_{\rm IF}(\bm{p}_f,\sigma_f,\tau_f)\, \Gamma_{\textcolor{red}{\gamma^*}N}^\mu\, \frac{\slashed{p}_i + m_N}{p_i^2 - m_N^2 + i\epsilon}\, \bar{u}_{\rm IF}(\bm{p}_r,\sigma_r,\tau_r)\, \Gamma^\nu_{dNN}\, \chi_\nu(M_J),
\end{aligned}
\end{equation}
where $\Gamma_{\textcolor{red}{\gamma^*}N}^\mu$ is the effective electromagnetic vertex between the
virtual photon and bound nucleon, $\Gamma_{{dN}N}^\mu$ represents the vertex of the $d \rightarrow
NN$ transition, and $\chi_\nu(M_J)$ is the polarization vector of the deuteron. Two features of this
fully covariant expression are worth emphasizing. First, the EM vertex
$\Gamma^\mu_{\textcolor{red}{\gamma^*}N}$ is never
sandwiched between two on-shell spinors: the intermediate nucleon enters through a propagator rather
than as an external on-shell state. Second, as discussed in Chapter~\ref{chap:lightfront}, no
nuclear wave function appears explicitly in this representation. The de Forest prescription instead
finds its natural home in \emph{noncovariant} treatments---such as the IF and FF---where intermediate 
nucleons are described by on-shell spinors and the nuclear wave function does appear explicitly.

The advent of high-energy eA experiments, in which electrons transfer large momenta to bound
nucleons and yield final-state nucleons with momenta above a few GeV
\cite{Fomin:2017ydn,Boeglin:2015cha,Arrington:2011xs}, has motivated renewed interest in formulating
off-shell EM currents directly within the LF framework. References~\cite{Vera:2018orr,Vera:2021rnw}
carry out precisely this
construction. Using effective LF diagrammatic rules to sum all time-orderings of
Fig.~\ref{fig:e_d_scattering}---a step required to correctly account for the LF off-shellness of the
intermediate nucleon---they obtain an off-shell EM current operator of the form
\begin{equation}
    \hat{J}_{N}(0) = \sum_{\sigma \sigma' \tau} \int [dp][dp']\, \sqrt{2p'^+}\sqrt{2p^+}\, J_N(\bm{p}',\sigma',\tau;\,\bm{p},\sigma,\tau)\, \hat{a}^\dagger(\bm{p}',\sigma',\tau)\, \hat{a}(\bm{p},\sigma,\tau),
\end{equation}
with single-particle matrix element
\begin{equation}
    J_N(\bm{p}',\sigma',\tau;\,\bm{p},\sigma,\tau) = \bar{u}(\bm{p}',\sigma',\tau) \Gamma^\mu_{\textcolor{red}{\gamma^*}N}\!\left(1 + \frac{\slashed{\Delta}_{p}}{2m_N}\right)\!u(\bm{p},\sigma,\tau),
\end{equation}
where $u(\bm{p},\sigma,\tau)$ is a LF spinor and the EM vertex is parametrized as
\begin{equation}
    \Gamma_{\textcolor{red}{\gamma^*}N}^\mu = \gamma^\mu F_1 + i\sigma^{\mu\nu}q_\nu\, \frac{\kappa F_2}{2m_N} + q^\mu F_3.
\end{equation}
Here $\sigma^{\mu\nu} \equiv \frac{i}{2}[\gamma^\mu,\gamma^\nu]$ is the Pauli tensor, $\kappa$
is the anomalous magnetic moment of the nucleon, $F_1$, $F_2$, $F_3$ are the elastic form
factors, taken to be functions of $Q^2$ alone, with any off-shell dependence neglected, and $q^\mu \neq p'^\mu-p^\mu$. 
Furthermore, the $F_3$ term does not contribute to the cross section by virtue of the gauge invariance of the leptonic current. 
The off-shell information is carried by the operator $\slashed{\Delta}_p$ appearing
alongside $\Gamma^\mu_{\textcolor{red}{\gamma^*}N}$:
\begin{equation}\label{EM_offshell_factors}
    \slashed{\Delta}_p = \frac{\gamma^+}{2}\bigl(p^-_{\rm c} - p^- \bigr),
    \qquad
    p^- = \frac{m_N^2 + (\bm{p}^\perp)^2}{p^+},
\end{equation}
where $p^-$ is the on-shell LF energy and $p^-_{\rm c}$ (subscript ${\rm c}$ for
``conserved'') is the LF energy that the nucleon would carry if energy were conserved at each vertex. The difference $p^-_c - p^-$ is precisely the LF
off-shellness of the bound nucleon.

Refs. \cite{Vera:2018orr,Vera:2021rnw} compared their LF predictions against the widely used de
Forest prescriptions. In essentially all kinematic cases examined, the LF approach predicts smaller
off-shell effects than de Forest at $Q^2 >1~\mathrm{GeV}^2$. More importantly, the LF approach
predicts a pronounced suppression of these effects with increasing $Q^2$, which is understood
intuitively as a decreasing sensitivity of the hard scattering process to the off-shellness of the
target nucleon.

\section{Inclusive Quasi-Elastic Electron--Deuteron Scattering}
Equipped with the LF off-shell EM current of the preceding section, we now
compute inclusive electron--deuteron scattering in the plane-wave impulse
approximation (PWIA). The deuteron nuclear tensor reads
\begin{equation}\label{deuteron_tensor}
\begin{aligned}
    W_d^{\mu\nu} = \frac{1}{4\pi m_d}\cdot\frac{1}{3}\sum_{M_J}\,\frac{1}{2}\sum_{\substack{\sigma_1 \sigma_2 \\ \tau_1 \tau_2}} \int [dp_1][dp_2]\; &\bra{\Psi_d}\hat{J}^{\mu}_N(0)\ket{p_1,p_2}\bra{p_1,p_2}\hat{J}^{\nu}_N(0)\ket{\Psi_d} \\
    &\times (2\pi)^4\,\delta^4(q^\mu + P_d^\mu - p_1^\mu - p_2^\mu),
\end{aligned}
\end{equation}
where the $\tfrac{1}{3}\sum_{M_J}$ averages over the three deuteron polarizations, the factor of
$\tfrac{1}{2}$ accounts for the indistinguishability of the two-nucleon final state, and we drop
the momentum label for nuclear states at rest from this point on.

The current matrix element receives two contributions, corresponding to the
virtual photon striking nucleon~1 or nucleon~2 of the deuteron:
\begin{equation}
\begin{aligned}
    \braket{p_1,p_2|\hat{J}_N^\mu(0)|\Psi_d} = \sum_{\sigma_i}\biggl[\,
    &\frac{m_d}{p_1^+ - q^+}\, J_N\!\bigl(\bm{p}_1,\sigma_1,\tau_1;\,\bm{p}_1-\bm{q},\sigma_i,\tau_1\bigr)\\[-2pt]
    &\quad\times \psi_d\!\left(\tfrac{2(p_1^+ - q^+)}{P_d^+},\bm{p}_1^\perp,\sigma_i,\sigma_2,\tau_1,\tau_2\,\bigg|\,M_J\right) \\
    -\;&\frac{m_d}{p_2^+ - q^+}\, J_N\!\bigl(\bm{p}_2,\sigma_2,\tau_2;\,\bm{p}_2-\bm{q},\sigma_i,\tau_2\bigr)\\[-2pt]
    &\quad\times \psi_d\!\left(\tfrac{2(p_2^+ - q^+)}{P_d^+},\bm{p}_2^\perp,\sigma_i,\sigma_1,\tau_2,\tau_1\,\bigg|\,M_J\right)
    \biggr].
\end{aligned}
\end{equation}
Squaring this matrix element produces interference terms between the two contributions, which are
routinely neglected in calculations of this type, but are kept here.

\begin{figure}
    \centering
    \includegraphics[width=1\linewidth]{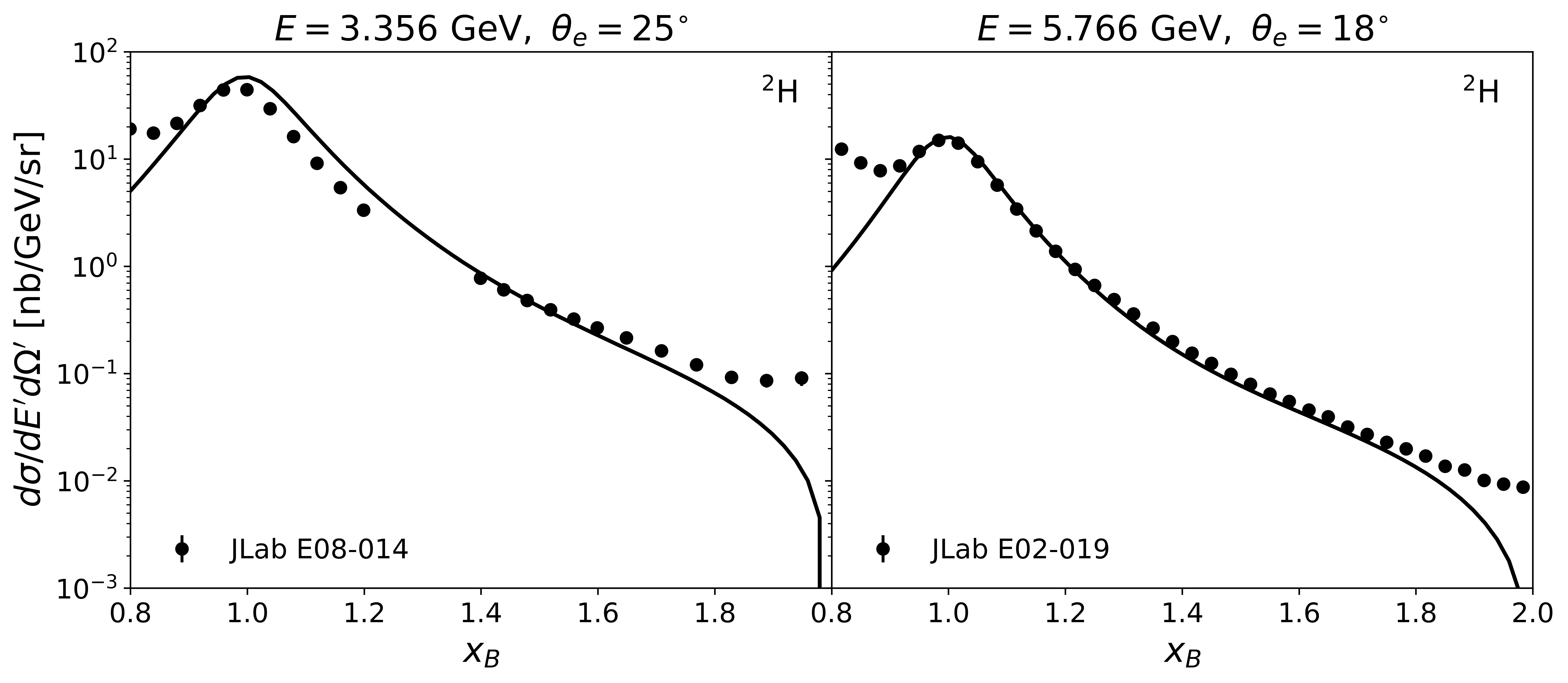}
    \caption{Inclusive electron–deuteron cross section as a function of the Bjorken variable $x_B$, at the JLab E08-014 kinematics $E = 3.356$ GeV, $\theta_e = 25^\circ$ (left) and the JLab E02-019 kinematics $E = 5.766$ GeV, $\theta_e = 18^\circ$ (right). Black points are the experimental data; the solid black curves are our light-front calculation using the AV18 NN interaction.}
    \label{fig:H2_cs_pw}
\end{figure}

Figure~\ref{fig:H2_cs_pw} presents the inclusive electron–deuteron cross section as a function of
the Bjorken variable $x_B$, compared to JLab data at two kinematic settings: $E = 3.356$ GeV,
$\theta_e = 25^\circ$ from JLab E08-014~\cite{Zhang:2025nst} (left panel) and $E = 5.766$ GeV,
$\theta_e = 18^\circ$ from JLab E02-019~\cite{Fomin:2011ng} (right panel). The approximate
LF deuteron wave function obtained from the procedure outlined in
Chapter~\ref{chap:lightfront} reproduces the E02-019 data well, owing to its larger $Q^2$, which
provides higher spatial resolution of the high-momentum tail of the deuteron than the more modest
$Q^2$ of E08-014. The residual discrepancy near the QE peak in the E08-014 panel reflects
physics absent both from the plane-wave impulse approximation and from the AV18 interaction itself —
principally two-body meson-exchange currents and the low-energy tail of the $\Delta(1232)$
excitation, neither of which is generated by a one-body current acting on a purely nucleonic NN
potential without explicit meson-exchange or $\Delta$ degrees of freedom. Both contributions fall in
importance at the higher $Q^2$ of E02-019. Overall, both datasets are well described using the AV18
NN interaction, supporting the use of the minimal-relativity AV18 potential as an approximate
LF NN potential — one capable of capturing the high-momentum two-nucleon dynamics necessary
for calculations developed later in this dissertation.

\section{Summary}

Benchmarking the LF nuclear structure calculation presented in this dissertation requires comparing
to observables that have high enough momentum transfer to probe the short-distance nuclear
structure---relevant for high-energy processes involving quarks at Jefferson Lab and the
Electron-Ion Collider---but also low enough energy transfer to avoid breaking up the nucleons,
probing the quarks inside. Inclusive QE electron-nucleus scattering at
$Q^2>1$~GeV$^2$, $x_B > 1$, and $0.3<\nu<1$~GeV provides exactly such an observable.

Because the electromagnetic interaction is weak, it is accurately modeled through the exchange of a
single virtual photon. By utilizing the Ward identity and the symmetry of the unpolarized nuclear
tensor, the cross section factorizes into transverse and longitudinal response functions, as
expressed in Eq.~\eqref{cs_response_func}. These response functions isolate and carry all the
relevant nuclear structure information. Extracting the actual structure in the QE regime requires a
theoretically consistent treatment of the electromagnetic probe interacting with a bound, off-shell
nucleon. Rather than relying on legacy noncovariant extrapolations like the de Forest prescription,
which fundamentally clash with covariant Feynman diagram calculations by inserting on-shell spinors,
this work employs the effective LF off-shell EM current of Refs.~\cite{Vera:2018orr, Vera:2021rnw}.
This approach correctly utilizes internal propagators and predicts a natural suppression of
off-shell effects at high $Q^2$.

With this off-shell current and the LF nuclear structure developed later in this work, we can
accurately calculate QE cross sections within the plane-wave impulse approximation. Applying
this framework to deuterium provides a critical baseline test. We explicitly retain the subtractive
interference between the individual nucleon contributions when squaring the current matrix elements.
Our calculation for deuterium supports that the approximate LF wave function obtained from the
procedure outlined in Chapter~\ref{chap:lightfront}, utilizing the AV18 interaction, works well. The
successful reproduction of higher-$Q^2$ JLab data confirms that this framework captures the
high-momentum two-nucleon dynamics necessary for the calculations developed in the subsequent
chapters, with the expected minor deviations at lower $Q^2$ correctly attributable to omitted
two-body meson-exchange currents and $\Delta$-excitations.

% !TEX root = ../uwthesis.tex
\chapter{Density Functional Theory}\label{chap:dft}
Density Functional Theory (DFT) is one of the most popular ab-initio approaches towards studying
many-body systems. Developed by Hohenberg and Kohn in 1964 \cite{Hohenberg:1964zz}, and later
expanded upon by Kohn and Sham one year later \cite{Kohn:1965zzb}, DFT was originally utilized to
study many-electron atomic systems, but has seen great utility in many fields such as chemistry,
nuclear physics, and condensed matter physics. The basic idea is that the ground-state energy of a
many-body system can be represented in terms of the interacting ground-state density alone.
Furthermore, under the Kohn-Sham (KS) scheme, the interacting ground-state density can be reproduced
by the ground-state density of an auxiliary non-interacting system. Thus, one reduces the many-body
problem from solving a coupled N-body Schrodinger equation, to solving for the spectra of a one-body
Schrodinger equation. This makes DFT calculations comparatively simple to implement, often very
accurate, and computationally feasible for systems involving large particle number. For a more
comprehensive review and further reading, see Refs. \cite{RevModPhys.87.897, doi:10.1142/9872}.

\section{Coulombic Density Functional Theory}
DFT rests on two fundamental theorems, proved by Hohenberg and Kohn for electromechanical many-body
systems in an external potential $v_{ext}(\textbf{r} )$\cite{Hohenberg:1964zz}:

\begin{enumerate}
    \item The ground state energy for a given external potential, $E_v$, is a unique functional of
the electron density $\rho(\textbf{r})$,

        \begin{center}\label{density_functional}
            $E_v[\rho] = F_{HK}[\rho] + \int d^3r \,  v_{ext}(\textbf{r}) \,\rho(\textbf{r})$,
        \end{center}

    where the functional $F_{HK}$ does not depend on $v_{ext}$ nor on the particle number of the
system. It only depends on the form of the interaction between the particles, and hence is
universal.

    \item The electron density that minimizes the energy of the energy functional is the true
electron density, corresponding to the full solution of the Schrodinger equation.
\end{enumerate}

The fundamental issue facing us is that the Hohenberg-Kohn (HK) theorems only show the existence of
$F_{HK}$, but offer no guidance in constructing the functional. This task is put into better context
under the KS Scheme \cite{Kohn:1965zzb}, which decomposes $F_{HK}$ by:

\begin{equation}\label{KS_scheme}
    F_{HK}[\rho] = T_s[\rho] + E_H[\rho] + E_{xc}[\rho]
\end{equation}

\noindent where $T_s$ is the kinetic energy of the non-interacting system, $E_H$ is the Hartree
energy, and in this context, is the energy corresponding to the classical Coulomb potential, and
$E_{xc}$ is the exchange-correlation energy which, by definition, contains all the many-body quantum
effects not included in $T_s$ and $E_H$. One can perform a variation of $E_v$ with respect to 
$\rho$ under a fixed particle number, $N$, constraint:

\begin{equation}\label{variation_energy_functional}
     \frac{\delta}{\delta \rho(\textbf{r})} \left\{   E_v[\rho] - \mu \left( \int d^3r' \rho(\textbf{r}') - N \right) \right\} = 0,
\end{equation}

\noindent where $\mu$ is the Lagrange multiplier associated to constant particle number. With Eq.
(\ref{KS_scheme}) this yields,

\begin{equation}\label{variational_equation}
    \frac{\delta T_s[\rho]}{\delta \rho(\textbf{r})} + V_{KS}(\textbf{r}) = \mu,
\end{equation}

\begin{equation}\label{DFT_potential}
    V_{KS}(\textbf{r}) = v_{ext}(\textbf{r}) + \frac{\delta(E_H[\rho] + E_{xc}[\rho])}{\delta \rho}.
\end{equation}

\noindent Consider a completely different problem: a system of non-interacting electrons moving in
an external potential $V(\mathbf{r})$. If we applied the exact same variational principle to that
system's energy functional, we would obtain Eq. (\ref{variational_equation}) but with
$V_{KS}(\mathbf{r}) \rightarrow V(\mathbf{r})$. Hence, Eq. (\ref{variational_equation}) is satisfied
by the density obtained from solving the single particle Schrodinger equation for electrons moving
in an external potential $V_{KS}(\mathbf{r})$,

\begin{equation}
    \left[ -\frac{\nabla^2}{2m} + V_{KS}(\textbf{r}) \right] \phi_i(\textbf{r}) = \epsilon_i  \, \phi_i(\textbf{r}),
\end{equation}

\noindent where

\begin{equation}\label{density_decomposition}
    \rho(\textbf{r}) = \sum_{i=1}^N | \phi_i(\textbf{r})|^2,
\end{equation}

\noindent and the N-body wavefunction is a Slater determinant of the N lowest-lying single-particle
orbitals. Since $V_{KS}(\mathbf{r})$ also depends on the electron density, these equations must be
evaluated self-consistently. In practice, $V_{KS}(\mathbf{r})$ is either determined from
first-principles, which has been done in non-relativistic Coulombic systems, or phenomenologically
with parameters optimized by fitting to a set of data. The ladder is the most widely used way for
DFT applications to nuclear systems.

\section{Nuclear Density Functional Theory}
For DFT applications to nuclear systems, one finds that the first HK theorem does not hold. The
proof of the first theorem is based on the existence of $v_{ext}(\mathbf{r})$. However, nuclei are
self-bound objects, meaning that the system is not confined due to an external potential, but due to
the inter-nucleonic interactions contained in the $F_{HK}$ functional. It has been shown that for
self-bound systems, the HK theorems do hold for the intrinsic density
\cite{Engel:2006qu,Giraud:2008zz}. This means that nuclei that can be described using DFT all come
from the same universal energy functional, $F_{HK}$. However, the independent, single-particle,
description obtained from the KS scheme is lost because the intrinsic density depends on relative
coordinates. In practice, one artificially pins the nucleus to the origin by constructing energy
density functionals that depend on absolute coordinates, breaking translational invariance. Doing so
allows one to regain the independent-particle description of the system, but at the cost of losing
the fundamental backing from the intrinsic density HK theorems. Regardless, such approaches to
nuclear structure have seen remarkable success in describing a plethora of nuclear observables and
dynamics.

\section{Light-Front Density Functional Theory}
The concepts of DFT are general and can be applied to relativistic frameworks where energy density
functionals can be naturally obtained from field-theoretic Lagrangians. Historically,
field-theoretic descriptions of nuclear systems began with the Walecka model \cite{WALECKA1974491}.
Originally developed to study neutron stars, the Walecka model is a phenomenological model of
nucleons interacting through the exchange of scalar and vector mesons which mimic nuclear attraction
and repulsion respectively. The extension of the model to finite nuclear systems was later developed
by Horowitz and Serot, called Relativistic Mean-Field Theory (RMF) \cite{Horowitz:1981xw}. Using
Dyson's equations for the dressed single-particle propagator, Horowitz and Serot developed a set of
self-consistent equations where nucleon densities act as current sources that generate the meson
fields, which in turn influence the individual nucleon dynamics. In what follows we recast this
framework in the language of Kohn–Sham DFT and extend it to the LF, with a further generalization to
density-dependent couplings.

The starting point is a phenomenological field-theoretic Lagrangian, $\mathcal{L}$, describing
nucleons interacting through the exchange of isoscalar-scalar $\sigma$-mesons, isoscalar-vector
$\omega$-mesons, isovector-vector $\rho$-mesons, and photons.
% Equation (1) - Total Lagrangian density
\begin{equation}
\mathcal{L} = \mathcal{L}_N + \mathcal{L}_m + \mathcal{L}_{int},
\label{eq:lagrangian_total}
\end{equation}
$\mathcal{L}_N$ is the Dirac Lagrangian describing free nucleons
% Equation (2) - Free nucleon Lagrangian
\begin{equation}
\mathcal{L}_N = \bar{\psi}(i\gamma_\mu \partial^\mu - m_N)\psi,
\label{eq:lagrangian_nucleon}
\end{equation}
$\mathcal{L}_m$ corresponds to the free meson and photon fields,
% Equation (3) - Free meson and electromagnetic field Lagrangian
\begin{equation}
\begin{split}
\mathcal{L}_m =&\ \frac{1}{2}\partial_\mu \sigma \partial^\mu \sigma - \frac{1}{2}m_\sigma^2 \sigma^2 - \frac{1}{4}\Omega_{\mu\nu}\Omega^{\mu\nu} + \frac{1}{2}m_\omega^2 \omega_\mu \omega^\mu \\
&- \frac{1}{4}\vec{R}_{\mu\nu} \cdot \vec{R}^{\,\mu\nu} + \frac{1}{2}m_\rho^2 \vec{b}_\mu \cdot \vec{b}^{\,\mu} - \frac{1}{4}F_{\mu\nu}F^{\mu\nu},
\end{split}
\label{eq:lagrangian_meson}
\end{equation}
where the field tensors are given by
% Equation (4) - Omega field tensor
\begin{equation}
\Omega_{\mu\nu} = \partial_\mu \omega_\nu - \partial_\nu \omega_\mu,
\label{eq:omega_tensor}
\end{equation}
% Equation (5) - Rho field tensor
\begin{equation}
\vec{R}_{\mu\nu} = \partial_\mu \vec{b}_\nu - \partial_\nu \vec{b}_\mu,
\label{eq:rho_tensor}
\end{equation}
% Equation (6) - Electromagnetic field tensor
\begin{equation}
F_{\mu\nu} = \partial_\mu A_\nu - \partial_\nu A_\mu.
\label{eq:em_tensor}
\end{equation}
The interaction terms between the meson/electromagnetic fields are encapsulated in
$\mathcal{L}_{int}$
% Equation (7) - Interaction Lagrangian
\begin{equation}
\mathcal{L}_{int} = -g_\sigma \bar{\psi}\psi\sigma - g_\omega \bar{\psi}\gamma^\mu \psi \omega_\mu - \frac{1}{2}g_\rho \bar{\psi}\gamma^\mu \vec{\tau}\psi \cdot \vec{b}_\mu - e\bar{\psi}\gamma^\mu \frac{1+\tau_3}{2}\psi A_\mu,
\label{eq:lagrangian_int}
\end{equation}
In the above expressions, $m_N$ is the nucleon mass, while $m_m$ and $g_m$ ($m = \sigma, \omega,
\rho$) denote the meson masses and meson--nucleon coupling constants. The electromagnetic field is
denoted $A^\mu$, with coupling $e$, and the $\rho$-meson field is denoted $\vec{b}^{\, \mu}$.

RMF treats the meson mediators as classical fields. Because the meson fields are not quantized,
quantum-mechanical exchange and correlation interactions are neglected here; hence, RMF is
intrinsically a Hartree-level theory. Furthermore, by restricting our focus to static,
spherically-symmetric nuclei, the pion fields do not contribute to the ground state due to parity
conservation. These meson fields phenomenologically capture distinct aspects of the nucleon-nucleon
interaction: the $\sigma$-meson simulates intermediate-range attraction, the $\omega$-meson provides
short-range repulsion, and the $\rho$-meson accounts for the isospin dependence. It is well known
that pions play a critical role in the nuclear force, providing the long-range attraction and the
short-range tensor force. Although the underlying assumptions of the RMF framework prevent the pion
fields from contributing explicitly to the ground state, the specific combination of the remaining
effective meson fields is designed to implicitly absorb and replicate their overall physical
effects.

From Noether's Theorem, the conserved four-momenta follows from the energy momentum tensor
$\mathcal{T}^{\mu\nu} = \sum_n \frac{\partial \mathcal{L}}{\partial(\partial_\mu\phi_n)}
\partial^\nu\phi_n - g^{\mu\nu}\mathcal{L}$, with $\phi_n$ running over all fields:
\begin{equation}\label{EMT}
    \begin{gathered}
        \mathcal{T}^{\mu\nu} = \bar{\psi} i \gamma^\mu \partial^\nu \psi - \Omega^{\mu \alpha} \partial^\nu \omega_\alpha - F^{\mu \alpha} \partial^\nu A_\alpha \\ -  \vec{R}^{\mu \alpha} \cdot \partial^\nu \vec{b}_\alpha + \partial^\mu \sigma \partial^\nu \sigma- g^{\mu \nu} \mathcal{L}\\
        -\frac{1}{2}g^{\mu \nu} \left[ (\partial_{\alpha}\omega^\alpha )^2 +  (\partial_{\alpha}\vec{b}^{\,\alpha})^2 + (\partial_{\alpha}A^\alpha)^2 \right],
    \end{gathered}
\end{equation}
where the terms in the square brackets are all equal to zero by the equations of motion. The
conserved LF momenta on surfaces of constant $x^+$ are
\begin{equation}\label{noether_charge}
    P^\mu = \frac{1}{2} \int dx^- d\bm{x}^\perp \mathcal{T^{+ \mu}}.
\end{equation}

In developing LF DFT, one must carefully select a viable energy density functional. The natural
choice of the light-front Hamiltonian, $P^-$, is fundamentally flawed in this context; given the
light-front dispersion relation $P_A^- = \frac{m_A^2 + (\bm{P}_A^\perp)^2}{P^+_A}$, the system can
trivially minimize its energy by driving the longitudinal momentum $P_A^+ \rightarrow \infty$. The
linear combination $\frac{1}{2}(P^+ + P^-)$ is more promising. It avoids the unphysical $P^+_A
\rightarrow \infty$ minimization limit, and connected to the rest mass condition on the LF $P^+_A =
P^-_A = m_A$ \cite{Blunden:1999gq}. We adopt this combination as our energy functional throughout.

At this stage, we can introduce several critical simplifications by leveraging the physical
constraints of our specific approach. First, because we focus on static, ground-state nuclei, the
meson and photon fields exhibit no dependence on LF time, $x^+$. Second, by restricting our analysis
to spherically symmetric, even-even nuclei, the perpendicular components of the meson and photon
fields vanish due to the time-reversal invariance of the nuclear ground state, and the plus and
minus components of the vector fields are rendered equal. Hence, we remove the Lorentz indicies from
the vector meson and photon fields, labeling them with subscript 0. Finally, the charged $\rho$
meson fields do not contribute; this arises because we operate within the Hartree
approximation—thereby neglecting exchange-correlation effects—and because the nucleus has a definite
total charge.

On the $x^+=0$ hypersurface, we quantize the nucleon fields into field operators by expanding them
in terms of single-particle orbitals indexed by $j$,
\begin{equation}\label{field_operator}
    \hat{\psi}(x) = \sum_j \left[ \psi_j(\bm{x}) \,\hat{a}_j + \phi_j(\bm{x})  \, \hat{b}_j^\dagger \right],
\end{equation}
where $\hat{a}_j$ ($\hat{b}_j^\dagger$) is the annihilation (creation) operator for a nucleon
(anti-nucleon) in state $j$, and $\psi_j(\bm{x})$ ($\phi_j(\bm{x})$) is the corresponding
single-particle wavefunction. The creation and annihilation operators obey fermionic anti-commutation relations
\begin{equation}\label{creation_annihilation_operators}
    \{ \hat{a}_i, \hat{a}_j^{\dagger}\} = \delta_{ij} \;\;\; \mbox{and} \;\;\; \{ \hat{a}_i, \hat{a}_j\} = \{ \hat{a}_i^{\dagger}, \hat{a}_j^{\dagger}\} = 0,
\end{equation}
and the nucleon and anti-nucleons anti-commute. The ground state of the nucleus with baryon number
$A$ is constructed by a product of the $A$-lowest lying nucleon single particle states, i.e. the
Slater determinant
\begin{equation}\label{gs_wf}
    \ket{\Psi_A^{MF}} = \prod_{i=1}^A \hat{a}_i^{\dagger} \ket{\Omega} \;\;\; \mbox{with} \;\;\; \braket{\Psi_A^{MF}|\Psi_A^{MF}} = 1.
\end{equation}
Note that our nuclear model contains no anti-nucleons and that the nuclear state is built on top of
the interacting vacuum $\ket{\Omega}$. When evaluating matrix elements in the nuclear ground state,
one must normal-order with respect to the trivial LF vacuum in order to remove the zero-point
energy. Doing so yields two terms: one corresponding to the positive-energy nucleons and the other
to contributions from the Dirac sea, the latter of which we discard \cite{Reinhard:1989zi}. Building
the nucleus exclusively from the positive-energy solutions---i.e.\ the physical nucleons---together
with this prescription for evaluating matrix elements constitutes the no-sea scheme, which
explicitly omits the dynamical effects of anti-nucleons. Although it is commonly referred to in the
literature as the ``no-sea approximation,'' in the present context it should not be regarded as an
approximation at all; it rests on no underlying physical or mathematical justification and is better
understood as a foundational modeling choice \cite{doi:10.1142/9872}.

Applying the simplifications above together with the quantization just introduced, the Hamiltonian
density evaluated with respect to the nuclear ground state is
\begin{equation}\label{LF_hamiltonian_density}
    \mathcal{H}_{\rm LF}^{RMF} = \frac{1}{2}\mathcal{T}^{+-} + \frac{1}{2}\mathcal{T}^{++} = \mathcal{H}_{\psi} + \mathcal{H}_{\sigma} + \mathcal{H}_{\omega} + \mathcal{H}_{\rho} + \mathcal{H}_{A} + \mathcal{H}_\text{int},
\end{equation}
with
\begin{equation}\label{minus_decomposition}
    \begin{gathered}
        \mathcal{H}_{\psi} = \sum_{i=1}^A\bar{\psi_i}(i\bm{\gamma}^{\perp} \cdot \bm{\partial}^{\perp} + i\gamma^3 \partial^+ + m_N) \psi_i \\
        \mathcal{H}_{\sigma} = \frac{1}{2}[(\bm{\partial}^{\perp} \sigma)^2 + (\partial^+ \sigma)^2 + m_\sigma^2 \sigma^2]\\
        \mathcal{H}_{\omega} = -\frac{1}{2}[(\bm{\partial}^{\perp} \omega_0)^2 + (\partial^+ \omega_0)^2 + m_\omega^2 \omega_0^2]  \\
        \mathcal{H}_{\rho} = -\frac{1}{2}[(\bm{\partial}^{\perp} b_0)^2 + (\partial^+ b_0)^2 + m_\rho^2 \,b_0^2]  \\
        \mathcal{H}_A = -\frac{1}{2}[(\bm{\partial}^{\perp} A_0)^2 + (\partial^+ A_0)^2] \\
        \mathcal{H}_{\text{int}} = g_\sigma \rho_s \sigma + g_\omega \rho_B \omega_0 + g_\rho \rho_3 b_0 + e \rho_p A_0.
    \end{gathered}
\end{equation}
where we have introduced the scalar density $\rho_s$, the baryon density $\rho_B$, the isovector
density $\rho_3$, and the proton density $\rho_p$
% Equation (9) - Isoscalar-scalar density
\begin{equation}
\rho_s(\bm{x}) = \sum_{i=1}^{A} \bar{\psi}_i(\bm{x})\psi_i(\bm{x}),
\label{eq:scalar_density}
\end{equation}

% Equation (10) - Isoscalar-vector current
\begin{equation}
\rho_B(\bm{x}) = \sum_{i=1}^{A} \bar{\psi}_i(\bm{x}) \gamma^0\psi_i(\bm{x}),
\label{eq:isoscalar_vector_current}
\end{equation}

% Equation (11) - Isovector-vector current
\begin{equation}
\rho_3(\bm{x}) = \sum_{i=1}^{A}\bar{\psi}_i(\bm{x}) \gamma^0 \frac{\tau_3}{2} \psi_i(\bm{x}),
\label{eq:isovector_vector_current}
\end{equation}

% Equation (12) - Electromagnetic current
\begin{equation}
\rho_p(\bm{x}) = \sum_{i=1}^{A} \bar{\psi}_i(\bm{x}) \gamma^0 \frac{1+\tau_3}{2} \psi_i(\bm{x}).
\label{eq:em_current}
\end{equation}
It is important to remember that when evaluating matrix elements of the field operators with respect
to the nuclear ground state, you must normal order with respect to the free vacuum. The LF energy
density functional is obtained by integrating the Hamiltonian density over the LF spatial volume,
\begin{equation}\label{eq:energy_functional}
    E^{RMF}_{\rm LF}[\psi,\bar{\psi},\sigma,\omega,\rho_3,A] = \frac{1}{2}\int dx^-\, d\bm{x}^\perp\, \mathcal{H}_{\rm LF}^{RMF}(x^-,\bm{x}^\perp),
\end{equation}

As discussed in Appendix~\ref{appendix:dirac}, only the $+$ projection of the nucleon field,
$\psi_i^+ \equiv \Lambda^+\psi_i$, is dynamical on the LF; the $-$ projection is a constraint solved
for in terms of $\psi_i^+$. We therefore impose the normalization on the $+$ components only,
\begin{equation}\label{eq:normalization}
    \sum_{i=1}^A \frac{1}{2}\int dx^-\, d\bm{x}^\perp\, |\psi_i^+(\bm{x})|^2 = A.
\end{equation}
Minimizing $E^{RMF}_{\rm LF}$ with respect to $\psi_i^\dagger$ subject to
Eq.~\eqref{eq:normalization} introduces a set of Lagrange multipliers $p_i^-$ that play the role of
single-particle LF energies. The stationarity condition reads
\begin{equation}\label{eq:variation}
    \frac{\delta \mathcal{H}^{RMF}_{\rm LF}}{\delta \psi_i^\dagger(\bm{x})} - p_i^-\,\Lambda^+ \psi_i(\bm{x}) = 0.
\end{equation}
Performing the functional derivatives on $\mathcal{H}_{\rm LF}^{RMF}$ yields the LF Dirac equation
\begin{equation}\label{eq:sp_dirac_RMF}
    \left[\,i\bm{\alpha}^{\perp} \!\cdot \bm{\partial}^{\perp} + i\alpha^3\partial^+ + \gamma^0m_N + \Sigma_H(\bm{x})\,\right]\psi_i(\bm{x}) = p_i^- \Lambda^+ \psi_i(\bm{x}),
\end{equation}
with the Hartree self-energy
\begin{equation}\label{eq:hartree_selfenergy}
    \Sigma_H(\bm{x}) = g_\sigma\,\gamma^0\sigma(\bm{x}) + g_\omega\,\omega_0(\bm{x}) + \tfrac{1}{2}g_\rho\tau_3\,b_0(\bm{x}) + \tfrac{1}{2}e(1+\tau_3)A_0(\bm{x}),
\end{equation}
where $\alpha^\mu = \gamma^0 \gamma^\mu$. The meson and photon fields obey constrained Klein-Gordon
equations
\begin{equation}\label{eq:meson_kg}
    \begin{gathered}
        \left[ \bm{\partial}^{\perp} \cdot \bm{\partial}^{\perp}  + \partial^+ \partial^+  - m_\sigma^2 \,\right]\sigma(\bm{x}) = g_\sigma \rho_s(\bm{x}), \\
        \left[ \bm{\partial}^{\perp} \cdot \bm{\partial}^{\perp} + \partial^+ \partial^+ - m_\omega^2 \, \right]\omega_0(\bm{x}) = -g_\omega \rho_B(\bm{x}), \\
        \left[ \bm{\partial}^{\perp} \cdot \bm{\partial}^{\perp} + \partial^+ \partial^+ - m_\rho^2 \, \right]b_0(\bm{x}) = -g_\rho \rho_3(\bm{x}), \\
        \left[ \bm{\partial}^{\perp} \cdot \bm{\partial}^{\perp} + \partial^+ \partial^+ \, \right]A_0(\bm{x}) = -e \rho_p(\bm{x}). \\
    \end{gathered}
\end{equation}
Equations (\ref{eq:sp_dirac_RMF}) and (\ref{eq:meson_kg}) are solved numerically and
self-consistently. A notable computational difficulty in the LF approach is that the Dirac
single-particle Hamiltonian must be solved in cylindrical coordinates, with states labeled by their
energy and the total angular momentum projection quantum number $m_j$. Ref.~\cite{Blunden:1999gq}
carried out this calculation and verified that rotational invariance is restored by checking that
the $m_j$ states exhibit the correct degeneracy. In the same paper, along with
Ref.~\cite{Smith:2002ci}, the authors further develop a procedure for recasting the LF Dirac
single-particle Hamiltonian into a spherically symmetric form. However, they state the manipulations
as a way to use IF DFT to 'approximate' LF DFT results. We now discuss this procedure and show that
this is not an approximation.

\section{Connecting IF and LF Formulations of DFT}
To recast equation~\eqref{eq:sp_dirac_RMF} into a form that makes the underlying spherical symmetry
of the nucleus manifest, we perform two successive steps.

\paragraph{Step 1 — Replace $x^-$ with the spatial $z$ coordinate.} Since we solve for the states of
the Hamiltonian at $x^+=0$, $x^- = -2z$. In other words, the LF longitudinal direction can be mapped
onto the ordinary spatial coordinate $z$. The LF plus-derivative becomes
\begin{equation}\label{eq:deriv_change}
    \partial^+ \equiv 2\,\frac{\partial}{\partial x^-} = -\frac{\partial}{\partial z} = \partial^3,
\end{equation}
where in the last equality we used $\partial^3 = -\partial/\partial z$ in the mostly-minus metric.
We further relabel the spatial argument $\bm{x}\to\bm{x}_{\rm IF}$, as discussed in the conventions
and notations in the beginning of the dissertation.

\paragraph{Step 2 — Phase redefinition.} Writing
\begin{equation}\label{eq:phase}
    \psi_i(\bm{x}_{\rm IF}) = \Psi_i(\bm{x}_{\rm IF})\,e^{ip_i^- z/2},
\end{equation}
the action of $i\alpha^3\partial^3$ on $\psi_i$ produces two terms,
\begin{equation}\label{eq:phase_action}
    i\alpha^3\partial^3\psi_i(\bm{x}_{\rm IF}) = i \alpha^3 \left[\, \partial^3\Psi_i(\bm{x}_{\rm IF}) - i\tfrac{p_i^-}{2} \Psi_i(\bm{x}_{\rm IF})\,\right] e^{ip_i^- z/2}.
\end{equation}
Moving the second term in Eq. (\ref{eq:phase_action}) to the right-hand side of
Eq.~\eqref{eq:sp_dirac_RMF} collapses it to $\tfrac{p_i^-}{2}\Psi_i(\bm{x}_{\rm IF})$.

\paragraph{Resulting set of equations.} Putting the two steps together, the set of LF RMF equations
takes the form
\begin{equation}\label{eq:dirac_IF}
    \left[\,i\bm{\alpha}^{\perp}\!\cdot\bm{\partial}^{\perp} + i\alpha^3\partial^3 + \gamma^0 m_N + \Sigma_H(\bm{x}_{\rm IF})\,\right]\Psi_i(\bm{x}_{\rm IF}) = \frac{p_i^-}{2}\,\Psi_i(\bm{x}_{\rm IF}),
\end{equation}
\begin{equation}\label{eq:meson_k_IF}
    \begin{gathered}
        \left[ \bm{\partial}^{\perp} \cdot \bm{\partial}^{\perp}  + \partial^3 \partial^3  - m_\sigma^2 \,\right]\sigma(\bm{x}_{\rm IF}) = g_\sigma \rho_s(\bm{x}_{\rm IF}), \\
        \left[ \bm{\partial}^{\perp} \cdot \bm{\partial}^{\perp} + \partial^3 \partial^3 - m_\omega^2 \, \right]\omega_0(\bm{x}_{\rm IF}) = -g_\omega \rho_B(\bm{x}_{\rm IF}), \\
        \left[ \bm{\partial}^{\perp} \cdot \bm{\partial}^{\perp} + \partial^3 \partial^3 - m_\rho^2 \, \right]b_0(\bm{x}_{\rm IF}) = -g_\rho \rho_3(\bm{x}_{\rm IF}), \\
        \left[ \bm{\partial}^{\perp} \cdot \bm{\partial}^{\perp} + \partial^3 \partial^3 \, \right]A_0(\bm{x}_{\rm IF}) = -e \rho_p(\bm{x}_{\rm IF}). \\
    \end{gathered}
\end{equation}
in which spherical symmetry can now be exploited directly. In fact, Eqs.~\eqref{eq:dirac_IF} and
~\eqref{eq:meson_k_IF} are the exactly same equations that are solved in the original IF DFT
formulation, with $p_i^-/2$ as the single-particle energy of the nucleon \cite{Horowitz:1981xw}.
Hence, IF DFT calculations can be repurposed for LF DFT.

The claim of Ref.~\cite{Blunden:1999gq} was not that they are approximate, but rather that the
Fourier transform of the single-particle LF wavefunction \textemdash obtained from the IF wavefunction \textemdash has
non-zero support at $p^+=0$, which cannot occur for massive particles. On this basis, the authors
argued that the equivalence between IF and LF DFT is only approximate, attributing the discrepancy
to a breaking of the LF spectral condition. However, the non-zero support at $p^+=0$ does not signal
physics that the IF formulation fails to capture; it is instead a consequence of the
independent-particle nature of DFT models. The relation $p^+>0$ for massive particles holds only for
momentum eigenstates, and single-particle orbitals are not momentum eigenstates. The support at
$p^+=0$ is therefore to be expected: it would persist even if the LF--IF relation were never invoked
and the entire calculation were carried out directly on the LF. More precisely, single-particle
momentum distributions of the nucleus acquire support at $p^+=0$ because the nuclear state is not an
eigenstate of LF momentum---DFT calculations of nuclear structure pin the nucleus to the origin.
This interpretation is reinforced by studies of infinite nuclear matter on the light front, where
translational invariance is preserved, the nucleons are momentum eigenstates, and the support at
$p^+=0$ vanishes~\cite{Miller:1999mi, Miller:2000kv}. One can correspondingly expect that properly
projecting the nuclear state onto the zero-momentum eigenstate would remove the support at $p^+=0$,
leaving a node in the wavefunction. Methods for carrying out this projection, and for incorporating
it into the variational calculation, have been outlined in
Refs.~\cite{Peierls:1957er,PEIERLS1962154}; their application here is left to future work and lies
outside the scope of this dissertation.

It should be emphasized that this support is unrelated to the so-called zero modes of the quantized
Dirac fields---a long-standing and still unresolved problem in light-front quantization
\cite{Brodsky:1997de,Heinzl2001}. We avoid the zero modes by virtue of the no-sea scheme: the
single-particle wavefunctions from which our quantized field is built were obtained without
reference to anti-nucleon dynamics. Were anti-nucleons admitted into the nuclear ground state, the
zero modes could no longer be avoided. More fundamentally, the equivalence between the IF and LF
formulations of DFT rests on their shared underlying assumptions---static meson fields and the
no-sea scheme--which lead to identical equations.

\begin{figure}
    \centering
    \includegraphics[width=1\linewidth]{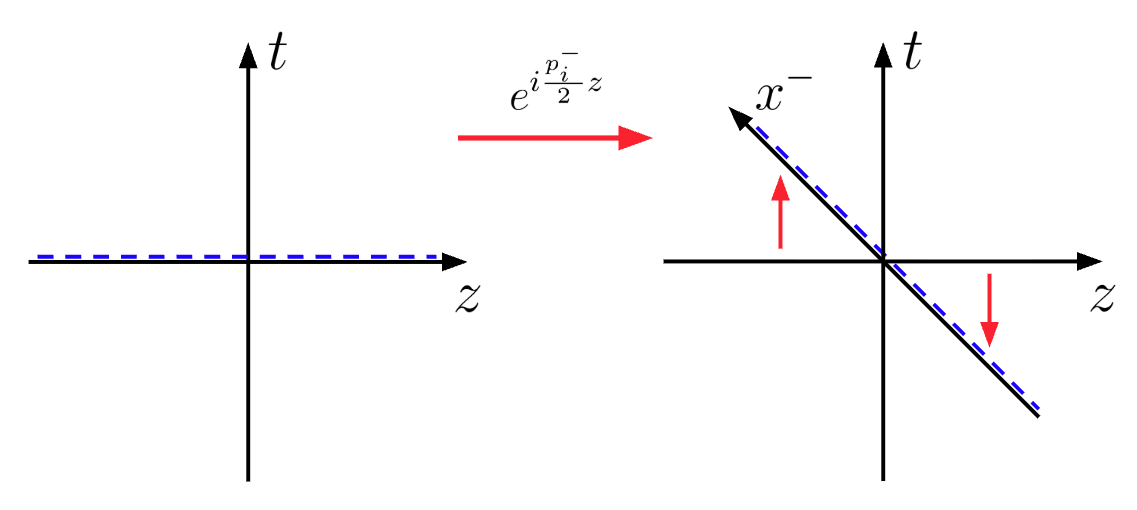}
    \caption{(color online) Schematic picture of converting instant form (IF) single-particle wavefunctions into light-front wavefunctions. The diagram on the left illustrates the IF wavefunction that is solved for at $t=0$ represented by the blue dashed line. The red arrows illustrate how the $z-$dependent time-evolution places the wavefunction on the surface $x^+=0$, yielding the analogous single-particle LF wavefunction.}
    \label{fig:et_to_lf}
\end{figure}

Furthermore, Eq.~\eqref{eq:phase} admits a natural geometric interpretation. In IF quantization,
states are defined on surfaces of constant $x^0=t$, and the Hamiltonian generates evolution from one
such surface to the next; in IF DFT the Dirac fields are quantized at $t=0$, hence the wavefunctions
solved for on that surface. LF quantization proceeds analogously, but on surfaces of constant $x^+$,
with the Dirac fields of LF DFT quantized at $x^+=0$. Equation~\eqref{eq:phase} bridges the two. It
takes the IF wavefunction defined at $t=0$ and applies a $z$-dependent time-evolution that
transports it onto the $x^+=0$ hypersurface, thereby yielding the corresponding LF wavefunction (see
Fig. \ref{fig:et_to_lf}). The relation between LF and IF wavefunctions has been examined previously
in Ref.~\cite{Miller:2009fc}. Focusing on two-body bound states, the authors showed that the IF and
LF wavefunctions can each be recovered from the Bethe--Salpeter wavefunction by integrating over
$k^0$ and $k^-$, respectively, but that neither can be obtained directly from the other. That a
direct map exists in our case is a consequence of the single-particle nature of the problem.

From here, one can obtain the single-particle Dirac equation and the Klein-Gordon equations for the
meson fields that characterize the original RMF formulation in Ref. \cite{Horowitz:1981xw}. However,
despite its successes, the original RMF formulation with constant meson-nucleon couplings failed to
reproduce several key properties of nuclear matter and finite nuclei. Most notably, it predicted a
nuclear-matter incompressibility of roughly 500--600 MeV, far in excess of the empirical value of $K
\approx 210$--$240$ MeV \cite{Garg:2018uam}, and yielded poor agreement with nuclear surface
properties and the equation of state away from saturation. Furthermore, the original results
underbind finite nuclei by roughly 2 MeV per nucleon. These deficiencies were traced to the absence
of medium dependence in the effective interaction. Dirac--Brueckner--Hartree--Fock (DBHF)
calculations, which resum ladder diagrams built from a realistic bare nucleon-nucleon interaction in
the nuclear medium \cite{Brockmann:1990cs}, demonstrated that the scalar and vector nucleon
self-energies acquire a pronounced density dependence through Pauli blocking, short-range
correlations, and implicit three-body effects---features that cannot be captured by fixed couplings.
This microscopic insight motivated the introduction of density-dependent meson-nucleon vertices
\cite{Fuchs:1995as,Typel:1999yq}, which phenomenologically encode the physics missing from a Hartree
description.

\begin{figure}[H]
    \centering
    \includegraphics[width=0.7\linewidth]{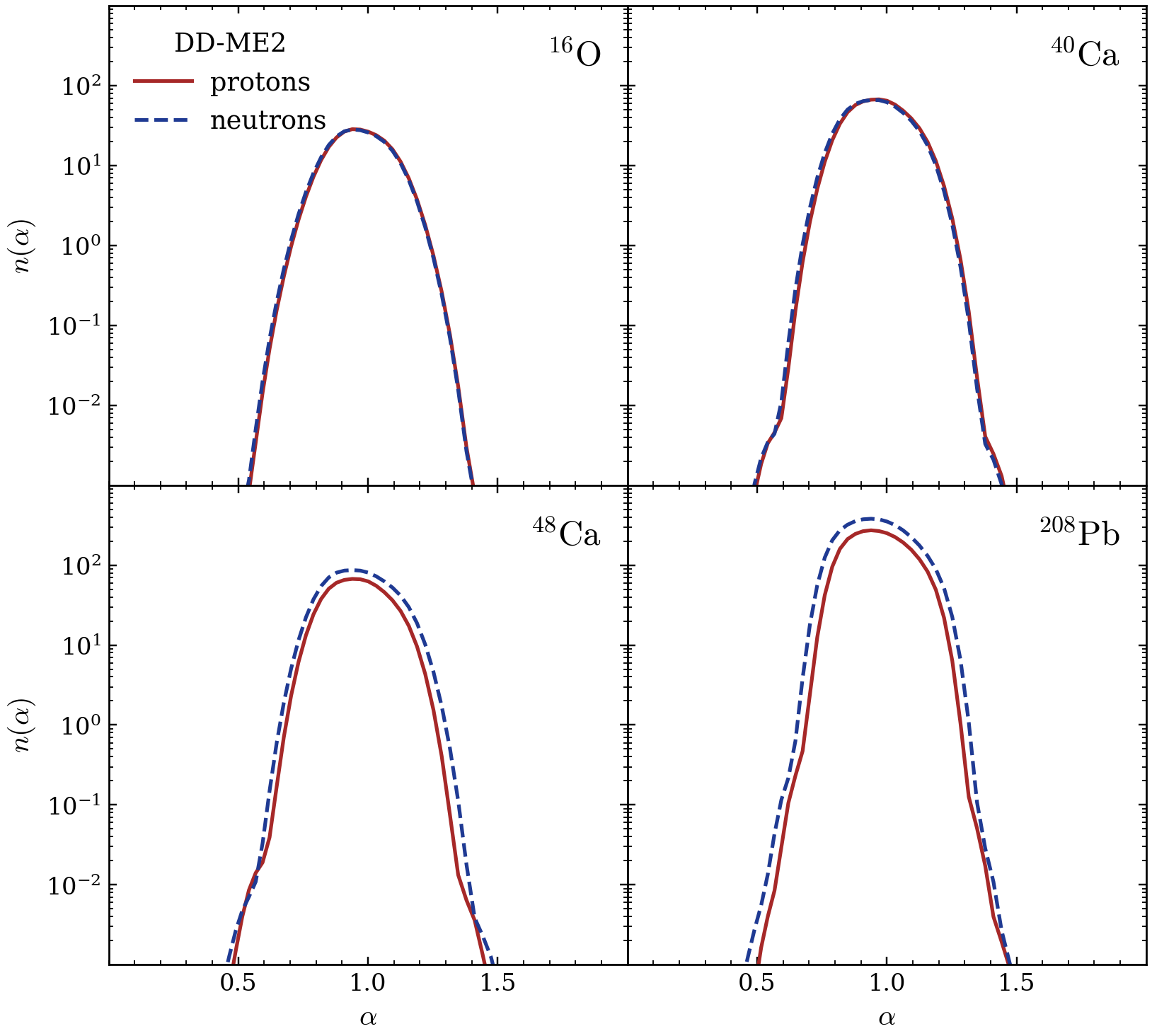}
\caption{(color online) Transverse-momentum-integrated light-front momentum distributions for $^{16}$O, $^{40}$Ca, $^{48}$Ca, and $^{208}$Pb versus $\alpha = A p^+/P_A^+$. $A$ is the mass number and $P^+_A$ is the nuclear plus momentum ($m_A$ in the rest frame). Solid red lines denote protons; dashed blue lines denote neutrons.}    \label{fig:LF_DFT_k_dist}
\end{figure}

Incorporating this density dependence into the meson coupling constants does not affect the
equivalence between the IF and LF formulations. The coordinate change of Step~1 is unaffected, and
the phase redefinition of Step~2 acts only on the wavefunctions, leaving the densities unchanged.
The equivalence therefore persists, and density-dependent LF DFT can be solved through its IF
counterpart. This equivalence allows existing IF codes to be employed directly. In this thesis we
use the DIRHB package \cite{Niksic:2014dra}, a set of Fortran codes that compute the ground-state
properties of even-even nuclei within relativistic self-consistent mean-field theory with
density-dependent couplings. For the effective interaction we adopt the DD-ME2 functional of Ref.
\cite{Lalazissis:2005de}, which reproduces total nuclear binding energies for approximately 200
nuclei to within 5 MeV. Figure~\ref{fig:LF_DFT_k_dist} presents the LF momentum distributions,
integrated over longitudinal momenta, for the four doubly magic nuclei considered in this paper. To
interpret these distributions qualitatively, consider the LF momentum distribution of a free proton,
which would be a delta function at $\alpha=1$. The spread in $\alpha$ in the nuclear case is
therefore due to Fermi motion inside the nucleus, while the offset of the peak from $\alpha=1$
reflects the effect of nuclear binding.

\section{Inclusive Electron-Nucleus Cross-Sections}
\begin{figure}
    \centering
    \includegraphics[width=0.8\linewidth]{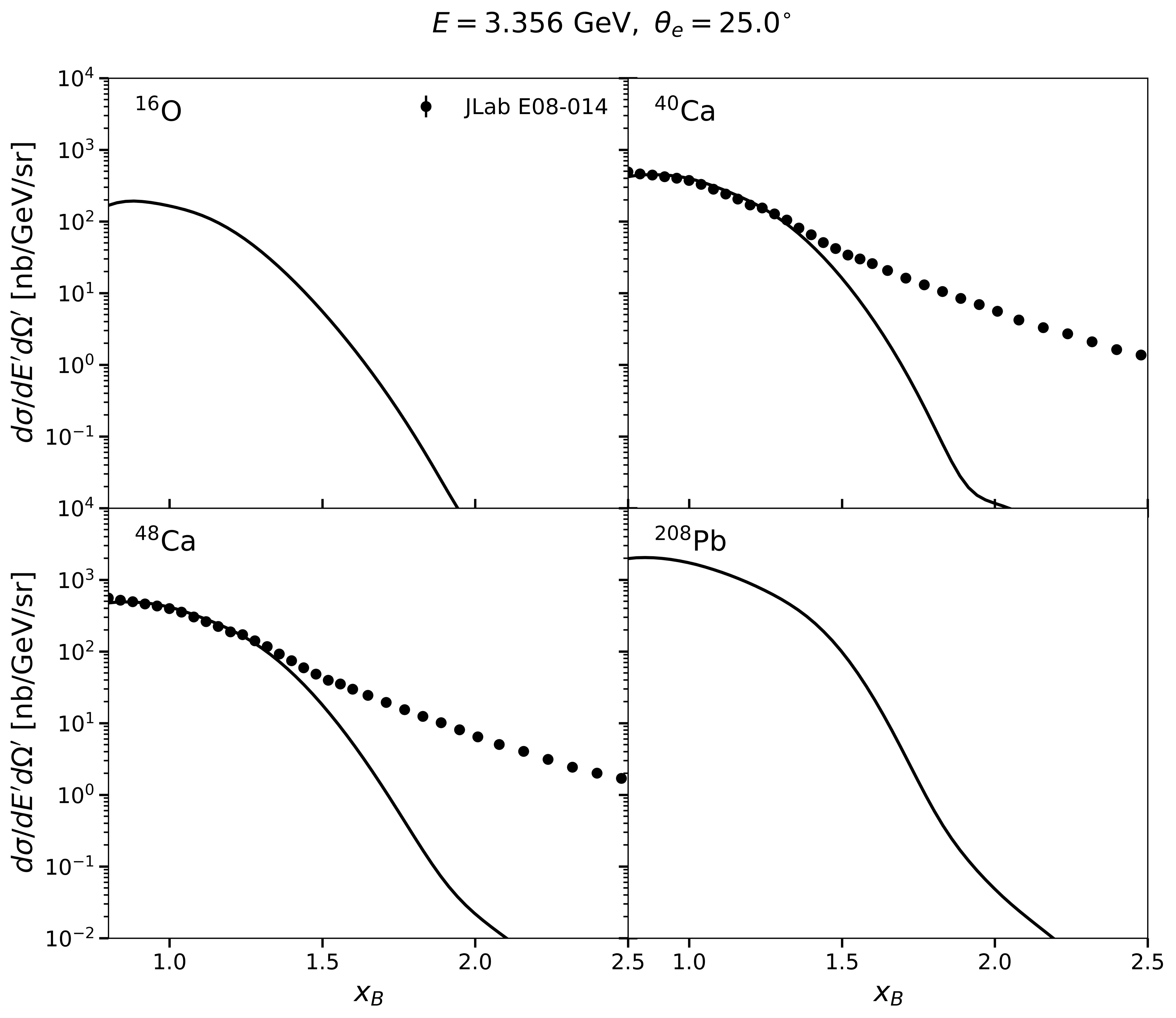}
    \caption{Inclusive electron–nucleus cross section at $E = 3.356$~GeV and $\theta_e = 25^\circ$ as a function of the 
    Bjorken variable $x_B$, for $^{16}\mathrm{O}$ (top left), $^{40}\mathrm{Ca}$ (top right), $^{48}\mathrm{Ca}$ (bottom left),
    and $^{208}\mathrm{Pb}$ (bottom right). The solid black curves are the mean-field contribution,
    evaluated with relativistic mean-field single-particle wave functions. Black points are the JLab
    E08-014 data, available for $^{40}\mathrm{Ca}$ and $^{48}\mathrm{Ca}$ at these kinematics.}
    \label{fig:MF_cs}
\end{figure}

Appendix~\ref{appendix:A-1 implementation} provides a detailed derivation of the inclusive electron-nucleus cross section 
calculated within the FF, utilizing LF DFT states to evaluate the intrinsic nuclear matrix elements. The resulting cross 
sections are displayed in Fig.~\ref{fig:MF_cs}. We find that the LF DFT framework successfully reproduces the data in the 
region $0.8 < x_B < 1.3$, capturing the quasi-elastic peak. However, a significant discrepancy 
emerges at $x_B > 1.3$, where the theoretical calculations systematically underestimate the cross section. This shortfall 
is directly attributable to the absence of short-distance nuclear structure---specifically, the high-momentum short-range 
correlations that are inherently missing from independent-particle models.

\section{Summary}
Mean-field approaches are among the most versatile and computationally accessible microscopic
descriptions of nuclei, capable of treating systems as small as $A = 16$. Modern
independent-particle models such as DFT have enjoyed remarkable success over the years: they
reproduce binding energies to within roughly 3~MeV, describe charge radii, underpin nuclear fission
calculations, and naturally encode nuclear shell structure. They are especially valuable for
high-energy nuclear reactions, where nuclear structure must be formulated relativistically under LF
quantization. In this regime \emph{ab initio} methods are not yet viable, owing to our limited
knowledge of LF NN potentials and our lack of experience solving the correlated
many-body Dirac equation. Mean-field theories circumvent these obstacles, providing accurate
descriptions of nuclei on the LF without requiring explicit knowledge of the bare NN and
three-nucleon forces. The results developed here can thus help bridge the low- and high-energy
nuclear physics communities.

This chapter has presented original work toward that goal. To date, only a single LF mean-field
study of finite nuclei exists \cite{Blunden:1999gq}; the major advances in IF DFT achieved since
then now make a modern light-front revisit possible \cite{doi:10.1142/9872}. We find that results
from IF DFT can be repurposed for LF DFT directly and naturally, without approximation. This
equivalence rests on their shared underlying assumptions which lead to identical equations. Using
the nuclear structure provided by LF DFT, we computed the inclusive electron-nucleus cross sections
for the high-energy kinematics of Jefferson Lab experiment E08-014. We find that, while mean-field
models correctly capture the quasi-elastic peak, they fail to reproduce the data for $x_B > 1.3$.
This shortfall reflects the absence of high-momentum configurations in our nuclear structure
calculation, and motivates the work of the next chapter.

% !TEX root = ../uwthesis.tex
\chapter{Similarity Renormalization Group}\label{chap:srg}

Renormalization Group (RG) techniques originated during the development of Quantum Electrodynamics 
in the 1940s \cite{Bethe:1947id, Dyson:1949bp}. At that time, RG methods were considered
mathematical tricks used to tame the infinities arising from higher-order loop integrals in
perturbative calculations. This approach was later refined and given physical interpretation by
Gell-Mann and Low, who discovered that the effective charge of an electron depends on the energy
scale at which it is probed \cite{Gell-Mann:1954yli}. In the late 1960s and early 1970s, Kenneth
Wilson generalized this framework by investigating the scale dependence of Hamiltonians themselves
\cite{PhysRevB.4.3174, RevModPhys.47.773}. He introduced the concept of ``Theory Space''—the
infinite-dimensional space of all possible Hamiltonians—and mapped how these theories are
mathematically connected through the flow of the RG. Ultimately, Wilson made mathematically precise
the idea that scale dictates our perception of the physical world.

In the 1990s, Glazek and Wilson sought to understand the success of constituent quark models, such
as the MIT bag model, by studying the RG evolution of the QCD Hamiltonian under light-front
quantization \cite{Glazek:1993rc, Glazek:1994qc}. For this type of problem, traditional Wilsonian
RG—which continuously integrates out momentum modes above a floating cutoff—is disadvantageous
because truncating the Hilbert space inherently alters the exact eigenvalue spectrum of the initial
Hamiltonian. To resolve this, Glazek and Wilson, and independently Wegner for condensed matter
applications \cite{Wegner:1994fdg}, developed continuous unitary RG techniques, which is known as 
the Similarity Renormalization Group (SRG). Instead of discarding high-energy modes, these RG flow equations systematically drive the bare
ultraviolet Hamiltonian toward an energy-diagonal form. Because these transformations are
strictly unitary, the original physical spectrum is exactly preserved. Glazek and Wilson were unable
to fully solve QCD using this technique because the multi-gluon interactions inherently induced by
the RG procedure were non-negligible and could not be truncated. However, the use of unitary flow
equations—particularly the form proposed by Wegner—found immense utility in low-energy nuclear
physics. In the nuclear domain, induced many-body operators are found to be suppressed; as the rank
of the interaction ($N$) increases, the corresponding $N$-body forces become increasingly
negligible. Because this allows the interaction space to be safely truncated, SRG techniques became a
critical breakthrough, directly enabling the success of ab initio methods like the No-Core Shell
Model \cite{Barrett:2013nh}.

Recently, and of particular importance to this work, the SRG has been utilized to reconcile
short-range correlation (SRC) phenomenology with independent-particle nuclear models. Historically,
nuclei have been predominantly studied at low energies, where the many-body system is well described
by the Schrödinger equation. This low-energy regime has seen incredible success through the
application of mean-field theories, which are often referred to as ``low-resolution'' pictures of the
nucleus. Despite their simplifications, these models are remarkably robust; frameworks like Density
Functional Theory provide some of our best microscopic descriptions, reproducing total nuclear
binding energies to within an accuracy of roughly 3 MeV \cite{Bulgac:2017bho}. However, the
low-resolution pictures cannot explain everything. Early work by Brueckner and collaborators
\cite{Brueckner:1955zzd} demonstrated that independent-particle models fail to describe high-energy,
high-momentum-transfer processes, such as $(e,e'p)$ knockout reactions. They concluded that
reproducing these reactions requires a nucleon momentum distribution with a significant
high-momentum tail, generated by the short-range nucleon-nucleon interaction. Today,
state-of-the-art numerical techniques like Quantum Monte Carlo, along with realistic two and
three-body nuclear potentials such as Argonne V18 and Urbana IX, give us the ability to calculate
the correlated many-body wavefunction of the nucleus \cite{Carlson:2014vla}. These calculations
confirm the existence of the high-momentum tails missing from independent-particle models, providing
what is now called the ``high-resolution'' picture of the nucleus.

However, the success of the high-resolution picture did not definitively resolve which framework is
correct, and theoretical tensions persisted. It was demonstrated that high-momentum observables
could be equally well reproduced within a low-resolution framework by accounting for initial- and
final-state interactions (see, e.g., early work by Greider \cite{PhysRev.114.786}, who utilized
optical-model scattering to explain high-energy observables without relying on short-range
interactions). Ultimately, it is the application of the SRG that formally bridges this conceptual
gap. By providing a mathematical framework to continuously transform the Hamiltonian and operators,
the SRG demonstrates that both the low- and high-resolution pictures represent the exact same
underlying physics evaluated at different RG scales, seamlessly reconciling the two frameworks.

The methods and applications of the SRG have been extensively documented across literature. For an
excellent pedagogical introduction, complete with explicit numerical implementations, see Ref.
\cite{Hergert:2016iju}; for broader foundational reading, see Refs. \cite{Binder:2013xaa,
Roth:2013fqa, Furnstahl:2013oba, Bogner:2009bt}. Rather than providing an exhaustive review of the
formalism, this chapter focuses on establishing the minimal set of theoretical fundamentals strictly
required to support the calculations in this work.

\section{Theory Fundamentals}

The basic idea of the SRG method is quite general. We wish to
construct a continuous unitary transformation, $\hat{U}(s)$, which `simplifies' the Hamiltonian by
making it more band-diagonal when represented as a matrix \cite{Wegner:1994fdg,Hergert:2016iju}.

\begin{equation}\label{srg_hamiltonian}
    \hat{H}(s) = \hat{U}(s) \,
    \hat{H}(s = 0) \, \hat{U}^{\dagger}(s).
\end{equation}

\noindent By convention $\hat{H}(s = 0)$ is the starting, bare Hamiltonian and $s$ is called the
flow parameter, which parameterizes $\hat{U}$. To elucidate, mathematically, if we split up the
Hamiltonian into diagonal and off-diagonal parts,

\begin{equation}\label{hamiltonian_split}
    \hat{H}(s) = \hat{H}_{d}(s) + \hat{H}_{od}(s),
\end{equation}

\noindent we want to construct $\hat{U}(s)$ such that,

\begin{equation}\label{SRG_hamiltonian_asymptotics}
\hat{H}(s) \underset{s \rightarrow \infty}{\longrightarrow} \hat{H}_{d}(s), \,\, \hat{H}_{od}(s) \underset{s \rightarrow \infty}{\longrightarrow} 0.
\end{equation}

To derive the differential equation for the $s$-dependent Hamiltonian, we take the derivative of Eq.
(\ref{srg_hamiltonian}) with respect to $s$ on both sides.

\begin{equation}\label{srg_flow_equation_derivation}
\begin{split}
    \frac{ d \hat{H}(s)}{d s} & = \frac{d\hat{U}(s)}{ds} \, \hat{H} \, \hat{U}^{\dagger}(s) +  \hat{U}(s) \, \hat{H} \, \frac{d\hat{U}^{\dagger}(s)}{ds} \\
    & =  \frac{d\hat{U}(s)}{ds} \, \hat{U}^{\dagger}(s)\, \hat{H}(s) + \hat{H}(s) \, \hat{U}(s) \, \frac{d\hat{U}^{\dagger}(s)}{ds}.
\end{split}
\end{equation}

\noindent Since $\hat{U}$ is unitary we have,

\begin{equation}\label{SRG_unitary_consition}
\frac{d}{ds} (\hat{U}(s) \hat{U}^{\dagger}(s)) = \frac{d}{ds} (\mathbb{I}) = 0 \Rightarrow \frac{d\hat{U}(s)}{ds} \hat{U}^{\dagger}(s) = - \hat{U}(s) \frac{d \hat{U}^{\dagger}(s)}{d s}.
\end{equation}

\noindent Defining the anti-Hermitian operator

\begin{equation}\label{SRG_generator}
    \hat{\eta}(s) \equiv \frac{d\hat{U}(s)}{ds} \hat{U}^{\dagger}(s) = - \hat{\eta}^{\dagger}(s),
\end{equation}

\noindent we can write down the SRG-flow equation as:

\begin{equation}\label{SRG_hamiltonian_flow_equation}
    \frac{d\hat{H}(s)}{ds} = [\hat{\eta}(s),\hat{H}(s)].
\end{equation}

\noindent We call $\hat{\eta}(s)$ the dynamical generator of the SRG-transformation. Eq.
(\ref{SRG_hamiltonian_flow_equation}) is a first-order linear differential equation that can be
calculated numerically. Hence, by calculating $\hat{H}(s)$ one can build $\hat{U}(s)$ directly by
using the eigenvectors of the evolved and initial Hamiltonians. There is no set procedure to
determine $\hat{\eta}(s)$, which generates an s-dependent Hamiltonian that obeys Eq.
(\ref{SRG_hamiltonian_asymptotics}).
For momentum-space SRG, Wegner showed that the off-diagonal part of the Hamiltonian flows to zero,
as $s \rightarrow \infty$, if we choose the generator to be

\begin{equation}\label{momentum_space_SRG_generator}
\hat{\eta}(s) = [\hat{T}(s=0), \hat{H}(s)],
\end{equation}

\noindent where $\hat{T}(s=0)$ is the bare kinetic energy operator \cite{Wegner:1994fdg}. It is
convenient to perform a change of variables, defining a new flow parameter $\lambda = s^{-1/4}$,
which has units of [$\mbox{fm}^{-1}$]. The flow parameter $\lambda$ provides a measure of the spread
of the band-diagonal in the SRG-evolved Hamiltonian \cite{Hergert:2016iju}. Under this change of
variables, the SRG flow equation is written as

\begin{equation}\label{SRG_hamiltonian_flow_equation_lambda}
\frac{d \hat{H}(\lambda)}{d\lambda} = -\frac{4}{\lambda^5} \, [\hat{\eta}(\lambda),\hat{H}(\lambda)].
\end{equation}

\noindent From here on, we will drop the $\lambda = \infty$ dependency on bare operators and
eigenstates.

\begin{figure}[H]
    \centering
    \includegraphics[width=0.9\linewidth]{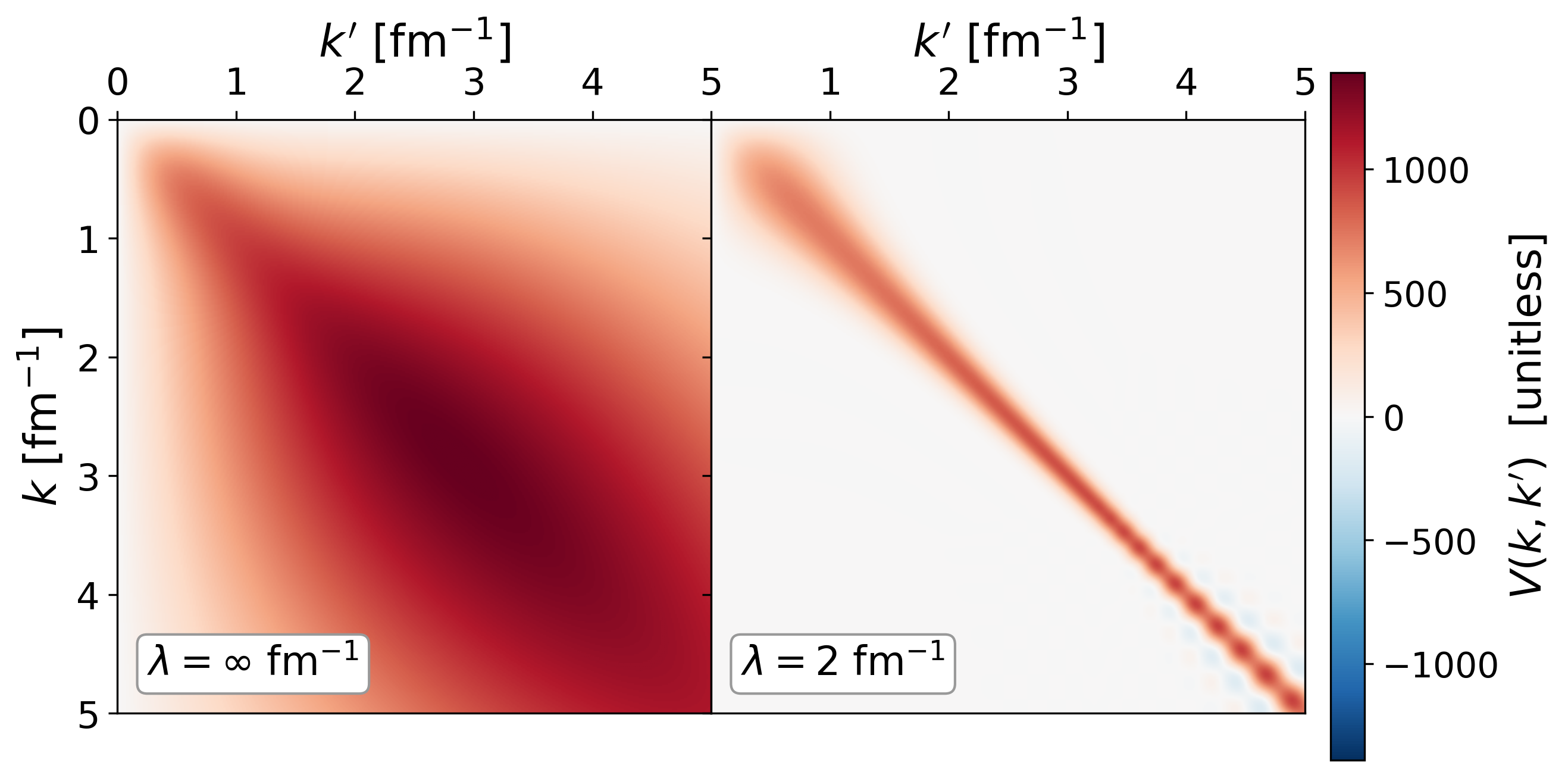}
    \caption{(color online) The SRG-transformed $^1P_1$ channel of the minimal relativity AV18 nucleon-nucleon potential.}
    \label{fig:srg_evolution}
\end{figure}

For a concrete example, Fig. \ref{fig:srg_evolution} shows the SRG transformation on the $^1P_1$
channel of the minimal relativity AV18 potential. $\lambda = \infty$ fm$^{-1}$ is the bare
potential, while $\lambda = 2$ fm$^{-1}$ is the SRG-evolved potential. Notice that the potential
becomes more band-diagonal as $\lambda$ approaches zero, meaning that the SRG flow equations under
the Wegner generator weakens the coupling between low- and high-momentum states.

\section{SRG Transformation of Operators: Connecting Low- and High- Resolution Pictures}

Now we discuss how the SRG can be used to bridge low- and high-resolution pictures of the nucleus.
This can be immediately seen by looking at the matrix element for the momentum distribution of a
single nucleon inside a nucleus,

\begin{equation}\label{lf_momentum_distribution}
    \begin{split}
        n_A^{\tau}(\bm{p}) & = \sum_{\sigma} \bra{\Psi_A} a^{\dagger}(\bm{p},\sigma,\tau) \, a(\bm{p},\sigma,\tau)\ket{\Psi_A} \\
        & = \sum_{\sigma} \bra{\Psi_A} \hat{n}^{\tau}(\bm{p},\sigma) \ket{\Psi_A}.
    \end{split}
\end{equation}

\noindent Where $\sigma$ is the spin projection, $\tau$ is the isospin projection, and
$\ket{\Psi_A}$ is the nuclear many-body state. Applying $\hat{U}^{\dagger}(\lambda) \hat{U}(\lambda)
= \mathbb{I}$ before and after the operator we find,

\begin{equation}\label{srg_momentum_distribution}
    \begin{split}
        n_A^{\tau}(\bm{p})  & = \sum_{\sigma} \bra{\Psi_A} \hat{n}^{\tau}(\bm{p},\sigma)\ket{\Psi_A} \\
        & =  \sum_{\sigma} \bra{\Psi_A} \hat{U}^{\dagger}(\lambda) \hat{U}(\lambda) \, \hat{n}^{\tau}(\bm{p},\sigma) \, \hat{U}^{\dagger}(\lambda) \hat{U}(\lambda) \ket{\Psi_A} \\
        & = \sum_{\sigma} \bra{\Psi_A(\lambda)} \hat{U}(\lambda) \, \hat{n}^{\tau}(\bm{p},\sigma) \, \hat{U}^{\dagger}(\lambda) \ket{\Psi_A(\lambda)} \\
        & = \sum_{\sigma} \bra{\Psi_A(\lambda)} \hat{n}^{\tau}_{\lambda}(\bm{p},\sigma) \ket{\Psi_A(\lambda)}.
    \end{split}
\end{equation}

\noindent Where $\hat{U}(\lambda)$ is the unitary SRG operator at flow parameter $\lambda$,
$\ket{\Psi_A(\lambda)}$ is the SRG-evolved nuclear ground state, and
$\hat{n}^{\tau}_{\lambda}(\bm{p},\sigma)$ is the SRG-evolved momentum distribution operator for
nucleon $\tau$. Looking back at Fig. \ref{fig:srg_evolution} we see that the SRG transformation
decouples low- and high-momentum states. Hence, as we SRG-evolve the Hamiltonian, it's
eigensolutions $\ket{\Psi_A(\lambda)}$ begin to reflect a dominantly “mean-field” or
"low-resolution" description of nuclei. However, since SRG transformations are unitary, the
high-resolution physics is shifted from the wavefunction to the operators. Hence, within the SRG
framework, low- and high-resolution pictures of nuclei are equivalent. Both can output the same
results as long as one is consistent with the SRG-scale of the reaction (operators) and structure
(states).

A crucial theoretical consideration is determining, a priori, whether the chosen probing operator is
truly the bare operator. Within the context of nuclear physics, the SRG transformation is driven
entirely by the strong nuclear interactions governing the initial state of the target. Therefore, if
the external probe—such as an electromagnetic or weak current—does not inherently contain nuclear
force dynamics, it constitutes the bare, high-resolution operator. For theoretical consistency, this
unevolved operator must strictly be evaluated using exact high-resolution many-body eigenstates.
However, as demonstrated by Eq. (\ref{srg_momentum_distribution}), when we apply the SRG
transformation to evolve the high-resolution nuclear state to a low-resolution scale, the unitary
transformation is formally shifted onto the operator. Consequently, the originally bare operator
absorbs the short-range nuclear dynamics, naturally inducing two-body and higher-rank many-body
currents which originate from nuclear forces. This has important implications for high-energy
approximations in nuclear scattering processes which we will discuss shortly.

\section{Application to Light-Front Physics}

The formal connection between the low- and high-resolution pictures of nuclei provided by the SRG
led Tropiano et al. to approximate high-resolution observables using shell-model wavefunctions
\cite{Tropiano:2024bmu}. Specifically, they substituted the exact evolved state
$\ket{\Psi_A(\lambda)}$ in Eq. (\ref{srg_momentum_distribution}) with an uncorrelated mean-field
nuclear wavefunction, $\ket{\Psi_A^{MF}}$. In principle, $\ket{\Psi_A(\lambda)}$ should be obtained
by solving the many-body Schrödinger equation associated with the evolved Hamiltonian; however, as
discussed previously, this low-resolution state reflects a dominantly independent-particle
description of the nucleus. Introducing this mean-field approximation breaks the formal SRG
invariance of the matrix element, inherently introducing a dependence on the resolution scale
$\lambda$. This scale dependence is the theoretical tradeoff for extracting correlated observables
from uncorrelated wavefunctions. Consequently, $\lambda$ becomes a parameter that must be
constrained. By focusing on contributions from the two-body nucleon-nucleon potential and neglecting
higher-rank SRG-induced interactions, Tropiano et al. demonstrated that setting $\lambda = 1.5
\text{ fm}^{-1}$ yielded high-resolution momentum distributions that showed excellent agreement with
exact high-resolution Quantum Monte Carlo calculations.

This operator-evolution approach to extracting high-resolution observables significantly simplifies
computational efforts by circumventing the need to explicitly solve the correlated nuclear many-body
problem. Within the domain of light-front nuclear physics, this methodology proves particularly
advantageous. Given the historical scarcity of full many-body calculations performed under
light-front quantization, relying on evolved operators paired with independent-particle
wavefunctions provides a highly practical and effective path forward. To accomplish this, we mirror
the procedure of Ref. \cite{Tropiano:2024bmu}. We restrict ourselves to incorporating the effects of
the two-body nucleon-nucleon interaction, we utilize the non-relativistic approximation of
light-front wavefunctions developed in Chapter \ref{chap:scattering} to construct our unitary SRG
transformation, and the light-front density functional theory model developed in Chapter
\ref{chap:dft} as input for a low-resolution wavefunction.

The two-body unitary light-front SRG transformation can be written as the inner product between
eigenstates of the un-evolved and SRG-evolved Hamiltonians,

\begin{equation}\label{two_body_LF_SRG}
    \hat{U}(\lambda) = \sum_{\beta} \int [dP] \, \ket{\beta \, , \lambda\, ;\bm{P}} \bra{\beta \, ;\bm{P}},
\end{equation}
where $\beta$ indexes the eigenvalues of the two-body light-front Hamiltonian,
$\bm{P}=(P^+, \bm{P}^\perp)$ are the plus and transverse momenta of the center of mass, and
$\lambda$ is the SRG-scale. To be explicit,

\begin{equation}\label{beta_sum}
    \begin{gathered}
        \sum_{\beta} = \sum_{\text{energy}} \sum_{J M_J} \sum_{T M_T} \sum_{S} \, \, , \\
        \sum_{\text{energy}} \xrightarrow{\text{bound states}} \sum_{M^2} \, \, ,\\
        \sum_{\text{energy}} \xrightarrow{\text{scattering states}} \frac{1}{2} \sum_{\substack{\sigma_1\tau_1 \\ \sigma_2 \tau_2}} \int  \frac{d\alpha d\bm{k}^\perp}{(2 \pi)^2 \alpha (2-\alpha)},
    \end{gathered}
\end{equation}
where $(\alpha, \bm{k}^\perp)$ are the intrinsic LF relative variables. The expansion of the two-body state in terms of anti-symmetrized plane wave nucleonic
states is defined by,

\begin{equation}\label{LF_2body_expansion}
\begin{aligned}
     \ket{\beta \, , \lambda\, ; \bm{P}} & =  \frac{1}{2}  \sum_{\substack{\sigma_1\tau_1 \\ \sigma_2 \tau_2}} \int [dp_1] [dp_2] \, (2 \pi)^3 2 P^+ \delta(p_1^+ + p_2^+ - P^+) \delta^{(2)}(\bm{p}_1^\perp + \bm{p}_2^\perp - \bm{P}^\perp) \\
     & \qquad\qquad\qquad\qquad \times \psi_{\beta}^{\lambda}(\alpha, \bm{k}^\perp,\sigma_1,\tau_1,  \sigma_2, \tau_2) \ket{\bm{p}_1,\sigma_1, \tau_1; \bm{p}_2,\sigma_2, \tau_2} \\
     &  =  \frac{1}{2}\sum_{\substack{\sigma_1\tau_1 \\ \sigma_2 \tau_2}} \int \frac{d\alpha d\bm{k}^\perp}{(2 \pi)^2 \alpha (2-\alpha)}
     \psi_{\beta}^{\lambda}(\alpha, \bm{k}^\perp,\sigma_1,\tau_1,  \sigma_2, \tau_2) \ket{p_1, p_2}.
\end{aligned}
\end{equation}

\noindent In Eq. (\ref{LF_2body_expansion}), $\psi_{\beta}^{\lambda}$ is the two-body light-front
wavefunction for eigenstate $\beta$, SRG-evolved to $\lambda$. In the above expression, $(\alpha,\bm{k}^\perp)$ are defined by

\begin{equation}\label{relative_lf_coordinates}
\begin{aligned}[c]
    p_1^+  &= \frac{\alpha}{2} P^+ , \\
    \bm{p}_1^\perp &= \frac{\alpha}{2} \bm{P}^\perp + \bm{k}^\perp,
\end{aligned}
\qquad
\begin{aligned}[c]
    p_2^+  &= \frac{2-\alpha}{2} P^+ ,\\
    \bm{p}_2^\perp &= \frac{2-\alpha}{2} \bm{P}^\perp - \bm{k}^\perp,
\end{aligned}
\end{equation}
One important observation can explain why SRG techniques are most suitable for relativistic physics
formulated using LF quantization. Notice that Eq. (\ref{two_body_LF_SRG}) has an integral
over the center-of-mass momenta. If we instead attempted to use SRG techniques formulated under
relativistic IF Quantization, we will find a computationally intractable problem. The
IF wavefunctions are not boost invariant, and depend on center-of-mass momenta. Hence, one
must re-calculate the relativistic two-body problem for every frame, making the construction of
$\hat{U}(\lambda)$ extremely difficult. On the LF, however, we don't encounter this problem
as our wavefunctions are boost invariant.

Using Eq. (\ref{LF_2body_expansion}), one can write Eq. (\ref{two_body_LF_SRG}) in second quantized
form
\begin{equation}
\begin{gathered}\label{two_body_LF_SRG_second_quatnizated}
    \hat{U}(\lambda) = \frac{1}{4}\sum_{\substack{\sigma_1 \sigma_2 \\ \sigma_3 \sigma_4}} \sum_{\substack{\tau_1  \tau_2 \\ \tau_3 \tau_4}} \int [dP] \frac{d\alpha d\bm{k}^\perp}{(2\pi)^3 \sqrt{\alpha (2 - \alpha)}} \frac{d\alpha'd\bm{k}'^\perp}{(2\pi)^3 \sqrt{\alpha' (2 - \alpha')}}  (P^+)^2 \\
    \times \braket{\alpha, \bm{k}^\perp, \sigma_1, \tau_1, \sigma_2, \tau_2|U(\lambda)|\alpha', \bm{k}'^\perp, \sigma_3, \tau_3, \sigma_4, \tau_4}\\
    \times \, \hat{a}^{\dagger}(\bm{p}_1,\sigma_1,\tau_1) \hat{a}^{\dagger}(\bm{p}_2,\sigma_2,\tau_2) \hat{a} (\bm{p}_4,\sigma_4,\tau_4) \hat{a}(\bm{p}_3,\sigma_3,\tau_3),
\end{gathered}
\end{equation}
where $p_3$ and $p_4$ follow the same relations as $p_1$ and $p_2$ respectively, but with $\alpha
\rightarrow \alpha'$ and $\bm{k}^\perp \rightarrow \bm{k}'^\perp$. The matrix element of
$\hat{U}(\lambda)$ between asymmetric two-nucleon states is
\begin{equation}
    \begin{aligned}
        & \braket{\alpha, \bm{k}^\perp, \sigma_1, \tau_1, \sigma_2, \tau_2|U(\lambda)|\alpha', \bm{k}'^\perp, \sigma_3, \tau_3, \sigma_4, \tau_4} = \\ & \qquad\qquad\qquad\qquad\qquad\qquad \sum_{\beta} \psi_{\beta}^{\lambda}(\alpha, \bm{k}^\perp,\sigma_1,\tau_1,  \sigma_2, \tau_2) \psi_{\beta}^{\dagger}(\alpha', \bm{k}'^\perp,\sigma_3,\tau_3,  \sigma_4, \tau_4).
    \end{aligned}
\end{equation}

\section{High-Resolution Light-Front Momentum Distributions}

It is instructive to re-write $\hat{U}(\lambda) = \hat{\mathbb{I}} + \delta \hat{U}(\lambda)$.
Replacing $\ket{\Psi_A(\lambda)}$ with $\ket{\Psi_A^{MF}}$, Eq. (\ref{srg_momentum_distribution})
becomes,

\begin{equation} \label{srg_mom_dist_approx}
    \begin{aligned}
        n_A^{\tau}(\bm{p}) = \sum_{\sigma} \Big[ & \bra{\Psi_A^{MF}} \hat{n}^{\tau}(\bm{p},\sigma) \ket{\Psi_A^{MF}} \\
         + \, & \bra{\Psi_A^{MF}} \delta\hat{U}(\lambda) \, \hat{n}^{\tau}(\bm{p},\sigma)\ket{\Psi_A^{MF}} \\
         + \,& \bra{\Psi_A^{MF}} \hat{n}^{\tau}(\bm{p},\sigma) \, \delta\hat{U}^{\dagger}(\lambda) \ket{\Psi_A^{MF}} \\
         + \,& \bra{\Psi_A^{MF}} \delta\hat{U}(\lambda) \, \hat{n}^{\tau}(\bm{p},\sigma) \, \delta\hat{U}^{\dagger}(\lambda) \ket{\Psi_A^{MF}} \Big]. \\
         &
    \end{aligned}
\end{equation}

\begin{figure}
    \centering
    \includegraphics[width=1\linewidth]{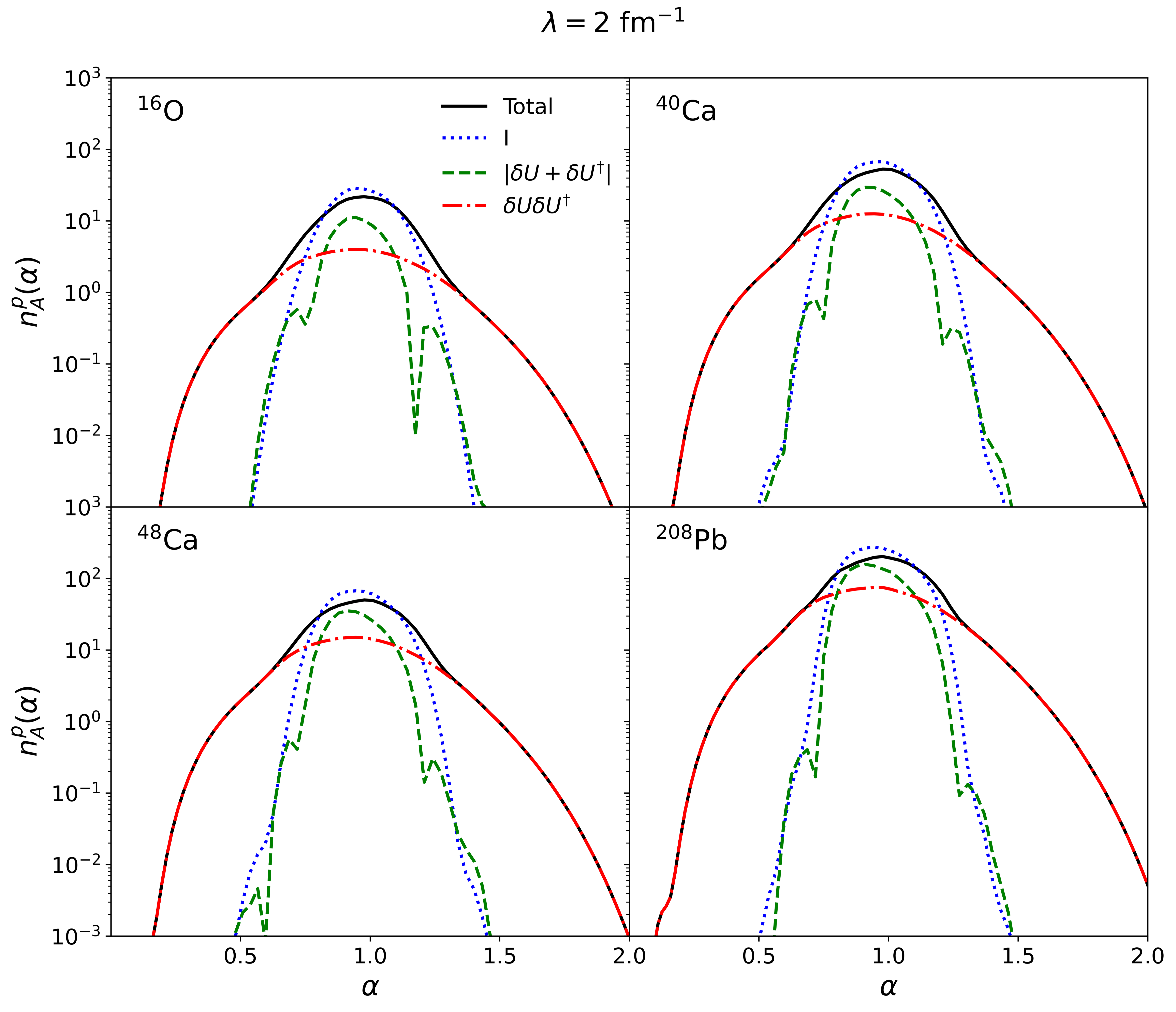}
    \caption{Light-front proton momentum distribution $n^p_A(\alpha)$ as a function
    of the longitudinal momentum fraction $\alpha = A\,p^+/M_A$, for
    $^{16}\mathrm{O}$ (top left), $^{40}\mathrm{Ca}$ (top right),
    $^{48}\mathrm{Ca}$ (bottom left), and $^{208}\mathrm{Pb}$ (bottom right). The
    solid black curve is the total distribution
    $n^p = I + \delta U + \delta U^{\dagger} + \delta U\,\delta U^{\dagger}$; the
    mean-field contribution $I$ (blue dotted), the two-body cross-term
    $|\delta U + \delta U^{\dagger}|$ (green dashed), and the two-body
    contribution $\delta U\,\delta U^{\dagger}$ (red dash-dot) are shown
    separately. SRG evolution scale $\lambda = 2~\mathrm{fm}^{-1}$.}
    \label{fig:lf_dist_4nuclei}
\end{figure}

\noindent Detailed derivations of each term are presented in Appendix \ref{SRG Implementation
Chapter}. Figure \ref{fig:O16_SRG_momentum_distribution} presents the approximated high-resolution
light-front momentum distribution of Oxygen-16. The second and third SRG-interference terms from Eq.
(\ref{srg_mom_dist_approx}) are represented by the green dashed line; the left and right panels are
included to illustrate the isolated behavior of these contributions. Qualitatively, these
interference terms deplete probability from the low-momentum mean-field region, while the $\delta U
\delta U^{\dagger}$ term subsequently redistributes it to higher momenta. This generates the
high-momentum tails characteristic of SRCs, consistent with both phenomenological understanding and
exact non-relativistic high-resolution calculations. To our knowledge, this represents the first
principled calculation of light-front momentum distributions exhibiting these high-momentum tails.
Figure \ref{fig:lf_dist_4nuclei} presents the approximated high-resolution light-front momentum
distributions for Oxygen-16, Calcium-40, Calcium-48, and Lead-208.

\begin{figure}
    \centering
    \includegraphics[width=1\linewidth]{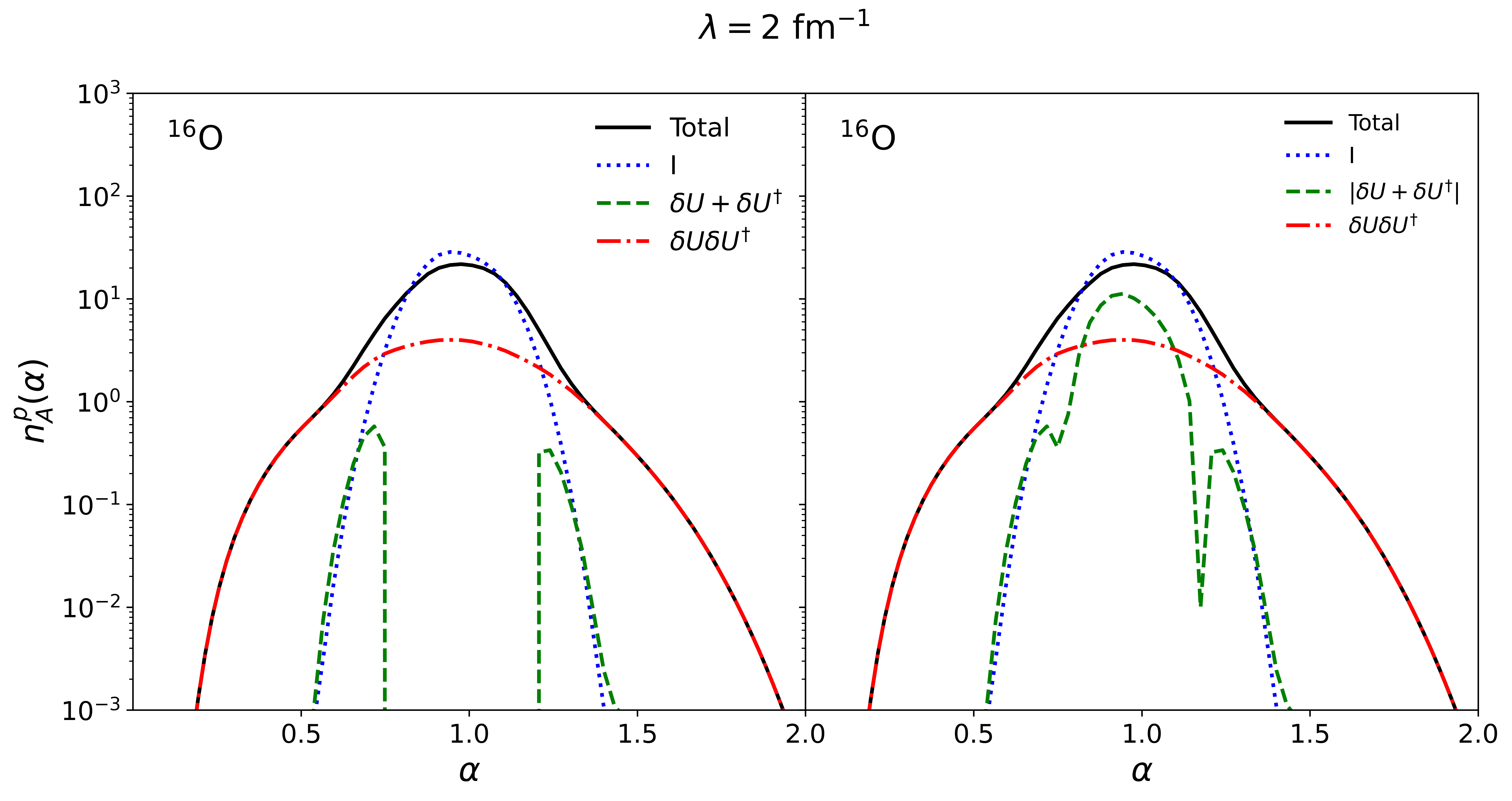}
    \caption{(color online) The approximated high-resolution light-front momentum distribution of Oxygen-16 using similarity renormalization group techniques. The left and right panels differ in their presentation of the green dashed line. The right figure presents the absolute value of the $\delta U + \delta U^\dagger$ term, while the left figure does not.}
    \label{fig:O16_SRG_momentum_distribution}
\end{figure}

\section{Inclusive Electron-Nucleus Scattering}
As briefly mentioned, the SRG framework carries crucial implications for our previous cross-section
calculations utilizing mean-field nuclear states. Our prior analysis relied on the plane-wave
impulse approximation—a high-energy approximation which assumes that the incident electron interacts
with only a single nucleon inside the nucleus. In formal terms, this restricts the interaction to a
bare, one-body current. Viewed through the perspective of the SRG, we now understand that employing
such a simplified, high-energy approximation for the probe implicitly requires the target to be
described by a high-resolution state. Consequently, it is unsurprising that uncorrelated mean-field
results fail to reproduce experimental results; there is an inherent mismatch between the reaction
mechanism and the nuclear structure. This discrepancy can be systematically resolved by formally
applying the SRG transformation to evolve the probing operators, generating higher-rank currents
that incorporate the nuclear dynamics missing from the low-resolution state.

Applying the SRG transformation systematically refines our previous mean-field calculations by 
depleting probability from the independent single-particle orbitals and redistributing it into 
short-range correlated (SRC) configurations. We implement this refinement in two steps. First, we 
update the one-body results of the previous chapter. Second, we incorporate 2N plus 
(A-2)-nuclear final-state configurations to explicitly capture the observable contributions from SRCs. 
Crucially, in evaluating these matrix elements, we only apply the SRG operator terms that correspond 
to the initial nuclear state. By intentionally neglecting the SRG evolution of the final scattering 
states, we isolate the specific impact of initial-state short-range correlations on the observable 
cross section.

\subsection{Updating LF DFT Results}

To incorporate the initial-state short-range correlations discussed above, we first update the one-body 
results derived in Appendix~\ref{appendix:A-1 implementation}. This is achieved by formally applying the 
SRG transformation operator, $\hat{U}(\lambda)$, to the intrinsic nuclear matrix element $M_i^A(\bm{p},\sigma,\tau)$. 
The transition from the bare matrix element to the SRG-evolved one is given by
\begin{equation}
    M_i^A(\bm{p},\sigma,\tau) =  \Braket{\Psi_i^{MF}|\hat{a}(\bm{p},\sigma,\tau)|\Psi_A^{MF}} \longrightarrow \Braket{\Psi_i^{MF}|\hat{a}(\bm{p},\sigma,\tau) \, \hat{U}(\lambda)|\Psi_A^{MF}}.
\end{equation}
By decomposing the unitary operator as $\hat{U}(\lambda) = \mathbb{I} + \delta\hat{U}(\lambda)$, we can 
separate the pure mean-field contribution from the SRG-induced corrections.

\begin{figure}
    \centering
    \includegraphics[width=1\linewidth]{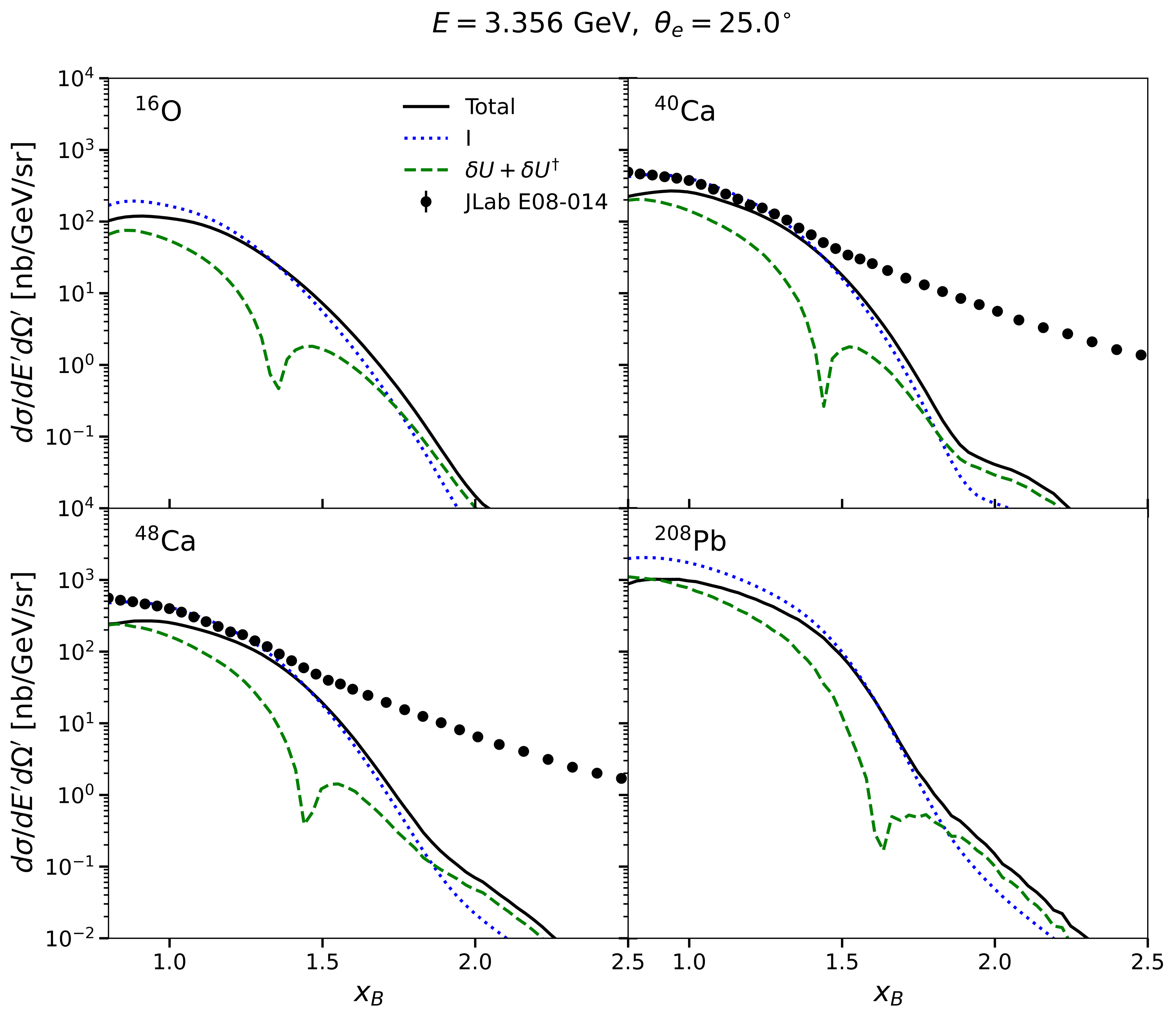}
    \caption{Same as Fig.~\ref{fig:MF_cs}, with the SRG operator
            corrections added. The solid black curve is the total
            $\mathbb{I} + \delta \hat{U} + \delta \hat{U}^{\dagger}$ contribution evaluated at an SRG scale of
            $\lambda = 2~\mathrm{fm}^{-1}$. The pure mean-field contribution (blue dotted) and the absolute 
            value of the one-body cross-term $|\delta \hat{U} + \delta \hat{U}^{\dagger}|$ (green dashed), which
            depledes the cross section near the quasi-elastic peak.}
    \label{fig:cs_4nuclei_A1}
\end{figure}

Figure~\ref{fig:cs_4nuclei_A1} displays the updated inclusive cross sections evaluated at an SRG scale 
of $\lambda = 2~\mathrm{fm}^{-1}$. The total updated one-body cross section (solid black curve) is composed 
of the bare mean-field baseline ($\mathbb{I}$, blue dotted curve) and the linear interference terms 
($\delta \hat{U} + \delta \hat{U}^{\dagger}$, green dashed curve). Consistent with the behavior of SRG 
evolution, these linear interference terms are negative in the quasi-elastic peak; they act to deplete probability 
from the low-momentum mean-field region. This depleted probability is precisely what gets redistributed to higher 
momenta via the two-nucleon final-state configurations, which we address in the following subsection.

\subsection{Inclusion of 2N + (A-2) Nuclear Final States}
SRG techniques allow us to extend our previous calculations. Because the SRG-evolved current
operator now contains two-body terms, the current can couple to two bound nucleons and promote them
into scattering states — that is, it introduces 2N plus (A-2)-nuclear final-state configurations
into our calculations. For such configurations, the terms in Eq. \eqref{nuclear_tensor} become

\begin{equation}
    \sum_X \rightarrow \frac{1}{2}\sum_{\substack{\sigma_1 \tau_1 \\ \sigma_2 \tau_2}} \sum_f \int [dP_{A-2}] [dP_{cm}'] \frac{d\alpha'd\bm{k}'^\perp}{(2 \pi)^3 \alpha' (2-\alpha')},
\end{equation}

\begin{equation}
\begin{gathered}
    \ket{X}\bra{X} \rightarrow \ket{\Psi_{A-2}^f,\bm{P}_{A-2}}\bra{\Psi_{A-2}^f,\bm{P}_{A-2}} \, \otimes \, \ket{ p_1, p_2 }\bra{ p_1, p_2 }, \\
    \ket{p_1, p_2} = \ket{\bm{p}_1, \sigma_1, \tau_1; \bm{p}_2, \sigma_2, \tau_2}, \\
    \bm{p}_1 = \left(\tfrac{\alpha'}{2} P_{cm}'^{+}, \, \tfrac{\alpha'}{2} \mathbf{P}_{cm}'^{\perp} + \bm{k}'^\perp\right), \\
    \bm{p}_2 = \left(\tfrac{2-\alpha'}{2} P_{cm}'^{+}, \, \tfrac{2-\alpha'}{2} \mathbf{P}_{cm}'^{\perp} - \bm{k}'^\perp\right),
\end{gathered}
\end{equation}
\noindent where $f$ denotes all possible (A-2)-nuclear final states. Utilizing the delta function in
light-front momenta, we integrate over $P_{A-2}$ and use $\bm{P}_{cm}' = \bm{P}_{cm} + \bm{q}$ to get

\begin{equation}
\begin{aligned}
    W_A^{\mu\nu} = & \, \frac{1}{2 m_A} \frac{1}{2P_{A-2}^+} \frac{1}{2} \sum_{\substack{\sigma_1 \tau_2 \\ \sigma_2, \tau_2}} \sum_f \int \frac{dP_{cm} d\bm{P}_{cm}^{\perp}}{(2 \pi)^3} \frac{d\alpha'd\bm{k}'^\perp}{(2 \pi)^3 \alpha' (2-\alpha')} \\
    & \times \braket{\Psi_A|J^\mu_A(0)|\Psi_{A-2}^f,\bm{P}_{A-2};p_1,p_2} \braket{\Psi_{A-2}^f,\bm{P}_{A-2};p_1,p_2|J^\nu_A(0)|\Psi_A} \\
    & \times \left[ \frac{1}{P^+_{cm} + q^+} \delta(q^- + P_A^- - (P_{A-2}^{f})^- - P_{cm}'^{-}) \right], \\
\end{aligned}
\end{equation}

\begin{equation}
\begin{gathered}
    (P_{A-2}^f)^- = \frac{M_f^2 + (\bm{P}_{cm}^{\perp})^2}{P_A^+ - P_{cm}^+}, \qquad P_{cm}'^- = \frac{4\frac{m^2 + (\bm{k}'^\perp)^{2}}{\alpha' (2 - \alpha')} + (\bm{P}_{cm}^\perp)^2}{P_{cm}^+ + q^+},\\
\end{gathered}
\end{equation}

\noindent The nuclear current matrix element is,

\begin{equation}
\begin{aligned}
    \braket{\Psi_{A-2}^f,\bm{P}_{A-2};p_1,p_2|J^\nu_A(0)|\Psi_A} & = \frac{1}{2}\sum_{\substack{s_1 t_1 \\ s_2 t_2}} \sum_{\sigma \sigma' \tau} \int [dp][dp'][dq_1][dq_2] \\
    & \times  \sqrt{2p^+} \sqrt{2p'^+} \sqrt{2q_1^+} \sqrt{2q_2'^+} \\
    & \times J_N^\nu(\bm{p}',\sigma',\tau;\bm{p},\sigma,\tau) \braket{p_1,p_2|\hat{a}^\dagger(\bm{p}',\sigma',\tau) \hat{a}(\bm{p},\sigma,\tau)|q_1, q_2} \\
    & \times \braket{\Psi_{A-2}^f,\bm{P}_{A-2}|\hat{a}(\bm{q}_2,s_2,t_2) \hat{a}(\bm{q}_1,s_1,t_1)|\Psi_A}.
\end{aligned}
\end{equation}

\noindent Following the same procedure as the (A-1) calculation in Appendix \ref{appendix:A-1
implementation}, we get

\begin{equation}
\begin{aligned}
    & \braket{\Psi_{A-2}^f,\bm{P}_{A-2};p_1,p_2|J^\nu_A(0)|\Psi_A} \approx \sqrt{2P_A^+} \sqrt{2P_{A-2}^+}\\
    & \times \sum_\sigma \Bigg[ \sqrt{\frac{p_2^+}{p_1^+ - q^+}} J_N^\nu(\bm{p}_1,\sigma_1,
    \tau_1;\bm{p}_1-\bm{q},\sigma,\tau_1) M_{ij}^A(\bm{p}_2,\sigma_2,\tau_2;\bm{p}_1-\bm{q},\sigma,\tau_1|\lambda)  \\
    & \qquad-\sqrt{\frac{p_1^+}{p_2^+ - q^+}} J_N^\nu(\bm{p}_2,\sigma_2,
    \tau_2;\bm{p}_2-\bm{q},\sigma,\tau_2)  M_{ij}^A(\bm{p}_1,\sigma_1,\tau_1;\bm{p}_2-\bm{q},\sigma,\tau_2|\lambda)  \Bigg],
\end{aligned}
\end{equation}

\begin{equation}
     M_{ij}^A(\bm{p}_2,\sigma_2,\tau_2;\bm{p}_1-\bm{q},\sigma,\tau_1|\lambda) =\braket{\Psi^{MF}_{ij}|\hat{a}(\bm{p}_2,\sigma_2,\tau_2) \hat{a}(\bm{p}_1-\bm{q},\sigma,\tau_1) \delta \hat{U}^{\dagger}(\lambda)|\Psi^{MF}_A},
\end{equation}

\noindent where we substitute the high-resolution (A-2)-nuclear states with their low-resolution
mean-field counterparts, approximating them as two-hole states relative to the target A-nucleus
state, i.e. $\ket{\Psi^{MF}_{ij}} = \hat{a}_j \hat{a}_i\ket{\Psi^{MF}_A}$. In doing so, we replace
$\sum_f \rightarrow \frac{1}{2} \sum_{ij}$. Furthermore, note that we use $\delta \hat{U}^\dagger(
\lambda)$ instead of $\hat{U}^\dagger(\lambda)$ in the above expression. This is because the $\mathbb{I}$
term is unphysical, as the unevolved current operator cannot promote two bound state nucleons into free plane-waves.
\begin{figure}
    \centering
    \includegraphics[width=1\linewidth]{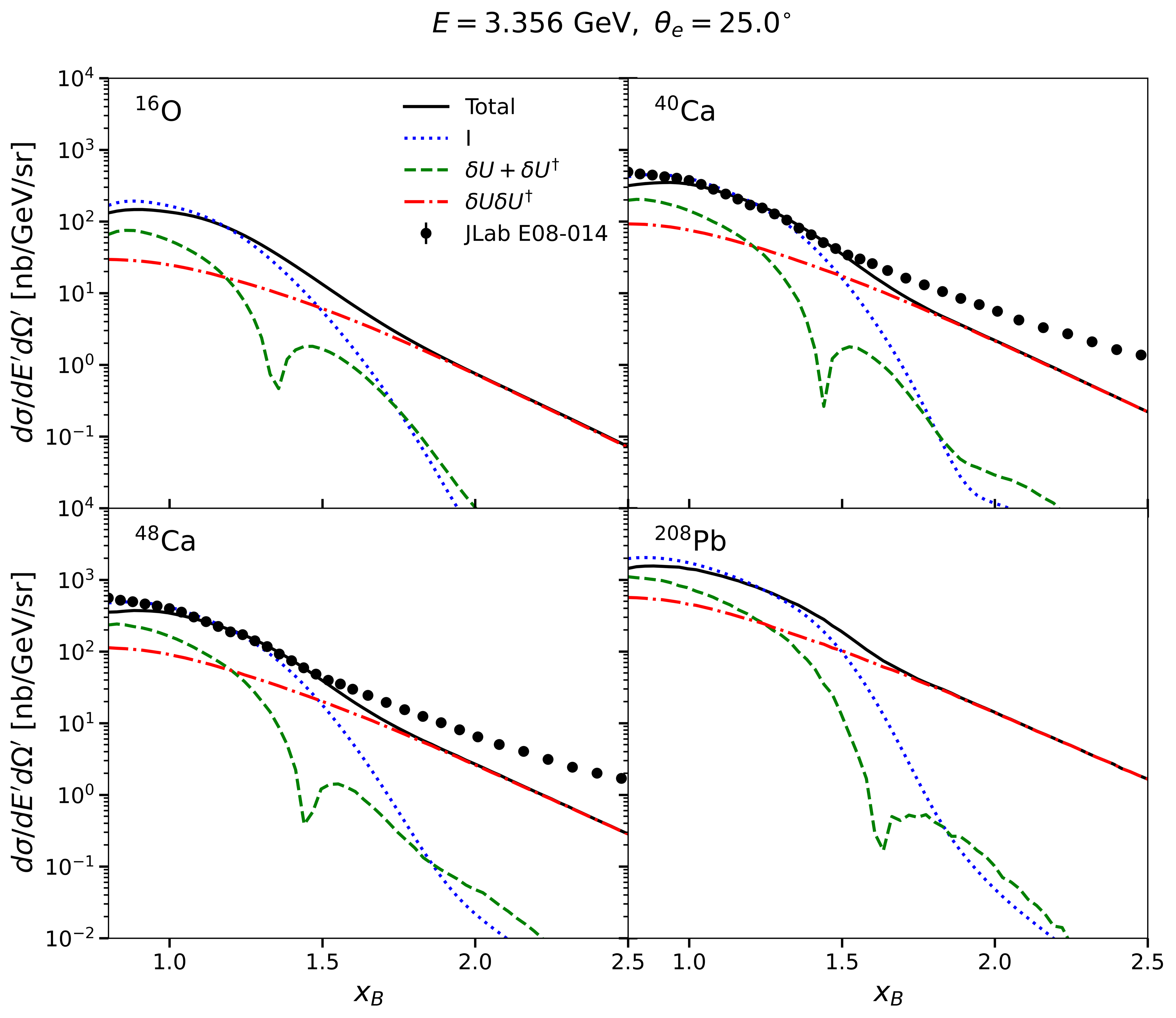}
    \caption{Same as Fig.~\ref{fig:cs_4nuclei_A1}, with the two-body SRG operator
    contribution $\delta U\,\delta U^{\dagger}$ (red dash-dot) added. The
    solid black curve now contains all SRG operator contributions through
    second order in the unitary transformation, $I + \delta U +
    \delta U^{\dagger} + \delta U\,\delta U^{\dagger}$, again at
    $\lambda = 2~\mathrm{fm}^{-1}$.}
    \label{fig:cs_4nuclei_A1_A2}
\end{figure}

Squaring this nuclear current matrix element to compute the nuclear tensor naturally generates the 
quadratic $\delta\hat{U}\,\delta\hat{U}^{\dagger}$ contribution to the cross section. Physically, 
this term corresponds to the virtual photon striking a nucleon that belongs to a short-range correlated pair, 
resulting in the emission of both the struck nucleon and its correlated partner, while leaving the residual 
nucleus in a two-hole (A-2) spectator state.

Figure~\ref{fig:cs_4nuclei_A1_A2} presents the total inclusive cross section incorporating these 2N + (A-2) 
final states. The isolated contribution from the quadratic $\delta U \,\delta U^{\dagger}$ term is 
represented by the red dash-dot curve. This term acts to explicitly redistribute the probability that was 
depleted from the quasi-elastic peak by the linear interference terms (green dashed curve). Consequently, 
this missing strength is shifted to higher values of missing momentum, manifesting as an enhancement in 
the cross section at $x_B > 1.3$. The total inclusive cross section---evaluated at $\lambda = 2~\mathrm{fm}^{-1}$ and incorporating all SRG 
operator contributions through second order in the unitary transformation ($\mathbb{I} + \delta U + 
\delta U^{\dagger} + \delta U \,\delta U^{\dagger}$)---is shown as the solid black curve. It 
should be noted that the specific scale parameter $\lambda = 2~\mathrm{fm}^{-1}$ was not determined through 
a rigorous fitting procedure, but was instead chosen phenomenologically to best reproduce the cross section 
in the intermediate $1.3 < x_B < 1.5$ region. While this systematic inclusion of initial-state SRCs successfully 
generates the high-momentum tails missing from the pure mean-field results, the theoretical calculation still 
underestimates the experimental data in the far tail ($x_B > 1.5$). Attempting to compensate for this deficit 
by further decreasing $\lambda$ to generate more high-momentum configurations proves unviable, as doing so 
leads to an overprediction of the data in the $1.3 < x_B < 1.5$ window. This residual shortfall indicates 
that initial-state correlations alone cannot fully account for the high-momentum reaction dynamics, confirming 
that the plane-wave impulse approximation remains insufficient in this regime.

\section{Summary}

This chapter demonstrated how the Similarity Renormalization Group reconciles
``low-resolution'' independent-particle models of the nucleus with ``high-resolution'' correlated
many-body pictures. By utilizing a continuous unitary transformation driven by the Wegner generator,
the bare Hamiltonian is systematically driven toward a band-diagonal form, decoupling low- and
high-momentum modes. Because the transformation is strictly unitary, the underlying physical
spectrum remains invariant; however, the short-range nuclear dynamics are formally shifted from the
highly correlated many-body state into the probing operators themselves.

This operator-evolution framework proves exceptionally powerful when formulated within LF
quantization. Because LF wavefunctions are boost-invariant, the relative and center-of-mass
coordinates cleanly separate---a fundamental feature that circumvents the intractable computational
scaling encountered when applying SRG techniques in the IF. By pairing SRG-evolved operators with
the uncorrelated wavefunctions of Chapter~\ref{chap:dft}, this work achieves the first relativistic
nuclear structure calculation of its kind to systematically incorporate short-range correlations
into a LF mean-field framework.

The practical utility of SRG techniques was demonstrated across two primary observables. First, by
evaluating the SRG-evolved one-body momentum distribution operator against a mean-field state, we
demonstrated the generation of high-momentum tails characteristic of SRCs. Second, the SRG framework
systematically resolves the theoretical mismatch inherent in standard plane-wave impulse
approximation calculations for inclusive electron-nucleus cross sections, which incorrectly pair
bare, one-body probes with low-resolution target states. By formally evolving the electromagnetic
current, we naturally induced the effective two-body interactions necessary to couple the virtual
photon to 2N plus (A-2)-nuclear final-state configurations. While this systematic inclusion of
short-range dynamics establishes a rigorous, computationally tractable baseline for accurately
modeling inclusive quasi-elastic electron-nucleus scattering at high energies, a discrepancy
remains. We find that although incorporating SRCs into the LF nuclear structure successfully
increases the cross-section support missing from mean-field results at $x_B > 1.5$, it still
underestimates the experimental data. This residual shortfall indicates that initial-state
correlations alone are insufficient, directly motivating the inclusion of final-state interactions
in the next chapter.

% !TEX root = ../uwthesis.tex
\chapter{Final State Interactions}\label{chap:fsi}

The inclusive cross section is the probability of detecting a scattered electron at given
kinematics, characterized by $x_B$ and $Q^2$, or equivalently by the scattered electron's energy and
angle. Experimentally, the electron interacts with the initial target and emerges into the detector,
while the rest of the system is left to evolve and re-interact unobserved. The yield of detected
electrons is therefore insensitive to the dynamics that occur after the initial electron-nucleus
scattering: whatever rescattering takes place inside the target system cannot retroactively alter
the electron that has already been measured. Theoretically, this translates into the statement that
the inclusive cross section can be written without any reference to the final state at all,
\begin{equation}\label{cs_without_final_state}
\begin{aligned}
    \frac{d\sigma}{dE'd\Omega'} \propto & \sum_X \braket{A|\hat{J}^\mu(0)|X} \braket{X|\hat{J}^\nu(0)|A} \delta(P_A^- + q^- - p_X^-) \\
    = & \braket{A|\hat{J}^\mu(0)\,\delta(P_A^- + q^- - \hat{P}^-)\,\hat{J}^\nu(0)|A}.
\end{aligned}
\end{equation}
\noindent The sum over final states in the first line is a book-keeping device for tabulating the
contributions of individual channels --- exactly the procedure followed throughout this thesis ---
but the second line shows that the inclusive observable is fundamentally a ground-state matrix
element of an operator built from the full Hamiltonian, with no final-state basis appearing
anywhere. Why, then, do the results of the previous chapter suggest missing physics from final-state
interactions (FSI)?

The answer is that the calculations performed so far have captured only a portion of the inclusive
cross section within each final-state channel. The resolution lies in recognizing what is required
to pass from the second line of Eq.~(\ref{cs_without_final_state}) back to the first. The collapse
of the sum over $|X\rangle$ into the operator $\delta(P_A^- + q^- - \hat{P}^-)$ is a spectral
decomposition, which is valid only when the states $|X\rangle$ are eigenstates of the very
Hamiltonian appearing inside the delta function. The energy-conserving delta function itself is a
consequence of overall four-momentum conservation, which derives from translation invariance
generated by the full Hamiltonian; it is therefore implicit at every stage that the initial and
final states are eigenstates of $\hat{P}^-$. The plane-wave calculations performed so far capture
only a portion of the full scattering amplitude, since the true continuum eigenstates are physical
scattering states, of which plane waves are merely the asymptotic piece; the remaining
$T$-matrix-driven contributions are generated order by order in light-front perturbation theory and
have simply not yet been included. Therefore, for inclusive scattering, FSI effects as
conventionally presented in the literature are not a separate dynamical mechanism that redistributes
the scattered-electron yield, but the missing theoretical contribution required to restore the
original inclusive observable towards its basis-independent form.

\section{Elastic Two-Nucleon Re-scattering}

In the $x_B > 1.5$ regime the dominant initial-state configurations are short-range correlated (SRC)
pairs. Such pairs are characterized by a large relative momentum and a small center-of-mass
momentum, so that the two correlated nucleons are close to one another while the pair as a whole is
spatially separated from the residual A-2 system. Consequently, when the virtual photon strikes one
member of the pair, the struck nucleon is far more likely to re-scatter from its correlated partner
than from the remaining A-2 nucleons as it leaves the nucleus. We model this final-state interaction
FSI by replacing the two-nucleon plane-wave final state with the scattering state generated by the LF
minimal-relativity AV18 potential,
\begin{equation}
\begin{gathered}
\ket{p_1,p_2}\bra{p_1,p_2}
\rightarrow \ket{\Psi_{NN}^{(-)},\bm{P}}\bra{\Psi_{NN}^{(-)},\bm{P}}, \\
\ket{\Psi_{NN}^{(-)},\bm{P}} = \ket{p_1,p_2}
+ \frac{1}{\hat{P}^+}\frac{1}{p_1^- + p_2^- - \hat{P}^- - i\epsilon}\,
\hat{P}^+\hat{T}^{(-)}\ket{p_1,p_2},
\end{gathered}
\end{equation}
where the $(-)$ superscript denotes the physical final-state scattering solution, $\hat{T}^{(-)}$ is
the half-off-shell $T$-matrix, and the limit $\epsilon \to 0^+$ is understood. The factor
$\hat{P}^+/\hat{P}^+$ is inserted because the $T$-matrix obtained from Eq.~\eqref{weinberg_eq} is
proportional to the matrix element of $\hat{P}^+\hat{T}^{(-)}$.

Rather than solving for $\hat{T}^{(-)}$ directly, we first compute the half-off-shell $K$-matrix.
The $K$-matrix generates standing-wave (principal-value) scattering states \cite{Taylor:1972pty};
because it is real and satisfies an integral equation that can be rendered non-singular, it is
numerically more convenient than the imaginary $T$-matrix. Partial-wave projection of
Eq.~\eqref{rel_lippmann_schwinger}, together with the standing-wave boundary condition, leads to the
coupled integral equations for the K-matrix,
\begin{equation}\label{K_matrix_eq}
    \left[\, \tilde{K}_{LL'}^{JTS}(k_{\rm off},p)
    = \tilde{V}_{LL'}^{JTS}(k_{\rm off},p)
    + \sum_{L''}\mathcal{P}\!\int \frac{dk'\,k'^2}{(2\pi)^3 E(k')}\,
    \frac{\tilde{V}_{LL''}^{JTS}(k_{\rm off},k')\,
    \tilde{K}_{L''L'}^{JTS}(k',p)}{4(p^2 - k'^2)} \,\right]_{\rm IF},
\end{equation}
where the labels $(J,T,S,L,L')$ specify the partial-wave channel and $\mathcal{P}$ denotes the
principal value.

We solve Eq.~\eqref{K_matrix_eq} numerically by discretizing the momentum integral on a
Gauss--Legendre grid $\{k_i, w_i\}$, which converts the integral equation into a matrix equation.
The kernel is singular at $k' = p$; we regulate it by principal-value subtraction, which exploits
the identity
\begin{equation}
    \mathcal{P}\!\int_0^\infty \frac{dk'}{p^2 - k'^2} = 0,
\end{equation}
so that
\begin{equation}
    \mathcal{P}\!\int_0^\infty dk'\,\frac{f(k')}{p^2 - k'^2}
    = \int_0^\infty dk'\,\frac{f(k') - f(p)}{p^2 - k'^2}.
\end{equation}
The subtracted integrand is finite at $k' = p$, with limiting value $-f'(p)/2p$, so the kernel of
Eq.~\eqref{K_matrix_eq} is smooth across the grid. After subtraction and discretization,
Eq.~\eqref{K_matrix_eq} takes the linear form $\left(\mathbb{I} - \tilde{V}\tilde{G}\right)\tilde{K}
= \tilde{V}$, where $\tilde{G}$ collects the (subtracted) grid weights and propagator factors; we
solve this system for $\tilde{K}$ by direct matrix inversion. Finally, the half-off-shell $T$-matrix
is recovered from $\tilde{K}$ through the (relativistic) Heitler relation,
\begin{equation}
\begin{gathered}
    \left[ \left( \tilde{T}_{LL'}^{JTS}(k_{\rm off},p) \right)^{(-)}  = \frac{\tilde{K}_{LL'}^{JTS}(k_{\rm off},p)}{\hat{\mathbb{I}}-i\rho_{\rm ps}(p)\tilde{K}_{LL'}^{JTS}(p,p)}\right]_{\rm IF} \\
    \left[ \rho_{\rm ps}(p) = \frac{p}{64\pi^2E(p)} \right]_{\rm IF}
\end{gathered}
\end{equation}
where $\rho_{\rm ps}$ is the on-shell phase-space factor fixed by the phase-space factors in
Eq.~\eqref{rel_lippmann_schwinger}. Finally, the LF analogue is obtained from Eq.
\eqref{eq:T_matrix_ET_LF}.

Incorporating FSI requires the struck nucleon to remain off-shell even
after interacting with the virtual photon. Consequently, the single-nucleon electromagnetic current
matrix element introduced in Chapter~\ref{chap:lightfront} must be generalized to account for this
final-state off-shellness:
\begin{equation}
    J_N(\bm{p}',\sigma',\tau;\,\bm{p},\sigma,\tau) = \bar{u}(\bm{p}',\sigma',\tau)\!\left(1 + \frac{\slashed{\Delta}_{p'}}{2m_N}\right)\! \Gamma^\mu_{\textcolor{red}{\gamma^*}N}\!\left(1 + \frac{\slashed{\Delta}_{p}}{2m_N}\right)\!u(\bm{p},\sigma,\tau),
\end{equation}
where the final-state off-shell operator, $\slashed{\Delta}_{p'}$, follows the exact same structure
as Eq.~\eqref{EM_offshell_factors}. We note that the recoil spectator nucleon involved in the
elastic rescattering is technically also off-shell. However, we neglect its corresponding off-shell
corrections here because they are kinematically suppressed. Specifically, these factors only become
significant in the integration region where $\alpha_{r} \equiv 2p_r^+/P^+ \to 0$, with $p_r$
the momentum of the SRC recoil nucleon. Physically, this limit describes a strongly virtual
spectator nucleon emerging from the deuteron vertex and instantaneously interacting with the struck
nucleon---a configuration whose contribution to the overall integral is negligible \cite{Vera:2021rnw}.

\section{Results and Discussion for Deuterium and Calcium-40}\label{sec:fsi_deuterium_CA40}

We apply the FSI formalism developed above to the deuteron. As a pure $A=2$ system, the deuteron
admits no residual $(A-2)$ spectators; the only final-state rescattering is the elastic
re-interaction between the two nucleons modeled here. We solve the K-matrix
equation~\eqref{K_matrix_eq} for the LF minimal-relativity AV18 potential in all $S$, $P$, and $D$ partial-wave
channels, retaining the coupled $^3S_1$--$^3D_1$ and $^3D_3$--$^3G_3$ channels, and construct
the LF half-off-shell $T$-matrix via Eq.~\eqref{eq:T_matrix_ET_LF}.

\begin{figure}
    \centering
    \includegraphics[width=1.0\linewidth]{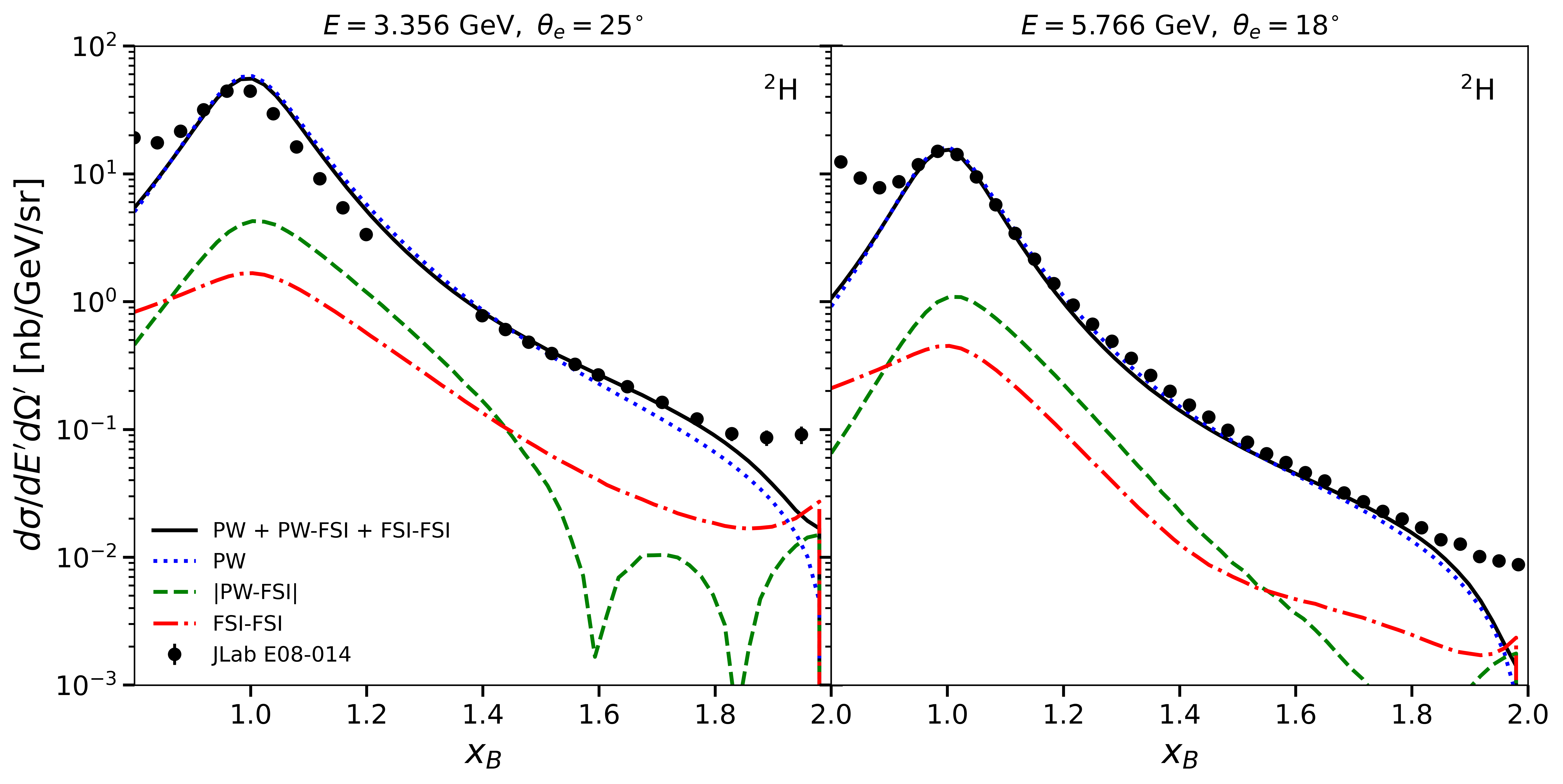}
    \caption{Inclusive electron–deuteron cross section as a function of the Bjorken
             variable $x_B$ at the JLab E08-014 kinematics ($E = 3.356$~GeV,
             $\theta_e = 25^\circ$, left) and the JLab E02-019 kinematics
             ($E = 5.766$~GeV, $\theta_e = 18^\circ$, right), decomposed into plane-wave
             and FSI contributions. The solid black curve is the full theory result,
             $\sigma_{\rm PW} + \sigma_{\rm PW-FSI} + \sigma_{\rm FSI-FSI}$; the blue dotted curve is the
             plane-wave piece; the green dashed curve is the PW–FSI interference $|\sigma_{\rm PW-FSI}|$; the red dash-dot curve
             is the FSI quadratic contribution $\sigma_{\rm FSI-FSI}$. Black points are the experimental
             data.}
    \label{fig:H2_FSI}
\end{figure}

Figure~\ref{fig:H2_FSI} shows the resulting inclusive cross sections compared to JLab data at 
two kinematic settings: $E = 3.356$ GeV, $\theta_e = 25^\circ$ from JLab E08-014~\cite{Zhang:2025nst} (left panel) 
and $E = 5.766$ GeV, $\theta_e = 18^\circ$ from JLab E02-019~\cite{Fomin:2011ng} (right panel), together 
with the individual FSI contributions. The interference term includes both cross terms from the squared matrix element, 
PW-FSI + FSI-PW; we label it simply as PW-FSI. Since it removes probability from the plane-wave channel by unitarity,
it is plotted as its absolute value. The quadratic term is positive and redistributes probability.

The central result is that the full cross section is nearly indistinguishable from the
plane-wave result alone across the entire $x_B$ range shown for both kinematics. The interference
correction lies one to two orders of magnitude below the cross section at most kinematics, as does the
quadratic correction.

We now extend the same calculation to $^{40}$Ca. We only apply FSI to the SRG-induced 2N + (A-2) final state term.
The FSI is again described by the two-nucleon $T$-matrix from the minimal-relativity AV18 potential. Results
are compared to JLab E08-014 data for $^{40}$Ca \cite{Zhang:2025nst}. The results for $^{40}$Ca are identical in 
character to those for the deuteron: the FSI corrections are negligible and the full cross section is again nearly 
indistinguishable from the plane-wave result alone.

\begin{figure}
    \centering
    \includegraphics[width=0.8\linewidth]{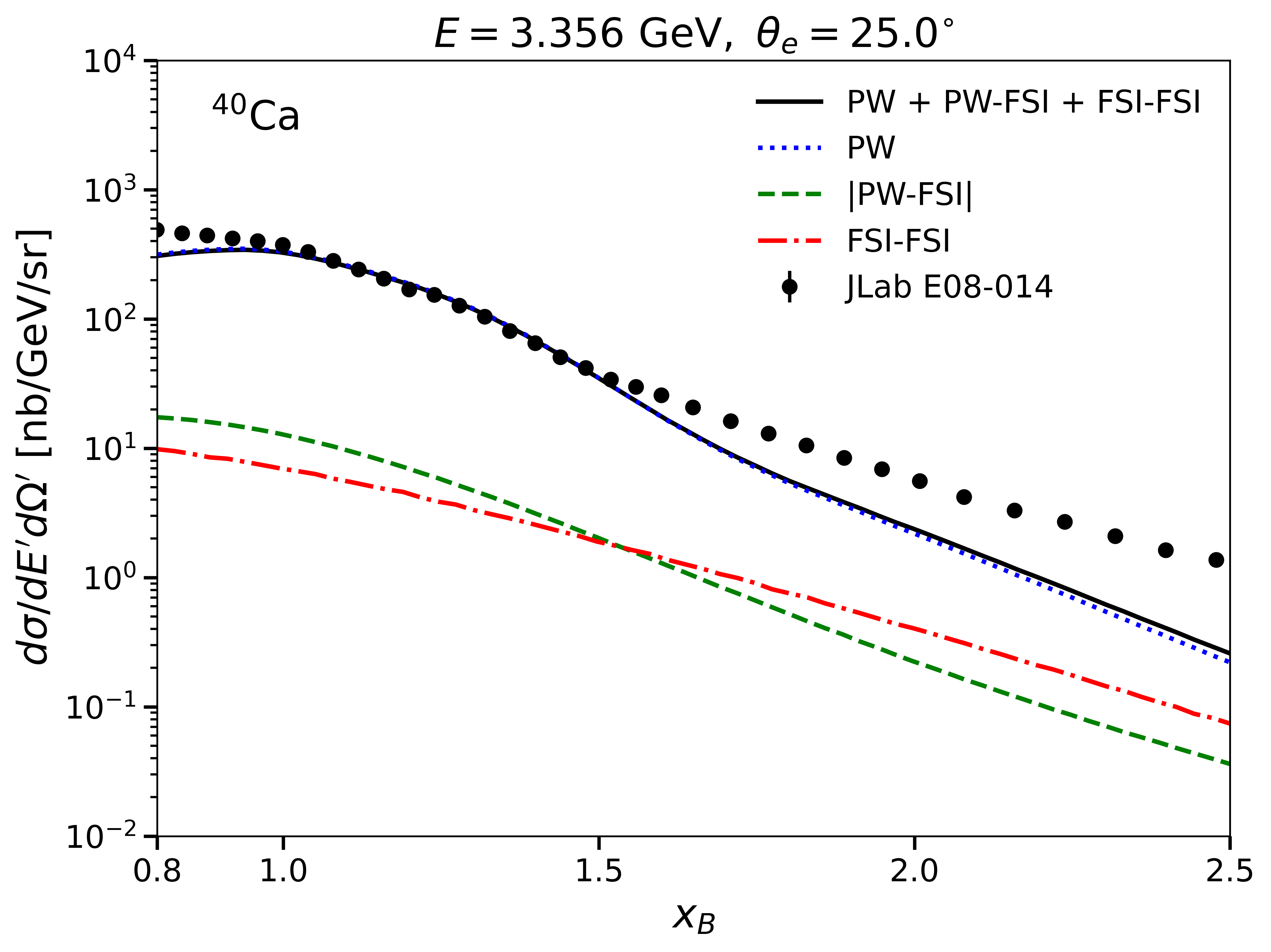}
    \caption{Inclusive electron-$^{40}$Ca cross section as a function of $x_B$ for JLab
    E08-014 kinematics. Curves as in Fig.~\ref{fig:H2_FSI}. Data (black points) from
    JLab E08-014. The `PW' term is the same as the `Total' term labeled in Fig. \ref{fig:cs_4nuclei_A1_A2}.}
    \label{fig:Ca40_FSI}
\end{figure}

Our results differ quantitatively from the similar elastic rescattering study of
Ref.~\cite{CiofidegliAtti:1994ys}. Both approaches employ elastic two-nucleon rescattering, so the
difference is attributable to details of the nuclear wavefunction and potential, as well as to the
relativistic light-front treatment of nuclear structure in the present work, compared to the
non-relativistic nuclear structure used in Ref.~\cite{CiofidegliAtti:1994ys}.

More broadly, the AV18 $T$-matrix captures only elastic
$NN$ degrees of freedom and is therefore missing the inelastic pionic channels that contribute to
the physical FSI. These include virtual-photon excitation of a nucleon to a $\Delta$-isobar,
with subsequent decay $\Delta \to N\pi$, and the photon scattering off a virtual pion exchanged
between nucleons and promoting it on-shell. The natural extension that incorporates these effects
phenomenologically, without requiring an explicit pion-production amplitude, is to replace the AV18
$T$-matrix with effective amplitudes built from experimental $NN$ cross sections, which
automatically absorb all inelastic channels. This approach is the Generalized Eikonal
Approximation, developed in
Refs.~\cite{Frankfurt:1996xx,Sargsian:2001ax,Sargsian:2009hf,Vera:2021rnw}, and constitutes the
natural follow-up to the present work.

These missing inelastic pionic contributions leave a clear empirical signature. While our
purely nucleonic model --- two-nucleon SRC configurations with elastic re-scattering --- reproduces
the deuterium data well for $x_B>1$ shown in Fig.~\ref{fig:H2_FSI}, the same
model falls short of the $^{40}$Ca data at large $x_B$ (Fig.~\ref{fig:Ca40_FSI}), where the
cross section is dominated by the SRC contribution. The growing discrepancy with nuclear mass
reflects the enhanced pionic FSI in the denser nuclear medium of heavier nuclei, which our
nucleonic framework does not capture.

This has a direct consequence for the interpretation of the SRC plateau in the experimental
ratio $\sigma(A)/\sigma(D)|_{x_B > 1.5}$. Because our theoretical cross section for heavy nuclei 
underpredicts the data while the deuterium cross section does not, the experimental plateau exceeds the purely nucleonic ratio
$\sigma_{\rm theory}(A)/\sigma_{\rm theory}(D)$. The standard identification of the plateau
value with nucleonic SRC abundance $a_2(A)$ \cite{Fomin:2017ydn} therefore overestimates the
purely nucleonic content: the experimental ratio is inflated by the pionic FSI channels that
our theory is missing, and which do not cancel between the heavy nucleus and deuterium. By the
same reasoning, theoretical and phenomenological extractions of the average number of SRC pairs
from momentum distribution ratios \cite{Weiss:2020bkp} inherit the same sensitivity, and the
pair counts extracted from experimental data should be understood as an upper bound on the
purely nucleonic SRC content.
% !TEX root = ../uwthesis.tex
\chapter{Conclusion}\label{chap:conclusion}

The advent of high-energy experiments at Jefferson Lab and the planned Electron-Ion Collider, both
of which aim to probe non-perturbative aspects of QCD using nuclear targets, places a new demand on
low-energy nuclear physics: the structure models that feed these programs must be relativistic.
Light-front quantization is the natural framework for this regime, but the path to relativistic
many-body calculations is not obvious. Correlated \textit{ab initio} approaches based on the
many-body Dirac equation are out of reach---the requisite two- and three-body relativistic
interactions have not been developed, and even the numerical machinery for solving the Dirac
many-body problem is in its infancy. Furthermore, the EIC's heavy-nuclei program compounds the
difficulty, pushing well beyond what even non-relativistic \textit{ab initio} methods can currently
address.

The strategy adopted in this dissertation is to be as conservative as possible with the existing,
successful machinery of IF nuclear structure, and to reformulate it over to the LF. The two key
developments along this line are presented.

\section*{Light-front density functional theory}

The first development, presented in Chapter~\ref{chap:dft}, is the formulation of LF density
functional theory and, more importantly, a demonstration of its formal equivalence to IF DFT. This
equivalence is exact: no approximation is required to map one onto the other. The practical
consequence is that existing and future developments in IF DFT transfers entirely to the LF. Applied
to inclusive electron-nucleus scattering at high momentum transfer and low energy transfer, LF DFT
reproduces the quasi-elastic peak well. It fails, however, at high Bjorken-$x_B$, in line with what
is already established from $A(e,e'p)$ measurements: mean-field theory does not describe the
high-momentum tail of the nuclear wave function.

\section*{Beyond mean field: SRG and short-range correlations}

Short-range correlation phenomenology identifies the missing strength at high $x_B$ with
two-nucleon configurations of high relative momentum and low center-of-mass momentum. The second
development of this program incorporates these correlations into the LF framework by using the
Similarity Renormalization Group. SRCs do increase the inclusive cross section in the expected
direction. They are, however, insufficient: even after including elastic nucleon-nucleon
rescattering as a final-state interaction, the predicted cross section still undershoots the data.
This suggests beyond purely nucleonic, elastic dynamics---toward inelastic channels as necessary 
ingredients in the high-$x_B$ regime, currently missing in phenomenology.

\section*{Implications for the SRC plateau}

The cross-section ratio $\sigma(A)/\sigma(D)\big|_{x_B > 1.5}$, which exhibits a characteristic
plateau, is standardly identified with the per-nucleon SRC probability $a_2(A)$~\cite{Fomin:2017ydn}.
This identification rests on the implicit assumption that the dominant non-nucleonic contributions
cancel between numerator and denominator. The present analysis shows this cancellation is
incomplete: the experimental cross section for nuclei contain inelastic 
contributions, whereas the deuterium cross section is well described by the
purely nucleonic model. The experimental ratio therefore exceeds the purely nucleonic ratio
$\sigma_{\rm theory}(A)/\sigma_{\rm theory}(D)$, and the conventional $a_2(A)$ overestimates the
purely nucleonic SRC content. The nucleonic theory developed here provides a lower bound on the
ratio; the gap between the experimental plateau and the theoretical prediction quantifies the inelastic
excess that inflates it. By the same reasoning, phenomenological extractions of SRC pair counts from
ratios of nuclear momentum distributions~\cite{Weiss:2020bkp} inherit the same sensitivity, and
those pair counts should be understood as upper bounds on the purely nucleonic content.

\section*{Code availability}

The codes developed for this dissertation are publicly available at
\url{https://github.com/dima-kim/Light-Front-Quasi-Elastic-Scattering}.

\section*{Outlook}

The most immediate continuation of this work is the incorporation of inelastic FSI through the
Generalized Eikonal
Approximation~\cite{Frankfurt:1996xx,Sargsian:2001ax,Sargsian:2009hf,Vera:2021rnw}. The GEA
replaces the AV18 $T$-matrix with effective scattering amplitudes built from experimental
nucleon-nucleon cross sections, absorbing all inelastic channels---pionic production
included---without requiring an explicit pion-production amplitude. Within the LF framework
developed here, the GEA is the minimal extension needed to test whether pionic FSI accounts
quantitatively for the deficit observed in $^{40}$Ca.

On a more fundamental level, the results of this dissertation highlight the importance of developing
light-front nucleon-nucleon potentials that incorporate pion and $\Delta$-isobar degrees of freedom
explicitly. The AV18 potential used here is an instant-form potential adapted to the light-front
through the minimal-relativity prescription---a procedure that does not generate the full
relativistic structure of the NN interaction. A genuinely light-front NN potential with explicit
pion-exchange and $\Delta$-isobar channels would encode the same pionic physics that the GEA
currently handles phenomenologically, but within a dynamically consistent relativistic framework.
Such a potential would improve the FSI treatment and simultaneously provide a more faithful starting
point for the SRG evolution that underlies the SRC physics of this dissertation, bringing the
entire program closer to a relativistically complete description of nuclear structure for the
high-energy experiments at JLab and the planned Electron-Ion Collider.

% ── appendices ────────────────────────────────────────
\appendix
% !TEX root = ../uwthesis.tex
\chapter{Quantization of Free Dirac Particle}\label{appendix:dirac}
Starting from the Dirac equation, we expand the contraction between the Dirac $\gamma$-matricies and
four derivative in terms of light-front coordinates,

\begin{equation} \label{dirac_eqn}
    (i \gamma_\mu \partial^\mu - m)\psi(x) = \left(i \left[ \frac{1}{2} \gamma^+ \partial^- + \frac{1}{2} \gamma^- \partial^+ - \bm{\gamma}^\perp \cdot \bm{\partial}^\perp \right] - m \right)\psi(x) = 0.
\end{equation}

\noindent Using light-front variables, one can develop a projection operator in Dirac space,

\begin{equation}
\begin{gathered}
    \Lambda^+ = \frac{1}{4}\gamma^- \gamma^+ = \frac{1}{2}\gamma^0 \gamma^+, \\
    \Lambda^- = \frac{1}{4}\gamma^+ \gamma^- = \frac{1}{2}\gamma^0 \gamma^-, \\
    \Lambda^+ + \Lambda^- = \mathbb{I}, \quad \Lambda^\pm \Lambda^\pm = \Lambda^\pm, \quad \Lambda^\pm \Lambda^\mp = 0.
\end{gathered}
\end{equation}

\noindent Defining $\psi(x) = \psi^+(x) + \psi^-(x)$, where $\psi^+(x) = \Lambda^+ \psi(x)$ and vice
versa, we project the Dirac equation using $\Lambda^+$ and $\Lambda^-$,
\begin{equation}\label{dirac_eq_proj}
    \begin{gathered}
        2i\partial^-\psi^+(x) = \left( -i \bm{\gamma}^\perp \cdot \bm{\partial}^\perp +m \right) \gamma^- \psi^-(x), \\
        2i\partial^+\psi^-(x) = \left( -i \bm{\gamma}^\perp \cdot \bm{\partial}^\perp +m \right) \gamma^+ \psi^+(x).
    \end{gathered}
\end{equation}
Notice that the second expression does not contain any LF time derivative terms, hence it is a
constraint equation. In other words, the dynamical field is $\psi^+(x)$, which can be used to solve
for $\psi^-(x)$. Thus, we impose equal LF time commutation relations on the dynamical + component of
the Dirac fields \cite{Chang:1972xt},
\begin{equation}\label{dirac_commutation}
    \{ \hat{\psi}^+(x^+,x^-,\bm{x}^\perp), \hat{\psi}^{+\dagger}(y^+,y^-,\bm{y}^\perp)  \}_{x^+ = y^+} = \Lambda^+ \delta(x^- - y^{-}) \delta^{(2)}(\bm{x}^\perp - \bm{y}^\perp).
\end{equation}
The mode expansion yields,
\begin{equation}\label{dirac_field_pw}
    \hat{\psi}(x) = \sum_\sigma \int \frac{dp^+d\bm{p}^\perp}{(2 \pi)^3 \sqrt{2p^+}} \left( \hat{a}(p) u(p) e^{-ip \cdot x} +  \hat{b}^\dagger(p)v(p) e^{+ip \cdot x} \right).
\end{equation}
Plugging Eq. \eqref{dirac_field_pw} into Eq. \eqref{dirac_commutation} yields the following
commutation relations for the creation and annihilation operators
\begin{equation}
    \{ \hat{a}(p), \hat{a}^\dagger(p') \} = \{ \hat{b}(p), \hat{b}^\dagger(p') \} = (2 \pi)^3 \delta(p^+ - p'^{+}) \delta^{(2)}(\bm{p}^\perp - \bm{p}'^\perp)\delta_{\sigma \sigma'}.
\end{equation}
The representation of the Dirac spinors can be obtained from plugging in Eq. \eqref{dirac_field_pw}
into Eq. \eqref{dirac_eq_proj}. In doing so, one finds that we have freedom in choosing what the
plus projection of the spinors are, while the minus projections are constrained. For instance, the
$u^-(p)$ spinor can be obtained from $u^+(p)$ through
\begin{equation}
    u^-(p) = \frac{1}{p^+}(\bm{\alpha}^\perp \cdot \bm{p}^\perp + \gamma^0m)u^+(p),
\end{equation}
same for $v^-(p)$ but with $-m$ in the above expression. Choosing the plus projections to be
proportional to eigenstates of $\Lambda^+$ we get the Brodsky-Lepage Dirac spinor representations
\cite{Lepage:1980fj,Brodsky:1997de}
\begin{align}
    u(\bm{p}, \sigma) &= \frac{1}{\sqrt{p^+}} \left( p^+ + \beta m + \bm{\alpha}^\perp \cdot \bm{p}^\perp \right) \times
    \begin{cases}
        \chi(\uparrow), & \text{for } \sigma = +1, \\
        \chi(\downarrow), & \text{for } \sigma = -1,
    \end{cases} \\
    v(\bm{p}, \sigma) &= \frac{1}{\sqrt{p^+}} \left( p^+ - \beta m + \bm{\alpha}^\perp \cdot \bm{p}^\perp \right) \times
    \begin{cases}
        \chi(\downarrow), & \text{for } \sigma = +1, \\
        \chi(\uparrow), & \text{for } \sigma = -1.
    \end{cases}
\end{align}
The two $\chi$-spinors, which are the eigenstates of $\Lambda^+$, are
\begin{equation}
    \chi(\uparrow) = \frac{1}{\sqrt{2}} \begin{pmatrix} 1 \\ 0 \\ 1 \\ 0 \end{pmatrix},
    \qquad
    \chi(\downarrow) = \frac{1}{\sqrt{2}} \begin{pmatrix} 0 \\ 1 \\ 0 \\ -1 \end{pmatrix}.
\end{equation}

% !TEX root = ../uwthesis.tex
\chapter{Inclusive Electron-Nucleus Cross Section for (A-1)-Nuclear Final States}\label{appendix:A-1 implementation}
Restricting ourselves to doubly-magic spherically symmetric ($J=0$) nuclear initial states, with
single nucleon plus (A-1)-nuclear final states, terms in Eq. (\ref{nuclear_tensor}) become

\begin{equation}
    \sum_X\int d\Pi_X \rightarrow \sum_{\sigma_1 \tau_1} \sum_f \int [dP_{A-1}][dp_1],
\end{equation}

\begin{equation}
    \ket{X}\bra{X} \rightarrow \ket{\Psi_{A-1}^f,\bm{P}_{A-1}}\bra{\Psi_{A-1}^f,\bm{P}_{A-1}} \otimes \ket{\bm{p}_1,\sigma_1,\tau_1}\bra{\bm{p}_1,\sigma_1,\tau_1},
\end{equation}

\noindent where $f$ denotes all the possible (A-1)-nuclear states. The nuclear tensor becomes

\begin{equation}
\begin{aligned}
    W_A^{\mu \nu} = & \, \frac{1}{4 \pi m_A} \sum_{\sigma_1 \tau_1} \sum_f \int [dP_{A-1}][dp_1] (2 \pi)^4 \delta^4(q^\mu + P_A^\mu - P_{A-1}^\mu - p_{1}^\mu)\\
    & \qquad\qquad\qquad\qquad    \times \bra{\Psi_A} \hat{J}^{\mu}_A (0) \ket{\Psi_{A-1}^f,\bm{P}_{A-1};\bm{p}_1,\sigma_1,\tau_1} \\
    & \qquad\qquad\qquad\qquad    \times \bra{\Psi_{A-1}^f,\bm{P}_{A-1};\bm{p}_1,\sigma_1,\tau_1} \hat{J}_A^{\nu}(0) \ket{\Psi_A} \\
    = & \, \frac{1}{2 m_A} \frac{1}{2 P_{A-1}^+}\sum_{\sigma_f \tau_f} \sum_f \int \frac{dp_i^+ d\bm{p}_{i}^{\perp}}{(2 \pi)^3} \left[ \frac{1}{p_i^+ + q^+}\delta(q^- + P_A^- - (P_{A-1}^f)^- - p_{1}^-) \right] \\
    & \qquad\qquad\qquad\qquad\qquad \times \bra{\Psi_A} \hat{J}^{\mu}_A (0) \ket{\Psi_{A-1}^f,\bm{P}_{A-1};\bm{p}_i+\bm{q},\sigma_f,\tau_f} \\
    & \qquad\qquad\qquad\qquad\qquad \times \bra{\Psi_{A-1}^f,\bm{P}_{A-1};\bm{p}_i+\bm{q},\sigma_f,\tau_f} \hat{J}_A^{\nu}(0) \ket{\Psi_A}, \\
\end{aligned}
\end{equation}

\noindent where in the second equality we integrated over $\bm{P}_{A-1}$, changed variables to
$\bm{p}_i = \bm{p}_1-\bm{q}$, and re-labeled $(\sigma_1, \tau_1) \rightarrow (\sigma_f, \tau_f)$.
Furthermore, we drop the $J=0$ label for the initial nuclear state. Since we work in the collinear
frame, where the virtual photon defines the negative $z$-axis,

\begin{equation}
    (P_{A-1}^f)^- = \frac{m_f^2 + (\bm{p_{i}^{\perp}})^2}{P_A^+ - p_i^+}, \qquad p_1^- = \frac{m^2 + (\bm{p_{i}^{\perp}})^2}{p_i^+ + q^+}.
\end{equation}

\noindent The nuclear current matrix element is

\begin{equation}
\begin{aligned}
    & \braket{\Psi_{A-1}^f,\bm{P}_{A-1};\bm{p}_i+\bm{q},\sigma_i,\tau_i| \hat{J}_A^{\nu}(0)| \Psi_A} \\
    & \qquad\qquad\qquad\qquad = \sum_{s_1 t_1} \sum_{\sigma \sigma' \tau} \int [dp][dp'][dq_1] \sqrt{2p^+} \sqrt{2p'^+} J_N^\nu(\bm{p}',\sigma',\tau;\bm{p},\sigma,\tau) \\
    & \qquad\qquad\qquad\qquad\quad \times \braket{\bm{p}_i+\bm{q},\sigma_f,\tau_f|\hat{a}^\dagger(\bm{p}',\sigma',
    \tau)\hat{a}(\bm{p},\sigma,
    \tau)|\bm{q}_1,s_1,t_1} \\
    & \qquad\qquad\qquad\qquad\quad \times \braket{\Psi_{A-1}^f, \bm{P}_{A-1};\bm{q}_1,s_1,t_1|\Psi_A} \\
    & \qquad\qquad\qquad\qquad = \sum_{s_1 t_1} \sum_{\sigma \sigma' \tau} \int [dp][dp'][dq_1] \sqrt{2p^+} \sqrt{2p'^+} \sqrt{2(p_i^+ + q^+)}(2q_1^+) \\
    & \qquad\qquad\qquad\qquad\quad \times \braket{0|\hat{a}(\bm{p}_i+\bm{q},\sigma_f,\tau_f)\hat{a}^\dagger(\bm{p}',\sigma',
    \tau)\hat{a}(\bm{p},\sigma,
    \tau)\hat{a}^{\dagger}(\bm{q}_1,s_1,t_1)|0} \\
    & \qquad\qquad\qquad\qquad\quad \times \, J_N^\nu(\bm{p}',\sigma',\tau;\bm{p},\sigma,\tau)  \braket{\Psi_{A-1}^f, \bm{P}_{A-1}|\hat{a}(\bm{q}_1,s_1,t_1)|\Psi_A}.
\end{aligned}
\end{equation}

\noindent Here $J_N^\nu(\bm{p}',\sigma',\tau;\bm{p},\sigma,\tau)$ is the single-nucleon
electromagnetic current matrix element defined in Chapter~\ref{chap:scattering}.

Using Wick's Theorem to evaluate $\braket{0|\hat{a}\hat{a}^\dagger\hat{a}\hat{a}^\dagger|0}$ 
yields two delta functions which we use to integrate over $q_1$ and $p'$. This gives

\begin{equation}
\begin{aligned}
    \braket{\Psi_{A-1}^f,\bm{P}_{A-1};\bm{p}_i+\bm{q},\sigma_i,\tau_i| \hat{J}_A^{\nu}(0)| \Psi_A}
    & = \sum_{\sigma} \int [dp]\sqrt{2p^+} J_N^\nu(\bm{p}_i+\bm{q},\sigma_f,\tau_f;\bm{p},\sigma,\tau_f)\\
    & \qquad\quad \times  \braket{\Psi_{A-1}^f, \bm{P}_{A-1}|\hat{a}(\bm{p},\sigma,\tau_f)|\Psi_A}.
\end{aligned}
\end{equation}

\noindent Now we focus on the nuclear matrix element term.

\begin{equation}\label{A-1_NME_decomp}
\begin{gathered}
    \braket{\Psi_{A-1}^f, \bm{P}_{A-1}|\hat{a}(\bm{p},\sigma,\tau_f)|\Psi_A} = \\\sqrt{2P_A^+}\sqrt{2P_{A-1}^+} (2 \pi)^3\delta(P_{A-1}^+ + p^+-P_A^+)\delta^{(2)}(\bm{P}_{A-1}^{\perp} + \bm{p}^{\perp}) M^A_f(\bm{p},\sigma,\tau_f),
\end{gathered}
\end{equation}

\noindent where $M^A_f$ is a quantity constructed from the overlap of intrinsic A- and (A-1)-nuclear
light-front wavefunctions. $M^A_f$ can be determined by looking at the matrix element of the
momentum distribution operator

\begin{equation}
\begin{gathered}
    \Braket{\Psi_A, \bm{P}_A'|\hat{a}^\dagger(\bm{p},\sigma,\tau)\hat{a}(\bm{p},\sigma,\tau)|\Psi_A,\bm{P}_A} = \\ 
    \qquad (2 \pi)^3 2P_A^+ \delta(P^+ - P'^+) \delta^{(2)}(\bm{P}^{\perp} - \bm{P}'^{\perp}) \, n(\bm{p},\sigma,\tau).
\end{gathered}
\end{equation}

\noindent By inserting a complete set of (A-1) states in between the creation and annihilation
operator and using Eq. (\ref{A-1_NME_decomp}) we find the following relation,

\begin{equation} \label{M_F^A true relation}
    n(\bm{p},\sigma,\tau) = \sum_f M_f^A(\bm{p},\sigma,\tau)^* M_f^A(\bm{p},\sigma,\tau).
\end{equation}

\noindent Up to this point, everything has remained exact and completely independent of any specific
nuclear model. The inherent model dependence is encapsulated within $M_f^A$, which we evaluate by
employing mean-field theory. Within the mean-field framework, we have the following relation:

\begin{equation}\label{M_F^A MFT relationn}
\begin{aligned}
    n^{MF}(\bm{p},\sigma,\tau) & =\Braket{\Psi_A^{MF}|\hat{a}^\dagger(\bm{p},\sigma,\tau)\hat{a}(\bm{p},\sigma,\tau)|\Psi_A^{MF}} \\
    &= \sum_i \Braket{\Psi_A^{MF}|\hat{a}^\dagger(\bm{p},\sigma,\tau) \Ket{\Psi_i^{MF}} \Bra{\Psi_i^{MF}}\hat{a}(\bm{p},\sigma,\tau)|\Psi_A^{MF}},
\end{aligned}
\end{equation}

\noindent comparing Eq. (\ref{M_F^A true relation}) and Eq. (\ref{M_F^A MFT relationn}) we find

\begin{equation}
\begin{gathered}
    \sum_f \rightarrow \sum_i \\
    M_f^A \rightarrow M_i^A \\
    M_i^A(\bm{p},\sigma,\tau) =  \Braket{\Psi_i^{MF}|\hat{a}(\bm{p},\sigma,\tau)|\Psi_A^{MF}}
\end{gathered}
\end{equation}

\noindent where we approximate the (A-1)-nuclear states as one-hole states relative to the target
A-nucleus state, i.e. $\ket{\Psi^{MF}_i} = \hat{a}_i\ket{\Psi_A^{MF}}$. The final result of the
nuclear current is

\begin{equation}\label{A1_current}
\begin{gathered}
    \braket{\Psi_{A-1}^f,\bm{P}_{A-1};\bm{p}_i+\bm{q},\sigma_f,\tau_f| \hat{J}_A^{\nu}(0)| \Psi_A} \approx \\
    \sum_{\sigma} \frac{1}{\sqrt{2p_i^+} }J_N^\nu(\bm{p}_i+\bm{q},\sigma_f,\tau_f;\bm{p},\sigma,\tau_f) M_i^A(\bm{p}_i,\sigma,\tau_f)
\end{gathered}
\end{equation}

% !TEX root = ../uwthesis.tex
\chapter{Derivation of Similarity Renormalization Group Terms}\label{SRG Implementation Chapter}

In this section, we provide explicit formulas for each term presented in Eq.
(\ref{srg_mom_dist_approx}). To avoid redundancy, we present a detailed step-by-step derivation
solely for the $\delta \hat{U}(\lambda) \,\hat{n}^{\tau}(p, \sigma) \, \mathbb{\hat{I}}$
interference term, as the underlying procedure straightforwardly generalizes to the remaining
components, which are subsequently outlined. The corresponding non-relativistic results are
documented in the appendix of Ref. \cite{Tropiano:2024bmu}, and the light-front adaptation presented
here closely follows that study.

Firstly, the SRG unitary transformation at flow parameter $\lambda$ in the plane-wave basis is given
by

\begin{equation}
    \begin{aligned}
        \hat{U}(\lambda) = \mathbb{\hat{I}} + \frac{1}{4}\sum_{\substack{\sigma_1 \sigma_2 \\ \sigma_3 \sigma_4}} \sum_{\substack{\tau_1  \tau_2 \\ \tau_3 \tau_4}} & \int [dP] \frac{d\alpha d\bm{k}^\perp}{(2\pi)^3 \sqrt{\alpha (2 - \alpha)}} \frac{d\alpha'd\bm{k}'^\perp}{(2\pi)^3 \sqrt{\alpha' (2 - \alpha')}}  (P^+)^2 \\
        & \times \braket{\alpha, \bm{k}^\perp, \sigma_1, \tau_1, \sigma_2, \tau_2|\delta U (\lambda)|\alpha', \bm{k}'^\perp, \sigma_3, \tau_3, \sigma_4, \tau_4}\\
        & \times \, \hat{a}^{\dagger}(p_1) \hat{a}^{\dagger}(p_2) \hat{a} (p_4) \hat{a}(p_3) \, + \, ...
    \end{aligned}
\end{equation}

\noindent where $(\alpha, \bm{k}^\perp)$ and $(\alpha', \bm{k}'^\perp)$ are the relative momenta
and $(P^+, \bm{P}^\perp)$ is the center of mass momenta. $\bm{p}_i = (p^+_i, \bm{p}_i^\perp)$ for
$i=1,2,3,4$ are defined as follows:

\begin{equation}
    \begin{gathered}
    \bm{p}_1 = \left(\frac{\alpha}{2} P^+, \frac{\alpha}{2} \bm{P}^\perp + \bm{k}^\perp \right), \qquad \bm{p}_2 = \left(\frac{2-\alpha}{2} P^+, \frac{2-\alpha}{2} \bm{P}^\perp - \bm{k}^\perp \right), \\
    \bm{p}_3 = \left(\frac{\alpha'}{2} P^+, \frac{\alpha'}{2} \bm{P}^\perp + \bm{k}'^\perp \right), \qquad \bm{p}_4 = \left(\frac{2-\alpha'}{2} P^+, \frac{2-\alpha'}{2} \bm{P}^\perp - \bm{k}'^\perp \right).
    \end{gathered}
\end{equation}

\noindent Also, the matrix element of $\delta \hat{U}$ is antisymmetrized, i.e.

\begin{equation}
\begin{aligned}
    & \braket{\alpha, \bm{k}^\perp, \sigma_1, \tau_1, \sigma_2, \tau_2|\delta U (\lambda)|\alpha', \bm{k}'^\perp, \sigma_3, \tau_3, \sigma_4, \tau_4}  \\
    & \qquad\qquad\qquad\qquad = -\braket{\alpha, \bm{k}^\perp, \sigma_1, \tau_1, \sigma_2, \tau_2|\delta U (\lambda)|2-\alpha', -\bm{k}'^\perp,\sigma_4, \tau_4, \sigma_3, \tau_3} \\
    & \qquad\qquad\qquad\qquad = -\braket{2-\alpha, -\bm{k}^\perp, \sigma_2, \tau_2, \sigma_1, \tau_1|\delta U (\lambda)|\alpha', \bm{k}'^\perp, \sigma_3, \tau_3, \sigma_4, \tau_4}.
\end{aligned}
\end{equation}

\noindent In this thesis, we focus on incorporating SRG-evolution at the two-body level only. We
will enforce this by using Wick's theorem to truncate all operators of rank $N>2$. For the momentum
distribution operator, we have four terms to consider:

\begin{enumerate}
    \item $\mathbb{\hat{I}} \,\hat{n}^{\tau}(\bm{p}, \sigma) \, \mathbb{\hat{I}}$
    \item $\delta \hat{U}(\lambda) \,\hat{n}^{\tau}(\bm{p}, \sigma) \, \mathbb{\hat{I}}$
    \item $\mathbb{\hat{I}} \,\hat{n}^{\tau}(\bm{p}, \sigma) \, \delta \hat{U^{\dagger}}(\lambda)$
    \item $\delta \hat{U}(\lambda) \,\hat{n}^{\tau}(\bm{p}, \sigma) \, \delta
\hat{U}^{\dagger}(\lambda)$
\end{enumerate}

\noindent The first term is just the bare momentum distribution operator. To begin evaluating the
second term, let us examine the string of creation and annihilation operators it involves:

\begin{equation}\label{SRG_interference_term_1}
    \hat{a}^{\dagger}(p_1) \, \hat{a}^{\dagger}(p_2) \, \hat{a} (p_4) \, \hat{a}(p_3) \,\hat{a}^{\dagger}(\bm{p},\sigma,\tau) \, \hat{a}(\bm{p},\sigma,\tau).
\end{equation}

\noindent By applying Wick's theorem, we can isolate operators of rank $N<2$,

\begin{equation}\label{interference_wicks_thm}
    \begin{aligned}
        \wick{a^{\dagger} a^{\dagger} a \c1 a \c1 a^{\dagger}  a} + & \wick{a^{\dagger} a^{\dagger} \c1 a a \c1 a^{\dagger}  a} \\
        &  = (2 \pi)^3 \delta(p_3^+ - p^+) \delta(\bm{p}_3^\perp - \bm{p}^\perp) \delta_{\sigma_3 \sigma} \delta_{\tau_3 \tau} \\
        & \qquad - (2 \pi)^3 \delta(p_4^+ - p^+) \delta(\bm{p}_4^\perp - \bm{p}^\perp)   \delta_{\sigma_4 \sigma} \delta_{\tau_4 \tau} \\
        & = \frac{2}{P^+} \Bigg[  (2 \pi)^3 \delta\left(\alpha' - \frac{2p^+}{P^+}\right) \delta\left(\frac{\alpha'}{2}\bm{P}^\perp + \bm{k}'^\perp -\bm{p}^\perp\right) \delta_{\sigma_3 \sigma} \delta_{\tau_3 \tau} \\
        & \qquad - (2 \pi)^3 \delta\left(2 - \alpha' - \frac{2p^+}{P^+}\right) \delta\left(\frac{2-\alpha'}{2}\bm{P}^\perp - \bm{k}'^\perp - \bm{p}^\perp\right) \delta_{\sigma_4 \sigma} \delta_{\tau_4 \tau} \Bigg].
    \end{aligned}
\end{equation}

\noindent First, we focus on the first delta function term in Eq (\ref{interference_wicks_thm}).

\begin{align}\label{srg_inter_first_term}
    \left[\delta \hat{U}(\lambda) \,\hat{n}^{\tau}(\bm{p}, \sigma) \, \mathbb{\hat{I}}\right]_1 = & \,\frac{1}{4}\sum_{\substack{\sigma_1 \sigma_2 \\ \sigma_3 \sigma_4}} \sum_{\substack{\tau_1  \tau_2 \\ \tau_3 \tau_4}} \int \frac{dP^+ d\bm{P}^\perp}{(2 \pi)^3} \frac{d\alpha d\bm{k}^\perp}{(2\pi)^3 \sqrt{\alpha (2 - \alpha)}} \frac{d\alpha'd\bm{k}'^\perp}{ \sqrt{\alpha' (2 - \alpha')}} \nonumber \\
    & \times \braket{\alpha, \bm{k}^\perp, \sigma_1, \tau_1, \sigma_2, \tau_2|\delta U (\lambda)|\alpha', \bm{k}'^\perp, \sigma_3, \tau_3, \sigma_4, \tau_4} \nonumber \\
    & \times \delta\left(\alpha' - \frac{2p^+}{P^+}\right) \delta\left(\frac{\alpha'}{2}\bm{P}^\perp + \bm{k}'^\perp -\bm{p}^\perp\right) \delta_{\sigma_3 \sigma} \delta_{\tau_3 \tau} \nonumber \\
    & \times\hat{a}^{\dagger}(p_1) \hat{a}^{\dagger}(p_2) \hat{a} (p_4) \hat{a}(p,\sigma,\tau) \nonumber \displaybreak \\
    = & \,\frac{1}{4}\sum_{\substack{\sigma_1 \sigma_2 \\ \sigma'}} \sum_{\substack{\tau_1  \tau_2 \\ \tau'}} \int \frac{dP^+ d\bm{P}^\perp}{(2 \pi)^3} \frac{d\alpha d\bm{k}^\perp}{(2\pi)^3 \sqrt{\alpha (2 - \alpha)}} \frac{1}{ \sqrt{p^+ (P^+ - p^+)}} \frac{P^+}{2} \nonumber \\
    & \times \Braket{\alpha, \bm{k}^\perp, \sigma_1, \tau_1, \sigma_2, \tau_2|\delta U (\lambda)|\frac{2p^+}{P^+}, \bm{p}^\perp - \frac{p^+}{P^+}\bm{P}^\perp, \sigma, \tau, \sigma', \tau'} \nonumber \\
    & \times\hat{a}^{\dagger}(p_1,\sigma_1,\tau_1) \hat{a}^{\dagger}(p_2,\sigma_2,\tau_2) \hat{a} (p_4,\sigma',\tau') \hat{a}(p,\sigma,\tau). 
\end{align}

\noindent Where in the second equality, we used the delta function to integrate over
$(\alpha',\bm{k}'^\perp)$ and re-labeled $(\sigma_4,\tau_4)\rightarrow(\sigma',\tau')$. In
evaluating the second delta function term in Eq. (\ref{interference_wicks_thm}), using the asymmetry
of $\delta \hat{U}(\lambda)$ and re-labeling $(\sigma_3,\tau_3)\rightarrow(\sigma',\tau')$, we find
that it is identical to Eq. (\ref{srg_inter_first_term}). Hence the final result is,

\begin{equation}\label{srg_inter_U}
\begin{aligned}
    \delta \hat{U}(\lambda) \,\hat{n}^{\tau}(\bm{p}, \sigma) \, \mathbb{\hat{I}} = & \,\frac{1}{2}\sum_{\substack{\sigma_1 \sigma_2 \\ \sigma'}} \sum_{\substack{\tau_1  \tau_2 \\ \tau'}} \int \frac{dP^+ d\bm{P}^\perp}{(2 \pi)^3} \frac{d\alpha d\bm{k}^\perp}{(2\pi)^3 \sqrt{\alpha (2 - \alpha)}} \frac{1}{ \sqrt{p^+ (P^+ - p^+)}} \frac{P^+}{2}\\
    & \times \Braket{\alpha, \bm{k}^\perp, \sigma_1, \tau_1, \sigma_2, \tau_2|\delta U (\lambda)|\frac{2p^+}{P^+}, \bm{p}^\perp - \frac{p^+}{P^+}\bm{P}^\perp, \sigma, \tau, \sigma', \tau'} \\
    & \times\hat{a}^{\dagger}(p_1) \hat{a}^{\dagger}(p_2) \hat{a} (p_4,\sigma',\tau') \hat{a}(p,\sigma,\tau). \\
\end{aligned}
\end{equation}

\noindent The third term is similar

\begin{equation}\label{srg_inter_Udag}
\begin{aligned}
    \mathbb{\hat{I}} \,\hat{n}^{\tau}(\bm{p}, \sigma) \, \delta \hat{U}^{\dagger}(\lambda) = & \,\frac{1}{2}\sum_{\substack{\sigma_3 \sigma_4 \\ \sigma'}} \sum_{\substack{\tau_3  \tau_4 \\ \tau'}} \int \frac{dP^+ d\bm{P}^\perp}{(2 \pi)^3} \frac{d\alpha' d\bm{k}'^\perp}{(2\pi)^3 \sqrt{\alpha' (2 - \alpha')}} \frac{1}{ \sqrt{p^+ (P^+ - p^+)}} \frac{P^+}{2}\\
    & \times \Braket{\frac{2p^+}{P^+}, \bm{p}^\perp - \frac{p^+}{P^+}\bm{P}^\perp, \sigma, \tau, \sigma', \tau' | \delta U^{\dagger}(\lambda) |\alpha', \bm{k}'^\perp, \sigma_3, \tau_3, \sigma_4, \tau_4} \\
    & \times  \hat{a}^{\dagger}(p,\sigma,\tau) \hat{a}^{\dagger}(p_2,\sigma',\tau') \hat{a} (p_4) \hat{a}(p_3). \\
\end{aligned}
\end{equation}

\noindent For the fourth term, looking at the string of creation and annihilation operators it
involves, we anticipate four terms at the two-body level:

\begin{itemize}
    \item $\wick{a^{\dagger} a^{\dagger} \c2 a \c1 a \,\, \c1 a^{\dagger} \c1 a \,\, \c1 a^{\dagger}
\c2 a^{\dagger} a a}$
    \item $\wick{a^{\dagger} a^{\dagger} \c2 a \c1 a \,\, \c1 a^{\dagger} \c1 a \,\, \c2 a^{\dagger}
\c1 a^{\dagger} a a}$
    \item $\wick{a^{\dagger} a^{\dagger} \c1 a \c2 a \,\, \c1 a^{\dagger} \c1 a \,\, \c1 a^{\dagger}
\c2 a^{\dagger} a a}$
    \item $\wick{a^{\dagger} a^{\dagger} \c1 a \c2 a \,\, \c1 a^{\dagger} \c1 a \,\, \c2 a^{\dagger}
\c1 a^{\dagger} a a}$
\end{itemize}

\noindent Again, utilizing the asymmetry of the $\delta U$ matrix element and re-labeling spin and
isospin projections we find that we only need to evaluate the first contraction above and multiply
the result by four. This gives

\begin{equation}\label{srg_UUdag}
\begin{gathered}
     \delta \hat{U}(\lambda) \,\hat{n}^{\tau}(\bm{p}, \sigma) \, \delta \hat{U}^{\dagger}(\lambda) = \\
     \frac{1}{4} \sum_{\substack{\sigma_1  \sigma_2 \sigma_3 \\ \sigma_4 \sigma \sigma'}} \sum_{\substack{\tau_1  \tau_2 \tau_3 \\ \tau_4 \tau'}} \int \frac{dP^+ d\bm{P}^\perp}{(2\pi)^3} \frac{d\alpha d\bm{k}^\perp}{(2\pi)^3 \sqrt{\alpha (2 - \alpha)}} \frac{d\alpha'd\bm{k}'^\perp}{(2\pi)^3 \sqrt{\alpha' (2 - \alpha')}} \frac{1}{p^+ (P^+ - p^+ )} \left( \frac{P^+}{2}\right)^2 \\
     \times \Braket{\alpha, \bm{k}^\perp, \sigma_1, \tau_1, \sigma_2, \tau_2|\delta U(\lambda)|\frac{2 p^+}{P^+}, \bm{p}^\perp - \frac{p^+}{P^+}\bm{P}^\perp, \sigma, \tau, \sigma', \tau'} \\
     \times \Braket{\frac{2 p^+}{P^+}, \bm{p}^\perp - \frac{p^+}{P^+}\bm{P}^\perp, \sigma, \tau, \sigma', \tau'|\delta U^{\dagger}(\lambda)|\alpha', \bm{k}'^\perp, \sigma_3, \tau_3, \sigma_4, \tau_4} \\
     \times \hat{a}^{\dagger}(p_1) \hat{a}^{\dagger}(p_2) \hat{a} (p_4) \hat{a}(p_3) .
\end{gathered}
\end{equation}

Now we evaluate all four terms with respect to the nuclear ground state

\begin{equation}
    \ket{\Psi_A^{MF}} = \prod_{i<F} \hat{a}_i^\dagger \ket{\Omega},
\end{equation}

\noindent where $a_i^\dagger$ creates a nucleon in single-particle orbital indexed by $i$,
$\ket{\Omega}$ is the interacting vacuum state, and $F$ refers to the Fermi surface. We perform a
change of basis, re-writing plane-wave creation and annihilation operators in terms of
single-particle orbitals

\begin{equation}\label{change_of_basis}
    \hat{a}(\bm{p},\sigma,\tau) = \sum_{i} \psi_{i}^+(\bm{p},\sigma,\tau) \,\hat{a}_{i}.
\end{equation}
Where $\psi^+_{i}$ is the overlap between the plus-projected single particle light-front
wavefunctions of nucleus A with eigenvalues $i$ and the LF Dirac spinor.
\begin{equation}
    \psi_{i}^+(\bm{p},\sigma,\tau) = \frac{1}{\sqrt{2p^+}}
    \bar{u}(\bm{p},\sigma,\tau)\gamma^+\psi_i(\bm{p}) = \sqrt{\frac{2}{p^+}}u^\dagger(\bm{p},\sigma,\tau)\psi_i^+(\bm{p})
\end{equation}
Contractions of $a_i$ with respect to the nuclear ground state are given by,

\begin{equation}\label{nuclear_contraction}
    \wick{\c1 a^\dagger_i \c1 a_j} = \braket{\Psi_A^{MF}|\hat{a}^\dagger_i \hat{a}_j|\Psi_A^{MF}} = \delta_{ij}
\end{equation}

\noindent for $i,j<F$. Using Eqs. (\ref{change_of_basis}, \ref{nuclear_contraction}) we immediately
get the following:

\begin{equation}
    \braket{\Psi_A^{MF}|\mathbb{\hat{I}} \,\hat{n}^{\tau}(\bm{p}, \sigma) \, \mathbb{\hat{I}}|\Psi_A^{MF}} = \sum_{i<F} |\psi_i^+(\bm{p},\sigma,\tau)|^2.
\end{equation}

\noindent The second (interference) term gives:
\begin{equation}
\begin{gathered}
     \braket{\Psi_A^{MF}|\delta\hat{U}(\lambda) \,\hat{n}^{\tau}(\bm{p}, \sigma) \, \mathbb{\hat{I}}|\Psi_A^{MF}} = \\
     \,\frac{1}{2}\sum_{\substack{\sigma_1 \sigma_2 \\ \sigma'}} \sum_{\substack{\tau_1  \tau_2 \\ \tau'}} \sum_{ij<F}\int \frac{dP^+ d\bm{P}^\perp}{(2 \pi)^3} \frac{d\alpha d\bm{k}^\perp}{(2\pi)^3 \sqrt{\alpha (2 - \alpha)}} \frac{1}{ \sqrt{p^+ (P^+ - p^+)}} \frac{P^+}{2}\\
    \times \Braket{\alpha, \bm{k}^\perp, \sigma_1, \tau_1, \sigma_2, \tau_2|\delta U (\lambda)|\frac{2p^+}{P^+}, \bm{p}^\perp - \frac{p^+}{P^+}\bm{P}^\perp, \sigma, \tau, \sigma', \tau'} \\
    \times \, \psi_i^{+\dagger}(p_1) \psi_j^{+\dagger}(p_2) \left[\psi_j^+(\bm{p}_4,\sigma',\tau') \psi_i^+(\bm{p},\sigma,\tau) - \psi_i^+(\bm{p}_4,\sigma',\tau') \psi_j^+(\bm{p},\sigma,\tau) \right]. \\
\end{gathered}
\end{equation}

\noindent The third (conjugate interference) term gives:
\begin{equation}
\begin{gathered}
    \braket{\Psi_A^{MF}|\mathbb{\hat{I}} \,\hat{n}^{\tau}(\bm{p}, \sigma) \, \delta \hat{U}^{\dagger}(\lambda)|\Psi_A^{MF}} = \\
    \frac{1}{2}\sum_{\substack{\sigma_3 \sigma_4 \\ \sigma'}} \sum_{\substack{\tau_3  \tau_4 \\ \tau'}} \sum_{ij<F}\int \frac{dP^+ d\bm{P}^\perp}{(2 \pi)^3} \frac{d\alpha' d\bm{k}'^\perp}{(2\pi)^3 \sqrt{\alpha' (2 - \alpha')}} \frac{1}{ \sqrt{p^+ (P^+ - p^+)}}\frac{P^+}{2}\\
    \times \Braket{\frac{2p^+}{P^+}, \bm{p}^\perp - \frac{p^+}{P^+}\bm{P}^\perp, \sigma, \tau, \sigma', \tau' | \delta U^{\dagger}(\lambda) |\alpha', \bm{k}'^\perp, \sigma_3, \tau_3, \sigma_4, \tau_4} \\
    \times \, \psi_{i}^{+ \dagger}(\bm{p},\sigma,\tau) \psi_{j}^{+ \dagger}(\bm{p}_2,\sigma',\tau') \left[ \psi_{j}^+(p_4) \psi_{i}^+(p_3) - \psi_{i}^+(p_4) \psi_{j}^+(p_3) \right]. \\
\end{gathered}
\end{equation}

\noindent The fourth (quadratic) term gives:
\begin{equation}\label{srg_UUdag_mf}
\begin{gathered}
     \braket{\Psi_A^{MF}|\delta\hat{U}(\lambda) \,\hat{n}^{\tau}(\bm{p}, \sigma) \, \delta \hat{U}^{\dagger}(\lambda)|\Psi_A^{MF}} = \\
     \frac{1}{4} \sum_{\substack{\sigma_1  \sigma_2 \sigma_3 \\ \sigma_4 \sigma \sigma'}} \sum_{\substack{\tau_1  \tau_2 \tau_3 \\ \tau_4 \tau'}} \sum_{ij < F} \int \frac{dP^+ d\bm{P}^\perp}{(2\pi)^3} \frac{d\alpha d\bm{k}^\perp}{(2\pi)^3 \sqrt{\alpha (2 - \alpha)}} \frac{d\alpha'd\bm{k}'^\perp}{(2\pi)^3 \sqrt{\alpha' (2 - \alpha')}} \frac{1}{p^+ (P^+ - p^+ )} \left( \frac{P^+}{2}\right)^2 \\
     \times \Braket{\alpha, \bm{k}^\perp, \sigma_1, \tau_1, \sigma_2, \tau_2|\delta U(\lambda)|\frac{2 p^+}{P^+}, \bm{p}^\perp - \frac{p^+}{P^+}\bm{P}^\perp, \sigma, \tau, \sigma', \tau'} \\
     \times \Braket{\frac{2 p^+}{P^+}, \bm{p}^\perp - \frac{p^+}{P^+}\bm{P}^\perp, \sigma, \tau, \sigma', \tau'|\delta U^{\dagger}(\lambda)|\alpha', \bm{k}'^\perp, \sigma_3, \tau_3, \sigma_4, \tau_4} \\
     \times \psi_{i}^{+ \dagger}(p_1) \psi_{j}^{+ \dagger}(p_2)\bigg[ \psi_{j}^{+}(p_4) \psi_{i}^{+}(p_3) - \psi_{i}^{+}(p_4) \psi_{j}^{+}(p_3) \bigg].
\end{gathered}
\end{equation}

For calculation of inclusive electron-nucleus scattering nuclear current matrix elements, we will
encounter two terms that involve the $\delta\hat{U}(\lambda)$ transformation. We write down the
results here.

\begin{equation}
\begin{gathered}
    \braket{\Psi_{i}^{MF}|\hat{a}(\bm{q}_1,s_1,t_1) \delta \hat{U}^\dagger(\lambda)|\Psi_A^{MF}} = \braket{\Psi_{A}^{MF}|\hat{a}_i^\dagger \hat{a}(\bm{q}_1,s_1,t_1) \delta \hat{U}^\dagger(\lambda)|\Psi_A^{MF}} = \\
    \frac{1}{2}\sum_{\sigma_2 \sigma_3 \sigma_4} \sum_{\tau_2 \tau_3  \tau_4}\sum_{j<F}\int \frac{dP^+ d\bm{P}^\perp}{(2 \pi)^3} \frac{d\alpha' d\bm{k}'^\perp}{(2\pi)^3 \sqrt{\alpha' (2 - \alpha')}} \frac{1}{ \sqrt{q^+ (P^+ - q^+)}}\frac{P^+}{2}\\
    \times \Braket{\frac{2q_1^+}{P^+}, \bm{q}_1^\perp - \frac{q_1^+}{P^+}\bm{P}^\perp, s_1, t_1, \sigma_2, \tau_2 | \delta U^{\dagger}(\lambda) |\alpha', \bm{k}'^\perp, \sigma_3, \tau_3, \sigma_4, \tau_4} \\
    \times \, \psi_{j}^{+ \dagger}(p_2) \left[ \psi_{j}^+(p_4) \psi_{i}^+(p_3) - \psi_{i}^+(p_4) \psi_{j}^+(p_3) \right]. \\
\end{gathered}
\end{equation}

\begin{equation}
\begin{gathered}
    \braket{\Psi_{ij}^{MF}|\hat{a}(\bm{q}_2,s_2,t_2) \hat{a}(\bm{q}_1,s_1,t_1) \delta \hat{U}^\dagger(\lambda)|\Psi_A^{MF}} = \\
    \braket{\Psi_A^{MF}|\hat{a}_i^\dagger \hat{a}_j^\dagger \hat{a}(\bm{q}_2,s_2,t_2) \hat{a}(\bm{q}_1,s_1,t_1) \delta \hat{U}^\dagger(\lambda)|\Psi_A^{MF}} = \\
    \frac{1}{2}\sum_{\sigma_3 \sigma_4} \sum_{\tau_3  \tau_4} \int \frac{d\alpha' d\bm{k}'^\perp}{(2\pi)^3 \sqrt{\alpha' (2 - \alpha')}} \frac{1}{ \sqrt{q_1^+ q_2^+}}\frac{q_1^+ + q_2^+}{2}\\
    \times \Braket{\frac{2q_1^+}{q_1^+ + q_2^+}, \bm{q}_1^\perp - \frac{q_1^+}{q_1^+ + q_2^+}(\bm{q}_1^\perp + \bm{q}_2^\perp), s_1, t_1, s_2, t_2 | \delta U^{\dagger}(\lambda) |\alpha', \bm{k}'^\perp, \sigma_3, \tau_3, \sigma_4, \tau_4} \\
    \times \, \left[ \psi_{j}^+(p_4) \psi_{i}^+(p_3) - \psi_{i}^+(p_4) \psi_{j}^+(p_3) \right]. \\
\end{gathered}
\end{equation}

% ==========   Bibliography

\nocite{*}   % include everything in the uwthesis.bib file

\bibliographystyle{unsrt}
\bibliography{uwthesis}

\end{document}